\documentclass [prb,preprint,superscriptaddress,floatfix,amsmath,amssymb] {revtex4-1}
\usepackage{amsmath}
\usepackage{graphicx}
\usepackage{hyperref}
\usepackage{color}
\usepackage{amssymb}
\usepackage{bm}

\newcommand{\beq} {\begin{equation}}
\newcommand{\eeq} {\end{equation}}
\newcommand{\bea} {\begin{eqnarray}}
\newcommand{\eea} {\end{eqnarray}}
\newcommand{\be} {\begin{equation}}
\newcommand{\ee} {\end{equation}}

\newcommand{\bo}{\bar \omega}
\DeclareMathOperator{\sgn}{sgn}

\begin{document}
\title {Interplay between superconductivity and non-Fermi liquid  at a quantum-critical point in a metal.\\
 I: The $\gamma$ model and its phase diagram at $T=0$.  The case $0 < \gamma <1$.}
\author{Artem Abanov}
\affiliation{Department of Physics, Texas A\&M University, College Station,  USA}
\author{Andrey V. Chubukov}
\affiliation{School of Physics and Astronomy and William I. Fine Theoretical Physics Institute,
University of Minnesota, Minneapolis, MN 55455, USA}
\date{\today}
\begin{abstract}
 Near a quantum-critical point in a metal a
 strong fermion-fermion interaction, mediated by a soft boson, acts in two different directions: it destroys fermionic coherence and it gives rise to an attraction in one or more pairing channels. The two tendencies compete with each other.  We analyze  a class of quantum-critical models,
  in which momentum integration  and the selection of a particular pairing symmetry  can be done explicitly, and the competition between non-Fermi liquid and pairing can be analyzed within an effective model with dynamical electron-electron interaction
 $V(\Omega_m) \propto 1/|\Omega_m|^\gamma$ (the $\gamma$-model). In this paper, the first in the series, we consider the case $T=0$ and $0<\gamma <1$. We argue that tendency to pairing is stronger, and the ground state is a superconductor. We argue, however, that a superconducting state
  is highly non-trivial as there exists a discrete set of topologically distinct solutions for the pairing gap $\Delta_n (\omega_m)$ ($n = 0, 1, 2..., \infty$).
   All solutions have
  the same spatial pairing symmetry, but differ in the time domain:
   $\Delta_n (\omega_m)$    changes sign $n$ times as a function of Matsubara frequency $\omega_m$.
      The $n =0$ solution $\Delta_0 (\omega_m)$ is sign-preserving and
      tends to a finite value at $\omega_m =0$, like in BCS theory.
        The $n = \infty$ solution
          corresponds to an  infinitesimally small  $\Delta (\omega_m)$, which oscillates  down to the lowest frequencies as  $\Delta (\omega_m) \propto  |\omega_m|^{\gamma/2} \cos \left[2\beta \log (|\omega_m|/\omega_0)\right]$, where $\beta  = O(1)$  and $\omega_0$ is of order of fermion-boson coupling.
    As a proof, we obtain the exact solution of the linearized gap equation at $T=0$ on the entire frequency axis for all $0<\gamma <1$,
     and an approximate solution of the non-linear gap equation.
     We argue that the presence of an infinite set of solutions opens up a new channel of gap fluctuations.
   We extend the analysis to the case where the pairing component of the interaction has additional factor $1/N$ and show that there exists a critical $N_{cr} >1$, above which superconductivity disappears, and the ground state becomes a non-Fermi liquid.
  We show that all solutions
    develop simultaneously once $N$ gets smaller than $N_{cr}$.
\end{abstract}
\maketitle

\section{ Introduction.}

 The interplay between superconductivity and pairing
 near a quantum-critical point (QCP) in a metal is a fascinating subject, which attracted  substantial attention in the correlated electron community after the discovery of superconductivity (SC) in
  the cuprates, heavy-fermion and organic materials,  and, more recently, Fe-pnictides and Fe-chalcogenides. (see, e.g., Refs.~\cite{scal,scal2,book1,review4,Norman_review,matsuda,Fernandes_2016,mack,Rice,review3,Sachdev2018,Coleman_book}).
   Itinerant QC models, analyzed analytically in recent years include, e.g.,  models of fermions in spatial dimensions $D \leq 3$ (Refs. \cite{senthil,max_last,raghu,*raghu2,*raghu3,*raghu4,*raghu5,Wang_H_18,Wang_H_17,Fitzpatrick_15}),
 two-dimensional (2D) models near  a spin-density-wave (SDW) and  charge-density-wave (CDW) instabilities (Refs. ~\cite{Millis1992,Altshuler1995a,Sachdev1995,acf,acs,*acs2,*finger_2001,*acn,Subir,*Subir2,*Sachdev2019,vojta,
wang,efetov,*efetov2,*efetov3,tsvelik,max2,sslee2,*sslee_2018,Bauer2015,ital,*ital2,*ital3,wang,*wang23,*wang_22,
khodas,*tsvelik,*tsvelik_1,*tsvelik_2,*tsvelik_3,Metzner2003,*DellAnna2006,metzner,*metzner_1,*metzner_2,tremblay_2,Wolfle2014}),
$2k_F$ density-wave instability~(Refs.\cite{2kf,*2kf2,*2kf3}), pair-density-wave instability
~\cite{Kim2008,Fradkin2007,*Fradkin2009,*Fradkin2014},  $q=0$ instabilities towards a Pomeranchuk order instability\cite{nick_b,*Oganesyan2001,*q=01,*q=02,
Metzner2003,*DellAnna2006,pepin,sslee,max,steve_sam,maslov1,*maslov2,*maslov3,*maslov4,Fradkin2016,triplet,*triplet2,*triplet3,review2,avi_1,*avi_2} and towards circulating currents~\cite{varma},
     2D fermions at a half-filled Landau level~\cite{PALee1989,*Monien1993,*Nayak1994,*Altshuler1994,*Kim_1994}, generic QC models with different critical exponents~\cite{raghu_15,moon_2,khvesh,She2009,*She2010,Wang2016,Kotliar2018,Wu_19,Wu_19_1,*Abanov_19,Emil2020},  Sachdev-Ye-Kitaev (SYK) and SYK-Yukawa models~\cite{Patel2019,Schmalian_19,Wang_19,schmalian_19a,Chowdhury_2020}, strong coupling limit of electron-phonon superconductivity,
     \cite{combescot,Bergmann,*Bergmann2,*ad,Marsiglio_88,*Marsiglio_91,Karakozov_91,Chubukov_2020b},  and even color superconductivity of quarks, mediated by gluon exchange\cite{son,*son2}.  These problems have also been studied using various numerical techniques~\cite{Yang2011,berg,*berg_2,*berg_3,*berg_4,kotliar,*kotliar2,tremblay_2,georges,*georges2}.

From theory perspective, the  key interest in the pairing near a QCP is due to the fact that an effective dynamic electron-electron interaction, $V(q, \Omega)$,
 mediated by a critical collective boson, which condenses at a QCP, provides strong attraction in one or more pairing channels and, at the same time,
  gives rise to non-Fermi liquid (NFL) behavior in the normal state.
  The two tendencies compete with each other: fermionic incoherence, associated with NFL behavior, destroys Cooper logarithm and reduces the tendency to pairing,  while the opening of a SC gap eliminates the scattering at  low energies  and reduces the tendency to NFL.
  To find the winner (SC or NFL), one needs to analyze  the set of integral equations for the fermionic self-energy, $\Sigma ({\bf k}, \omega)$, and the gap function, $\Delta ({\bf k}, \omega)$,
    for fermions with momentum/frequency $({\bf k}, \omega)$ and $(-{\bf k}, - \omega)$.

  We consider the subset of models in which collective bosons are slow modes compared to dressed fermions, for one reason or the other.
  In this situation, which bears parallels  with Eliashberg theory for
    electron-phonon interaction~\cite{eliashberg},  the self-energy and the pairing vertex can be approximated by their values at the Fermi surface (FS) and computed within one-loop approximation.
        The self-energy on the FS, $\Sigma ({\bf k}, \omega)$ is invariant under rotations from the point group of the underlying lattice. The rotational symmetry of the gap function  $\Delta ({\bf k}_F, \omega)$ and  the relation between the  phases of  $\Delta ({\bf k}_F, \omega)$ on different FS's in multi-band systems
      are model specific. Near a ferromagnetic QCP, the pairing interaction mediated by a soft boson is attractive in the p-wave channel.
      Near an antiferromagnetic  QCP, the strongest attraction is in $d-$wave channel for the case of a single FS and fermionic density near half-filling.
      For a nearly compensated metal with hole and electron pockets, as in Fe-based superconductors, the two attractive channels near an antiferromagnetic  QCP are $s^{+-}$ and $d-$wave.
       Near a $q=0$ nematic QCP, the pairing interaction, mediated by soft nematic fluctuations, is attractive in all channels:   $s-$wave, $p-$wave, $d-$wave, etc.
        In each particular case,  one has to project the pairing interaction into the proper irreducible channel, find the strongest one,  and solve for the pairing vertex within a given pairing symmetry.
   In principle, even after projection, one has to solve an infinite set of coupled equations in  momentum space as in a lattice system
    each irreducible representation contains an infinite set of eigenfunctions.  However, near a QCP the pairing is often confined to a narrow range on the FS around special "hot spots".
   In this situation, the momentum integration near the FS can be carried out exactly, and the set of coupled equations for the self-energy and the gap function reduce to two  one-dimensional (1D) equations for
    $\Sigma (\omega)$ and $\Delta (\omega)$, each with frequency-dependent effective  'local" interaction $V(\Omega)$ (Ref. \cite{acf}.
      The same holds  for the cases of s-wave pairing by a soft optical phonon, when momentum-dependencies of $\Sigma$ and $\Delta$  are not crucial and can be neglected,
        of non-s-wave pairing, when
        one eigenfunction gives the dominant contribution to the gap
     (e.g., $\cos k_x - \cos k_y$  for $d-$wave pairing in the cuprates), and for the case when  the fermionic density of states is peaked at particular ${\bf k}_F$ due to, e.g. van-Hove singularities ( see e.g., \onlinecite{review4} and references therein).

    Away from a QCP, the effective  $V(\Omega)$ tends to a finite value at $\Omega =0$. In this situation, the fermionic self-energy has a FL form at the smallest frequencies, the equation for $\Delta (\omega)$ is similar to that in a conventional Eliashberg theory for phonon-mediated superconductivity, and the only qualitative distinction  for electronically-mediated pairing is that $V(\Omega)$ by itself changes below $T_c$ due to feedback from fermionic pairing on collective modes.

\begin{figure}
	\begin{center}
\includegraphics[width=0.6\columnwidth]{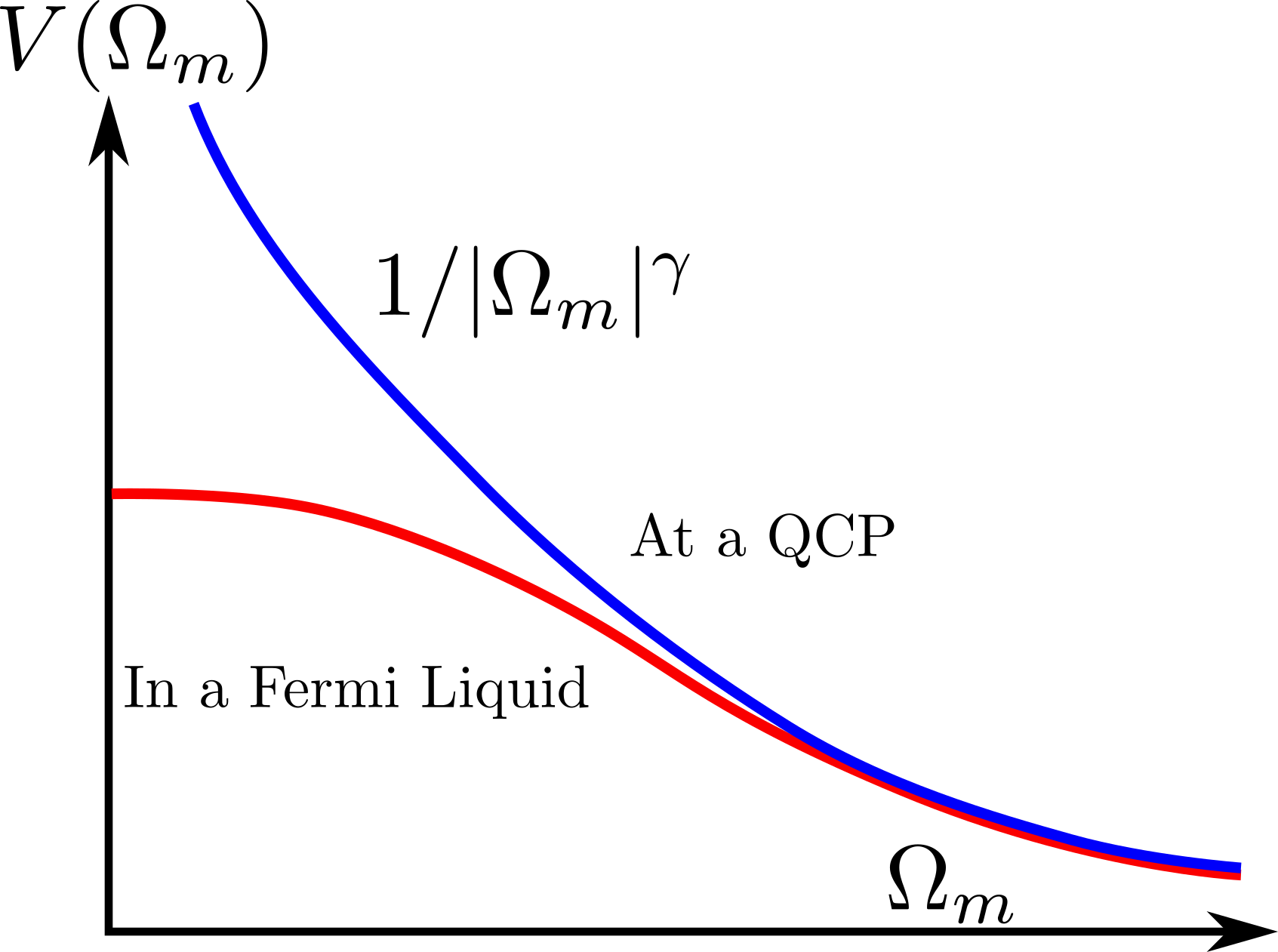}
		\caption{
 Frequency dependence of the effective interaction $V(\Omega_m)$, mediated by a soft boson, at $T=0$.
 Away from a QCP, $V(\Omega_m)$ tends to a finite value at $\Omega_m =0$.  Right at a QCP, the boson becomes massless, and $V(\Omega_m)$ diverges as $1/|\Omega_m|^\gamma$.}
\label{fig1}
	\end{center}
\end{figure}

     At a QCP, the situation becomes qualitatively different because the effective interaction $V(\Omega)$, mediated by a critical massless boson,  becomes a singular function of frequency: on Matsubara axis  $V(\Omega_m) \propto 1/|\Omega_m|^\gamma$ (Fig. \ref{fig1}). The exponent $\gamma>0$ depends on the model, ranging from small $\gamma = 0(\epsilon)$ in models in $D = 3 -\epsilon$ to $\gamma \leq 1$ in 2D models at SDW, CDW, and nematic QCP
      and in Yukawa-SYK model~\cite{Chubukov_2020a,review4,Chubukov_2020b,Schmalian_19,Wang_19}.
         The case $\gamma =2$ corresponds to fermions, interacting with a critical  Einstein phonon.
       The set of models with $V(\Omega_m) \propto 1/|\Omega_m|^\gamma$ has been nicknamed the $\gamma$-model, and we will use this notation.
        We present an (incomplete) set of models in Tables \ref{table_1}- \ref{table_3}.

\begin{figure}
	\begin{center}
\includegraphics[width=0.45\columnwidth]{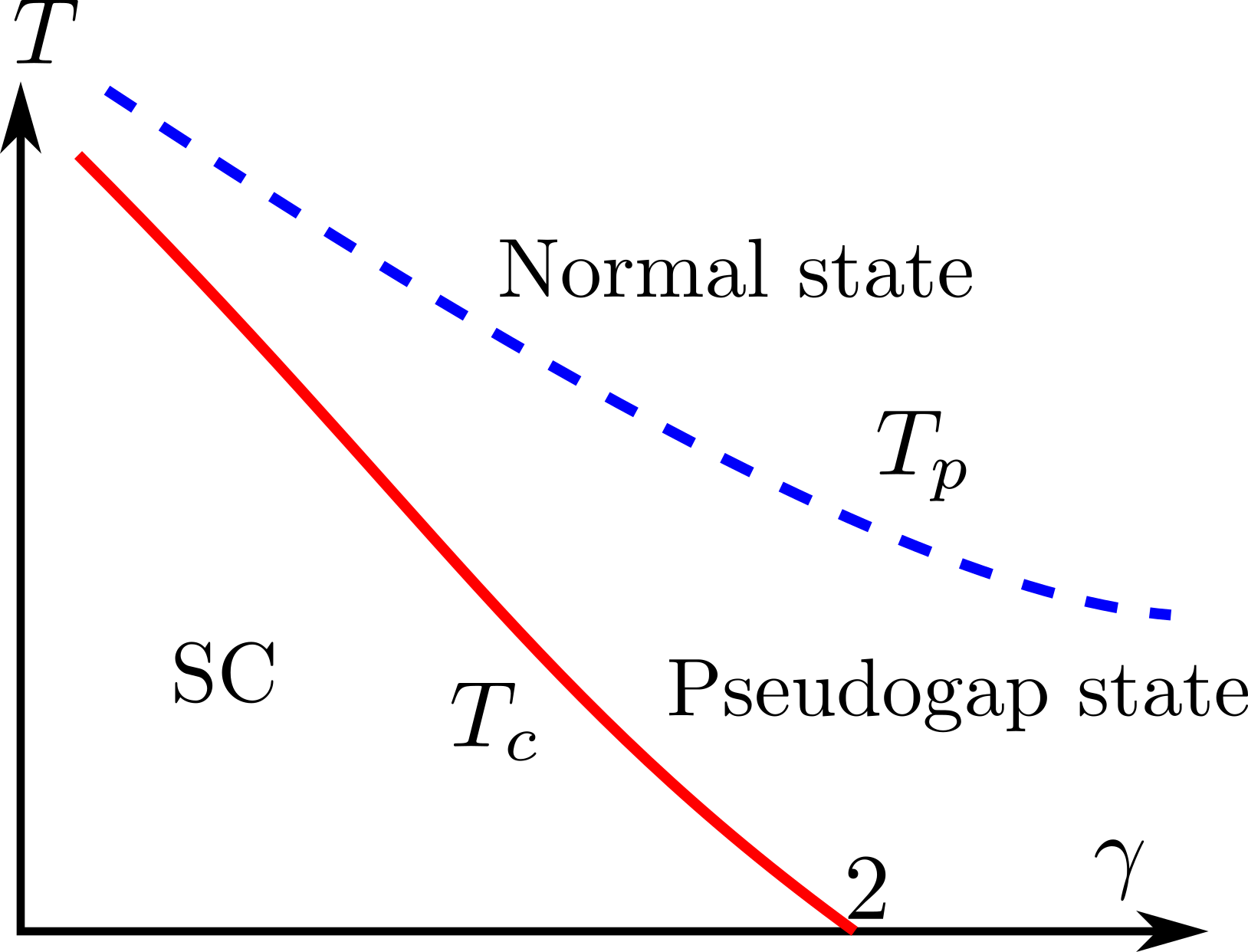}
		\caption{The  phase diagram of the $\gamma$-model. In this and subsequent papers we argue that for the pairing near a QCP the onset temperatures for the pairing and for long-range superconducting order, $T_p$ and $T_c$, differ. In between $T_p$ and $T_c$  the system displays pseudogap behavior.    The width of the pseudogap region increases with increasing $\gamma$ and extends to $T=0$ for  $\gamma \geq 2$.
 The behavior at larger $\gamma$ requires separate consideration~\cite{with_emil}. }
\label{fig1a}
	\end{center}
\end{figure}

       In this and subsequent papers, we  present comprehensive analysis of the competition between NFL and SC within the $\gamma-$model.  We define the dimensionless $V(\Omega_m)$ as $V(\Omega_m) = ({\bar g}/|\Omega_m|)^\gamma$, where ${\bar g}$ is the effective fermion-boson coupling. This ${\bar g}$ is the only parameter in the model with the dimension of energy, and we will see that it determines both the magnitude of the gap function $\Delta (\omega_m)$ and the upper limit
        of NFL behavior in the normal state, i.e., the scale below which $\Sigma (\omega_m) > \omega_m$.
 We show that the competition between NFL and SC  holds for all values of $\gamma$, but the physics and the computational analysis are different for the models with $\gamma <1$,
 $\gamma =1$,
$1 < \gamma <2$, $\gamma =2$, and
 $\gamma >2$, which we consider separately.
    We show that for all $\gamma$, a  NFL self-energy in the normal state does not prevent the formation of bound pairs of fermions at a non-zero onset temperature $T_p$.  However, we argue that $T_p$ is generally higher than the actual superconducting $T_c$, and at $T_c < T < T_p$  bound pairs remain incoherent.
In this $T$ range various  observables, e.g.,  the  spectral function and the density of states, display pseudogap behavior.  This holds for all $\gamma$, but the width of the pseudogap phase increases with $\gamma$.   We relate the pseudogap behavior to strong gap fluctuations, but argue that these fluctuations can be viewed as long-wavelength phase fluctuations only in some range ${\bar T}_p < T < T_p$. At smaller $T_c < T < {\bar T}_p$, pseudogap behavior is predominantly due to the existence of low-energy 'longitudinal" gap fluctuations, which change the functional form of $\Delta (\omega)$ and cause phase slips.  We  argue that
         longitudinal gap fluctuations develop a zero mode at $\gamma =2$ in which case SC order gets destroyed already at $T=0$.  Specifically, the presence of a zero mode implies that there exists an infinite number of solutions for $\Delta (\omega)$, all with the same condensation energy. In this situation, the ground state  is a mixture of different $\Delta (\omega)$, each with its own phase. We will argue that $T_c$ gradually vanishes at $\gamma  \to 2$ and remains zero at larger $\gamma$,  while $T_p$  and ${\bar T}_p$  stay finite.  We show the phase diagram in
         Fig. \ref{fig1a}.
 Our results present the 
scenario for the pseudogap, which holds even if  for a given solution for $\Delta (\omega)$ SC stiffness is larger than $T_c$, i.e., conventional phase fluctuations are weak. This scenario is complementary to the one in which the smallness of the superfluid stiffness is caused by the closeness to a Mott transition (see, e.g., Ref. \cite{Simard_2019} and references therein).

\subsection{A brief summary of the results of this work}

   In this paper, the first in the series, we analyze the $\gamma$-model at $T=0$ at $0 < \gamma < 1$.  For definiteness we focus on spin-singlet pairing. The analysis of spin-triplet pairing is  more involved because of different spin  factors in the self-energy and the pairing vertex~\cite{triplet,triplet2,triplet3}
In this paper we discuss even frequency pairing.
 The analysis of odd-frequency pairing~\cite{Balatsky_2020} is more involved and requires a separate consideration.

Our key goal is to understand the interplay between the competing tendencies toward pairing and toward NFL behavior. The first comes from the interaction in the particle-particle channel, the second from the interaction in the particle-hole channel.
 In the original $\gamma$-model, both interactions are given by the same $V(\Omega_m)$.  In order to separate the two tendencies we extend the model and introduce the knob to vary the relative strength of the interaction in the two channels. Specifically, we multiply the pairing interaction by $1/N$ and treat $N$ as a parameter.
   For $N >1$ ($N <1$) the tendency towards pairing decreases (increases)  compared to the one towards NFL ground state. The extension to integer $N >1$ can be formally justified by extending the model from $SU(1)$ to
 an $SU(N)$ global symmetry~\cite{raghu_15}. In our analysis, we use
 the extension to arbitrary $N \neq 1$ just as a computational trick to better  understand what happens in the physical case of $N =1$.

 \begin{figure}
	\begin{center}
		\includegraphics[width=0.6\columnwidth]{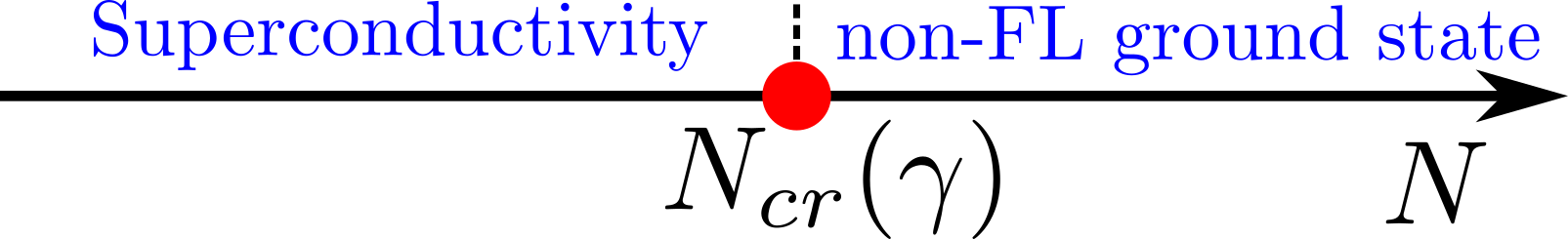}
		\caption{The $T=0$ phase diagram of the $\gamma$-model for $0<\gamma <1$ as a function of $N$. At smaller $N < N_{cr}$ the ground state is a superconductor. At larger $N > N_{cr}$ it is a NFL with no superconducting order. The value of $N_{cr}$ is larger than one.}
\label{fig3}
	\end{center}
\end{figure}
 The $T=0$ phase diagram of the $\gamma$-model for  $N \geq 1$ has been studied before~\cite{raghu_15,Wang2016}.   It was argued that for any $\gamma <1$ there exists a critical $N_{cr} >1$, separating a SC ground state at $N < N_{cr}$ and a normal, NFL ground state at $N > N_{cr}$ (see Fig.~\ref{fig3}). The physical case $N =1$ falls into the SC region.  A similar result has been recently found in the study of the pairing in the SYK-type model~\cite{schmalian_19a}.

 The conventional wisdom holds that  $\Delta (\omega)=0$ for $N > N_{cr}$,  is infinitesimally small for $N = N_{cr}$, and has a finite value
   for $N < N_{cr}$, i.e., that  the linearized gap equation has the solution at $N = N_{cr}$, and the non-linear gap equation has the solution for $N < N_{cr}$.
    We argue that in our case the conventional wisdom fails.  Namely, we prove that
  the linearized equation has a solution not only for $N = N_{cr}$, but also for all $N < N_{cr}$,  including the physical case of $N =1$. We obtain the exact solution of the linearized gap equation, $\Delta (\omega_m)$, for all $N < N_{cr}$ and all $0<\gamma \leq 1$. This $\Delta (\omega_m)$ oscillates at small $\omega_m \ll {\bar g}$ as $\Delta (\omega_m) \propto |\omega_m|^{\gamma/2} \cos\left(\beta_N \log ((|\omega_m|/{\bar g})^\gamma + \phi_N)\right)$,
   where $\beta_N$ and $\phi_N$ are particular functions of $N$ (for a given $\gamma$), e.g., $\beta_N \propto (N_{cr}-N)^{1/2}$  for $N \lesssim N_{cr}$.  At large $\omega_m \gg {\bar g}$,  $\Delta (\omega_m)$ decreases as $1/|\omega_m|^\gamma$ (see Fig. \ref{fig:fN}).

At small $\gamma$,  when  $\Delta (\omega_m)$ is a smooth function of frequency,
  the integral equation for $\Delta (\omega_m)$ can be approximated by  second-order differential equation, whose solution, $\Delta_{\text{diff}} (\omega_m)$, also exists for all $N \leq N_{cr}$ and can be expressed analytically as a combination of two complex-conjugated hypergeometric functions.
     This $\Delta_{\text{diff}} (\omega_m)$ also oscillates at small $\omega_m$ and decays as  $1/|\omega_m|^\gamma$ at large frequencies (see Fig. \ref{fig9}).
   We show that  $\Delta_{\text{diff}} (\omega_m)$ coincides with the exact $\Delta (\omega_m)$  to leading order in $\gamma$.

We then analyze the non-linear gap equation. We argue that it has an infinite, discrete set of solutions, specified by integer $n$, which ranges between $0$ and $\infty$. All solutions have the same spatial gap symmetry (i.e., $s$-wave, $d-$wave, etc).   A solution $\Delta_n (\omega_m)$ changes sign  $n$ times
  as a function of frequency. Each $\Delta_n (\omega_m)$ tends to a finite value at $\omega_m =0$,
     but the magnitude of $\Delta_n (0)$ progressively decreases with increasing $n$.
        At $n = \infty$, $\Delta_{\infty} (\omega_m)$ is the solution of the linearized gap equation, which  has an infinite number of sign changes due to $\cos(\beta_N  \log{(|\omega_m|/{\bar g})^\gamma} +\phi_N)$ oscillations running down to the smallest $\omega_m$.   The $n=0$  solution yields sign-preserving $\Delta_{0} (\omega)$.

  The existence of an infinite set of $\Delta_n (\omega)$,  each representing a local minimum of the Luttinger-Ward (LW) functional,  opens up a new channel of longitudinal gap fluctuations, which cause phase slips.
   As long as the set is discrete, there is a single global minimum of the LW functional. In our case, it corresponds to $n=0$. Then, at $T=0$ the system should display a SC order. Still, the existence of the infinite set of $\Delta_n (\omega_m)$ is a non-trivial aspect of the pairing at a QCP.
   In the next paper,  we analyze the $\gamma$-model for $0<\gamma <1$ at a finite $T$ and show explicitly that for any $N < N_{cr}$ there exists a discrete, infinite set of onset temperatures for the pairing, $T_{p,n}$, and the corresponding eigenfunctions $\Delta_n (\omega_m)$ change sign $n$ times as a function of Matsubara frequency. We argue  that at $T \to 0$, each $\Delta_n (\omega_m)$ approaches the $n-th$ solution of the non-linear integral gap equation at $T=0$.  We show that away from a QCP, only solutions with $n < n_{max}$ remain, where $n_{max}$ decreases with the deviation from a QCP.

We note in passing that a discrete set of  solutions for the gap function at $T=0$, with different {\it momentum} dependencies along the FS,  has been detected in the analysis of $\Delta ({\bf k}_F)$  near a nematic QCP,
 Refs. \cite{Yang2000,avi_1}
  There, however, the physics is different, and the number of solutions is finite at a QCP.  This is similar to our case away from a critical point.

 \begin{figure}
	\begin{center}
		\includegraphics[width=0.8\columnwidth]{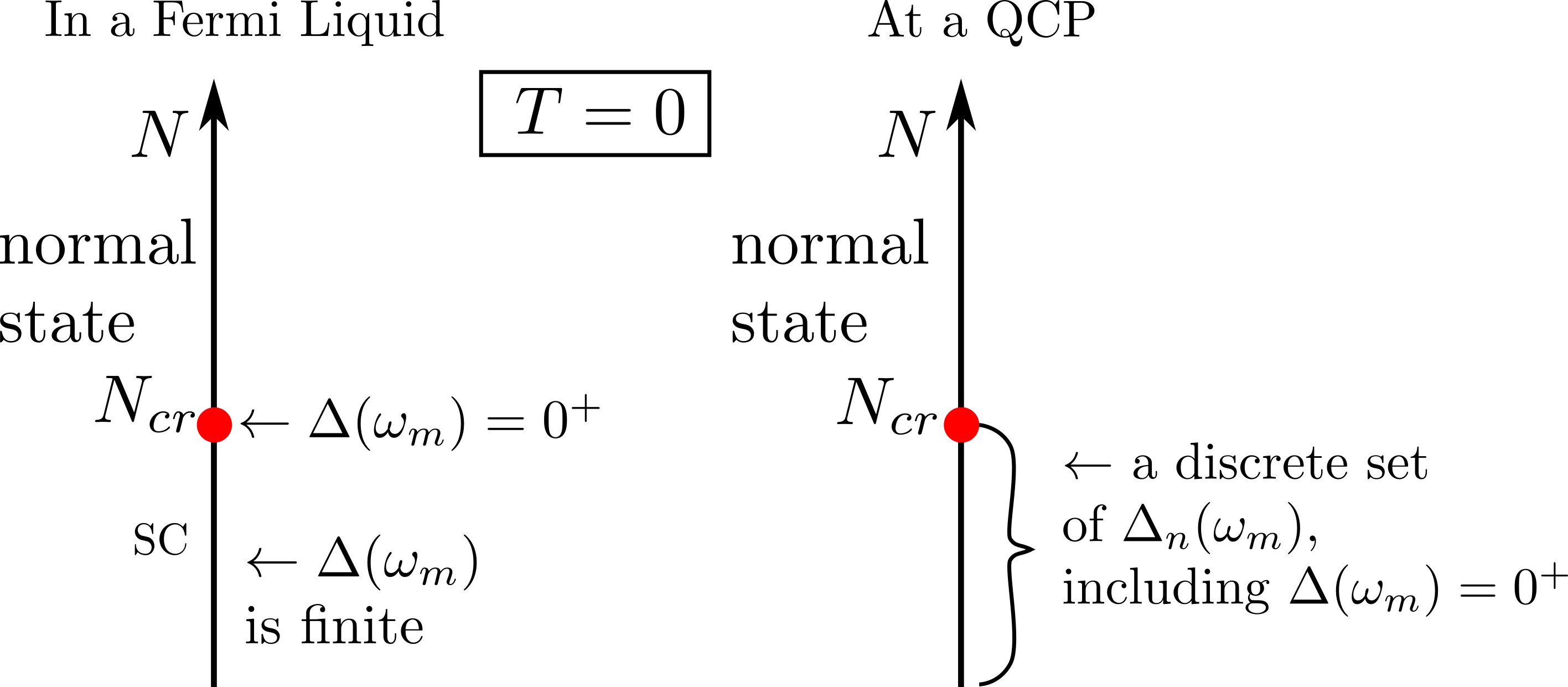}
		\caption{A phase diagram away from a QCP (left) and at a QCP (right). Away from a QCP, the gap equation  has a solution for infinitesimally small $\Delta (\omega_m)$  at $N = N_{cr}$, and  for a finite $\Delta (\omega_m)$ at $N < N_{cr}$, At a QCP, a solution with infinitesimally small gap function exists for all $N \leq N_{cr}$, and for $N < N_{cr}$ there is a discrete set of solutions for a finite $\Delta_n (\omega_m)$.  A gap $\Delta_n (\omega_m)$ changes sign $n$ times as a function of $\omega_m$.  }
\label{fig4}
	\end{center}
\end{figure}

\subsection{Relevance of the exact solution of the linearized gap equation}

The existence of an infinite  set of solutions of the non-linear gap equation is a direct consequence of a highly unusual result that at a QCP the linearized gap equation  has a solution not only at $N = N_{cr}$, but
 also for any $N < N_{cr}$ (see Fig. 4).  This  does not away from a QCP, where the linearized gap equation as a solution only at $N = N_{cr}$.  The  proof of the existence  of the solution for infinitesimally small $\Delta (\omega_m)$ for all $N \leq N_{cr}$, including the original $N=1$,  is, therefore, the central element of our analysis.

 \begin{figure}
	\begin{center}
		\includegraphics[width=0.8\columnwidth]{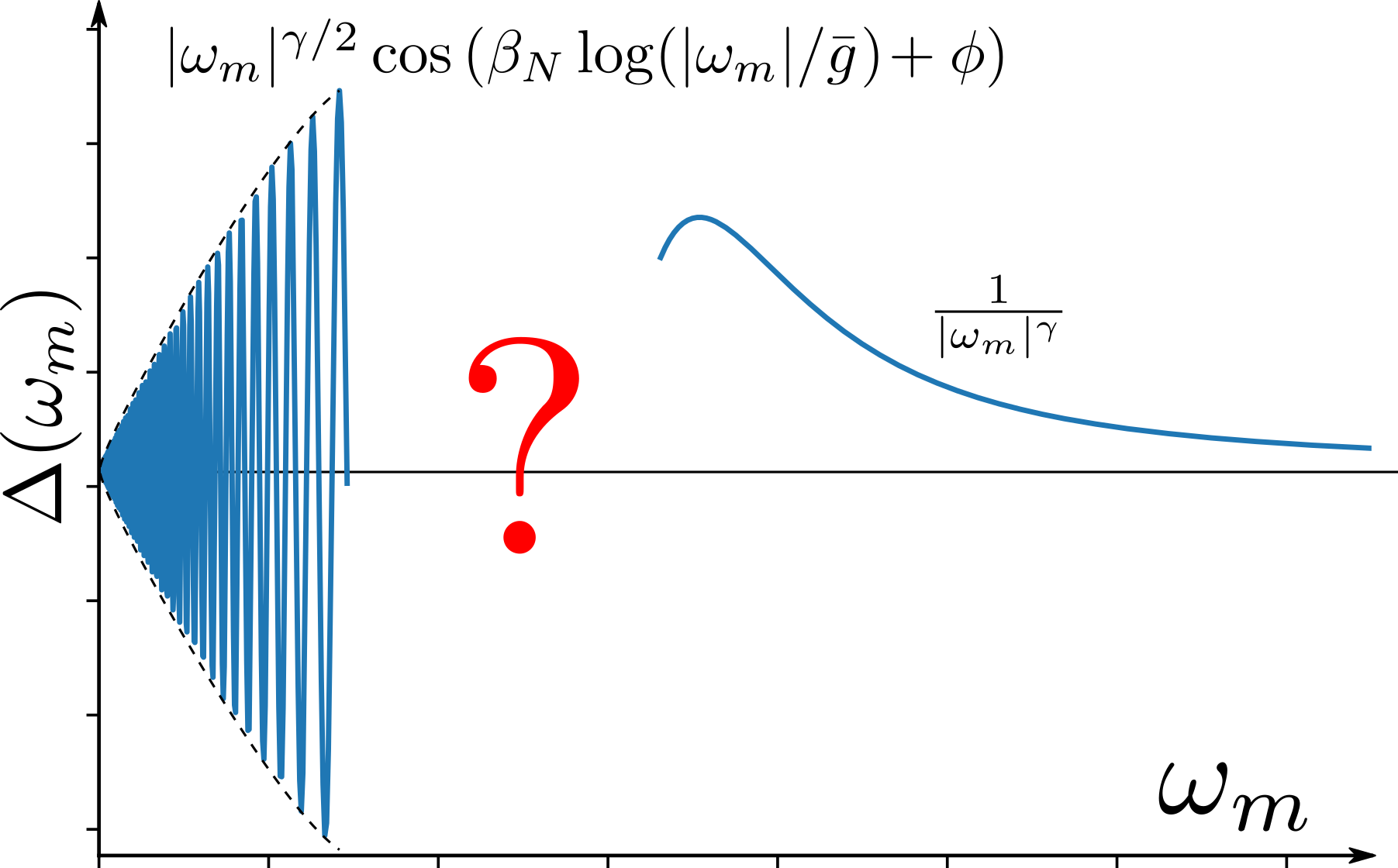}
		\caption{The gap equation for infinitesimally small $\Delta (\omega_m)$. The solution can be found separately at large and small frequencies. There is no guarantee, however, that for any $N \leq N_{cr}$ there exists the solution, which interpolates between the two limiting forms. }
\label{fig5}
	\end{center}
\end{figure}

 The linearized gap equation can be analyzed at frequencies much smaller and much larger than ${\bar g}$. In these two limits one can truncate the gap equation by  keeping only leading terms either in $\omega_m/{\bar g}$ or in ${\bar g}/\omega_m$. The truncated equation can be solved and yields $\Delta (\omega_m) \propto 1/|\omega_m|^\gamma$ at large $\omega_n$ and
  $\Delta (\omega_m) \propto |\omega_m|^{\gamma/2} \cos\left(\beta_N \log ((|\omega_m|/{\bar g})^\gamma + \phi\right)$ at small $\omega_m$,
     where the phase factor $\phi$
      is a free parameter.
        The generic task is then to verify whether by fixing $\phi$ one can find $\Delta (\omega_m)$, which smoothly interpolates between small and large $\omega_m$ (see Fig. \ref{fig5}).
        The verification would be straightforward if there was a finite frequency range where both forms were valid
 and could be made equal by fixing $\phi$.  In our case, however, the low-frequency and the high-frequency regimes do not overlap, so the only option is to
  solve the full equation.  We present the exact solution $\Delta_{\text{ex}} (\omega_m)$  and show that at small and large frequencies it reduces to known forms. The analytic expression for $\Delta_{\text{ex}} (\omega_m)$ is rather complex, but one can straightforwardly plot $\Delta_{\text{ex}} (\omega_m)$ for any input parameters (see Fig. \ref{fig:fN}).

 At small $\gamma$, where the interaction $V(\Omega_m) \propto 1/|\Omega_m|^\gamma$ is a slowly varying  function of frequency, the actual, integral gap equation can be approximated by  a differential equation.  The latter can be solved analytically, and the
 solution, $\Delta_{\text{diff}} (\omega_m)$,  is a hypergeometric function of $\omega$.
  Using the properties of a hypergeometric function at small and large values of the argument one can explicitly verify that $\Delta_{\text{diff}} (\omega_m)$ does interpolate between the known forms at small and large $\omega_m$ if one properly chooses the value of the phase $\phi$.

  We show that $\Delta_{\text{ex}} (\omega_m)$ and $\Delta_{\text{diff}} (\omega_m)$ coincide at small $\gamma$ and use the analytic form of $\Delta_{\text{diff}} (\omega_m)$ to analyze the structure of the gap function at different $N \leq N_{cr}$.
We demonstrate by the direct comparison that
at larger $\gamma$, $\Delta_{\text{ex}} (\omega_m)$ and $\Delta_{\text{diff}} (\omega_m)$ differ qualitatively.
 We argue that the series for the exact $\Delta_{\text{ex}} (\omega_m)$ contain non-local terms, not present for $\Delta_{\text{diff}} (\omega_m)$. These non-local terms are negligible at small $\gamma$, but must be kept for $\gamma = O(1)$, particularly near
  $\omega_m = \omega_{max}$, which separates oscillating behavior at smaller $\omega_m$ and $1/|\omega_m|^\gamma$ decay at larger $\omega_m$.

\subsection{Structure of the paper}

   The paper is organized as follows. In Sec.~\ref{sec:model} we briefly review the $\gamma$-model and extend it to $N \neq 1$.  We present the set of coupled Eliashberg equations  for the pairing vertex $\Phi (\omega_m)$ and the fermionic self-energy $\Sigma (\omega_m)$ and combine them into the equations for the gap function $\Delta (\omega_m)$ and the inverse quasiparticle  residue $Z(\omega_m)$.

 In Secs. \ref{sec:truncated} and \ref{sec:diff}  we study the linearized gap equation with infinitesimally small $\Delta (\omega_m)$. In Sec. \ref{sec:truncated} we analyze the truncated gap  equation at small and large $\omega_m$ and obtain the solutions, valid in the corresponding limits. Here we identify $N_{cr}$ and show that
the solution at small frequencies changes qualitatively between $N > N_{cr}$ and $N < N_{cr}$.
 In Sec. \ref{sec:diff} we consider the limit of small $\gamma$ and approximate the actual integral gap equation by  second-order differential equation.  We solve the
  differential equation for all $\omega_m$ and explicitly show how one can match low-frequency and high-frequency forms by fixing a single phase factor. We show that this only holds for $N \leq  N_{cr}$, while for $N > N_{cr}$ the potential solution does not  satisfy the condition of normalizability.
    In Sec. \ref{sec:exact_solution} we obtain the exact solution $\Delta_{\text{ex}} (\omega_m)$ of the full linearized gap equation.  We show that it exists for all $N < N_{cr}$.
  The solution, $\Delta_{\text{ex}} (\omega_m)$, oscillates at small frequencies and decays as $1/|\omega_m|^\gamma$ at large $|\omega_m|$.
  In Sec. \ref{sec:comparison} we analyze the structure of $\Delta_{\text{ex}} (\omega_m)$ and compare it with
   $\Delta_{\text{diff}} (\omega_m)$. We show that the two coincide at the smallest $\gamma$, but differ at $\gamma = O(1)$.  We argue that the difference is due to the fact that $\Delta_{\text{ex}} (\omega_m)$ contain non-local terms, not present in $\Delta_{\text{diff}} (\omega_m)$.
    In Sec. \ref{sec:nonlinear} we consider the non-linear gap equation. We first analyze the non-linear differential equation and argue that it has
    an infinite, discrete set of solutions $\Delta_{{\text{diff}},n} (\omega_m)$ for any $N < N_{cr}$. The index $n =0,1,2, ...$ specifies  the number of times $\Delta_{{\text{diff}},n} (\omega_m)$
      changes sign as a function of frequency. The solution with $n=0$ is sign-preserving. The solution  $n \to \infty$ coincides with the solution of the linearized differential equation.
       We conjecture that the actual, non-linear integral gap equation also possesses an infinite, discrete set of topologically distinct solutions $\Delta_n (\omega_m)$.   Sec. \ref{sec:summary} presents
        the summary of our results.

\subsection{Brief outline of subsequent papers}

     In the next paper in the series (Paper II, Ref. \cite{Paper_2})
      we consider the linearized gap equation for the same range  $0<\gamma <1$ at a finite $T$ and show that there exists an infinite discrete  set of critical temperatures $T_{p,n}$ for the pairing instability, all within the same pairing symmetry.  The corresponding eigenfunction $\Delta_n (\omega_m)$  changes sign $n$  times as a function of discrete Matsubara frequency $\omega_m = \pi T (2m+1)$. We argue
       that at $T \to 0$  these  finite $T$ solutions become $\Delta_n (\omega_m)$, which we find here in the $T=0$ analysis.

For $\gamma <1$,  the sign-preserving solution $\Delta_0 (\omega_m)$ has the largest condensation energy and the largest  $T_{p,0}$.
 Still,  the presence of an infinite set of $\Delta_n (\omega_m)$ at $T=0$ is not only highly unusual, but creates a new channel of "longitudinal" gap fluctuations.
 In  subsequent papers
 we will focus on the physical case $N=1$ and extend the analysis at $T=0$  to $\gamma >1$.  We will show that  the set  $\Delta_n (\omega)$ becomes more dense with increasing $\gamma$ and eventually becomes continuous  at $\gamma =2$. For this special $\gamma$,  all solutions of the non-linear gap equation with finite $n$ have equal condensation energy, and long-range superconducting  order  at $T=0$  gets destroyed upon averaging over these solutions, each with its own phase.  We will corroborate this by the analysis of the gap equation for the same $\gamma$ at finite $T$ and obtain the phase diagram, shown in  Fig. \ref{fig1a}.
Later, we will  show~\cite{with_emil} that the system behavior is rather special for larger $\gamma$, particularly for $\gamma >3$.

\section{$\gamma$-model.}
\label{sec:model}

 We  consider  itinerant fermions  at the onset
of a  long-range order in either spin or charge channel.  At the critical point, the propagator of a soft boson
 becomes massless and
 mediates singular interaction between fermions. We follow earlier works~\cite{acf,acs,moon_2,max,senthil,scal,efetov,max_last,raghu_15,haslinger,Wang2016,Kotliar2018}  and assume that this interaction is attractive in at least one pairing channel and that a pairing
  boson can be treated as slow mode compared to a fermion, i.e., at a given momentum $q$, typical fermionic frequency is much larger than  typical bosonic frequency. This is the case  for a conventional phonon-mediated superconductivity, where
    for $q \sim k_F$ a typical fermionic frequency is of order $E_F$, while typical bosonic frequency is of order Debye frequency $\omega_D$.  The ratio  $\delta_E =\omega_D/E_F$ is the  small parameter for Eliashberg
     theory of phonon-mediated superconductivity. This theory allows one to obtain a set of coupled integral
 equations for frequency dependent fermionic self-energy and the pairing vertex.  By analogy,  the theory of electronic superconductivity, mediated by  soft collective bosonic excitations in spin or charge channel, is also often called Eliashberg theory.  We will use this convention.

Within the Eliashberg  approximation,
 one can explicitly integrate over the momentum component perpendicular to the Fermi surface (for a given pairing symmetry) and reduce the
   pairing problem  to a set of coupled integral equations for frequency dependent self-energy $\Sigma (\omega_m)$
   and the pairing vertex $\Phi (\omega_m)$ with effective frequency-dependent dimensionless interaction $V(\Omega) = ({\bar g}/|\Omega|)^\gamma$.   This interaction gives rise to NFL form of the self-energy in the normal state and, simultaneously,
    gives rise to the pairing.

  At $T=0$, the coupled Eliashberg equations for the pairing vertex and the fermionic self-energy are, in Matsubara formalism,
    \bea \label{eq:gapeq}
    \Phi (\omega_m) &=&
     \frac{{\bar g}^\gamma}{2} \int d \omega'_{m} \frac{\Phi (\omega'_{m})}{\sqrt{{\tilde \Sigma}^2 (\omega'_{m}) +\Phi^2 (\omega'_{m})}}
    ~\frac{1}{|\omega_m - \omega'_{m}|^\gamma}, \nonumber \\
     {\tilde \Sigma} (\omega_m) &=& \omega_m
   +  \frac{{\bar g}^\gamma}{2}  \int d \omega'_{m}  \frac{{\tilde \Sigma}(\omega'_m)}{\sqrt{{\tilde \Sigma}^2 (\omega'_{m})  +\Phi^2 (\omega'_{m})}}
    ~\frac{1}{|\omega_m - \omega'_{m}|^\gamma}
\eea
 where ${\tilde \Sigma}(\omega_{m}) = \omega_m + \Sigma (\omega_m)$. In these equations, both $\Sigma (\omega_{m})$ and $\Phi (\omega_{m})$ are real functions.
  Observe that we define $\Sigma (\omega_m)$ with the overall plus sign and  without the overall factor of $i$. \footnote{In obtaining the Eliashberg equations we assumed that states away from the Fermi surface are irrelevant for the pairing and NFL behavior, and extended the integration over momenta to infinite limits. In this situation, fermionic self-energy
  $\Sigma (\omega_m)$ is real.  If the momentum integration over $k-k_F$ holds in finite and non-symmetric limits,  $\Sigma (\omega_m)$  has both real and imaginary components.  The same holds in SYK-type models at a non-zero chemical potential (see Ref. \protect\cite{Gu_2020} and references therein). }
  In the normal state ($\Phi \equiv 0$),
  \beq
  {\tilde \Sigma} (\omega_m) = \omega_m + \omega_0^\gamma |\omega_m|^{1-\gamma} \sgn{\omega_m},
\label{ss_111_a}
\eeq
 where $\omega_0 = {\bar g}/(1-\gamma)^{1/\gamma}$.   At small $\gamma$, $\omega_0 = {\bar g} e$.

The  superconducting gap function $\Delta (\omega_m)$ is defined as a real function
\beq
 \Delta (\omega_m) = \omega_m  \frac{\Phi (\omega_m)}{{\tilde \Sigma} (\omega_m)} = \frac{\Phi (\omega_m)}{1 + \Sigma (\omega_m)/\omega_m}.
  \label{ss_1}
  \eeq
   The equation for $\Delta (\omega_{m})$ is readily obtained from (\ref{eq:gapeq}):
   \beq
   \Delta (\omega_m) = \frac{{\bar g}^\gamma}{2}  \int d \omega'_{m}  \frac{\Delta (\omega'_{m}) - \Delta (\omega_m) \frac{\omega'_{m}}{\omega_m}}{\sqrt{(\omega'_{m})^2 +\Delta^2 (\omega'_{m})}}
    ~\frac{1}{|\omega_m - \omega'_{m}|^\gamma}.
     \label{ss_11}
  \eeq
   This equation contains a single function $\Delta (\omega_{m})$, but  for the cost that $\Delta (\omega_m)$ appears also in the r.h.s., which makes Eq. (\ref{ss_11}) less convenient for the analysis than Eqs. (\ref{eq:gapeq}).

  Eqs. (\ref{eq:gapeq})-(\ref{ss_11})  describe color superconductivity~\cite{son,*son2} and pairing in 3D ($\gamma = 0_+$, $ V (\Omega_m) \propto \log{|\omega_m|}$),  spin- and charge-mediated pairing in $D=3-\epsilon$ dimensions~\cite{senthil,max_last,raghu_15} and superconductivity in graphene~\cite{khveshchenko} ($\gamma = O(\epsilon) \ll 1$),  a 2D pairing ~\cite{2kf}  with  interaction peaked at $2k_F$ ($\gamma =1/4$),   pairing at a 2D nematic/Ising-ferromagnetic QCP~\cite{nick_b,steve_sam,triplet,*triplet2,*triplet3} ($\gamma =1/3$),   pairing at a 2D $(\pi,\pi)$ SDW QCP~\cite{Millis1992,acf,acs,wang} and an incommensurate CDW QCP~\cite{ital,*ital2,*ital3,wang_2,*wang_22,*wang23} ($\gamma =1/2$),
   dispersionless fermions randomly interacting with an Einstein phonon~\cite{Schmalian_19,Wang_19,schmalian_19a} and a spin-liquid model for the cuprates~\cite{tsvelik} ($\gamma =0.7$)
  a 2D pairing  mediated by an undamped  propagating boson ($\gamma =1$),  pairing in several Fe-based superconductors~\cite{kotliar} ($\gamma =1.2$) and even
  the strong coupling limit of phonon-mediated superconductivity for either dispersion-full~\cite{combescot,Bergmann,*Bergmann2,*ad,Marsiglio_88,*Marsiglio_91,Karakozov_91} or dispersion-less~\cite{Schmalian_19,Wang_19} fermions ($\gamma =2$).  The pairing models with parameter-dependent $\gamma$ have been analyzed as well (Refs. \onlinecite{Subir,moon_2}).   The case $\gamma =0$ describes a BCS superconductor.  We list some of the model in Tables \ref{table_1} - \ref{table_3}.

 A justification of Eliashberg theory for electronically mediated superconductivity (e.g., the reasoning to neglect vertex corrections) is case specific. For Ising-nematic fluctuations (the case $\gamma =1/3$), vertex corrections are small in ${\bar g}/E_F$, which is a small parameter of the theory~\cite{Metzner2003,*DellAnna2006,maslov3,avi_last,*avi_last_1}.
    In other cases, e,g., in SYK models, a small parameter for Eliashberg  approximation is $1/N$, where $N$ is the number of of fermionic flavors~\cite{Patel2019,Schmalian_19,Wang_19,schmalian_19a,Chowdhury_2020}.  For several 2D models, the corrections to Eliashberg approximation for the self-energy  in the normal state are logarithmically singular and in the absence of the pairing would change the system behavior at the smallest frequencies~\cite{acs,max2,sslee2,*sslee_2018}. One-loop logarithmic corrections just change $\gamma$ but keep the model intact (Ref. \cite{acs}), but higher-order corrections go beyond the $\gamma$ model
      (Refs. \cite{max2,sslee2,*sslee_2018}). Here, we assume that the onset temperature for the pairing, $T_p$, is larger, at least numerically, than the scale at which corrections to Eliashberg approximation become relevant, and stick with the Eliashberg theory.
    Recent Quantum Monte Carlo (QMC) calculations for superconductivity, mediated by antiferromagnetic spin fluctuations~\cite{berg_3} and Ising-nematic fluctuations~\cite{avi_last}, have found the onset of superconductivity at a temperature, almost identical to the one in the $\gamma$ model with $\gamma =1/2$ and $1/3$, respectively, without vertex corrections. Normal state QMC calculations for the models with antiferromagnetic and Ising-nematic fluctuations also found~\cite{avi_last,*avi_last_1} very good agreement with the behavior of the $\gamma$ models with $\gamma =1/2$ and $1/3$.  A good agreement has been also found between
     QMC calculations of superconducting $T_c$ for electron-phonon interaction and the $\gamma$-model with $\gamma =2$ (Ref. \cite{Chubukov_2020b}), for couplings smaller than the Fermi energy.

  \newcommand{\centered}[1]{\begin{tabular}{l} #1 \end{tabular}}
\begin{table}[ht]
\caption{Examples of pairing near quantum-critical point in $D=3$ and $D =3-\epsilon$.}
\begin{tabular}{|@{}l@{}|@{}c@{}|@{}c@{}|@{}c@{}|}
\hline
\hline
\centered{Model/Order\\}
&
\centered{$V(q,\Omega_{m})$\\}
&
\centered{$V(\Omega_{m})=\int d^{2}q V(q,\Omega_{m})$\\}
&
\centered{$\gamma $\\}
\\
\hline
\centered{
Ising-nematic order, \\
Ising FM, \\ }
&
\centered{$\frac{1}{q^{2}+\Gamma \frac{|\Omega_{m}|}{q}}$\\}
&
\centered{$\log \frac{1}{|\Omega_{m} |}$ \\}
&
\centered{$0+$ \\}
\\
\hline
\centered{
SDW/CDW order, \\
hot-spot models \\}
&
\centered{$\frac{1}{q^{2}+\Gamma |\Omega_{m}|}$\\}
&
\centered{$\log \frac{1}{|\Omega_{m} |}$ \\}
&
\centered{$0+$ \\}
\\
\hline
\centered{
Anisotropic Ising-nematic  \\}
&
\centered{$\frac{1}{q^{2+\epsilon }+\Gamma \frac{|\Omega_{m}|}{|q|}}$\\}
&
\centered{$\frac{1}{|\Omega_{m} |^{\frac{\epsilon }{3+\epsilon }}}$ \\}
&
\centered{$\frac{\epsilon }{3+\epsilon }$ \\}
\\
\hline
\centered{
Anisotropic SDW/CDW \\}
&
\centered{$\frac{1}{q^{2+\epsilon }+\Gamma |\Omega_{m}|}$\\}
&
\centered{$\frac{1}{|\Omega_{m}|^{\frac{\epsilon}{2+\epsilon }} }$ \\}
&
\centered{$\frac{\epsilon}{2+\epsilon }$ \\}
\\
\hline
\end{tabular}
\label{table_1}
\end{table}

\begin{table}[ht]
\caption{Examples of  pairing near quantum-critical point in $2D$.}
\begin{tabular}{|@{}l@{}|@{}c@{}|@{}c@{}|@{}c@{}|}
\hline
\hline
\centered{Model/Order\\}
&
\centered{$V(q,\Omega_{m})$\\}
&
\centered{$V(\Omega_{m})=\int dq V(q,\Omega_{m})$\\}
&
\centered{$\gamma $\\}
\\
\hline
\centered{
Ising-nematic order, \\
Ising FM, \\
fermions at $1/2$ filled Landau level \\ }
&
\centered{$\frac{1}{q^{2}+\Gamma \frac{|\Omega_{m}|}{q}}$\\}
&
\centered{$\frac{1}{|\Omega_{m} |^{2/3}}$ \\}
&
\centered{$2/3$ \\}
\\
\hline
\centered{
SDW/CDW order, \\
hot-spot models \\}
&
\centered{$\frac{1}{q^{2}+\Gamma |\Omega_{m}|}$\\}
&
\centered{$\frac{1}{|\Omega_{m} |^{1/2}}$ \\}
&
\centered{$1/2$ \\}
\\
\hline
\centered{
Undamped fermions \\}
&
\centered{$\frac{1}{q^{2}+\Omega_{m}^{2}}$\\}
&
\centered{$\frac{1}{|\Omega_{m} |}$ \\}
&
\centered{$1$ \\}
\\
\hline
\centered{
$Fe$-based superconductors \\}
&
\centered{\\}
&
\centered{$\frac{1}{|\Omega_{m}|^{1.2} }$ \\}
&
\centered{$1.2$ \\}\\
\hline
\end{tabular}
\label{table_2}
\end{table}

\begin{table}[ht]
\caption{Examples of pairing due to dispersionless phonons.}
\begin{tabular}{|@{}l@{}|@{}c@{}|@{}c@{}|@{}c@{}|}
\hline
\hline
\centered{Model/Order\\}
&
\centered{$V(q,\Omega_{m})$\\}
&
\centered{$V(\Omega_{m})=\int dq V(q,\Omega_{m})$\\}
&
\centered{$\gamma $\\}
\\
\hline
\centered{
Pairing by a soft \\
Einstein phonon \\}
&
\centered{$\frac{1}{\Omega_{m}^{2}}$\\}
&
\centered{$\frac{1}{\Omega_{m}^{2} }$ \\}
&
\centered{$2$ \\}
\\
\hline
\centered{
SYK-model with phonons. \\
Weak coupling \\}
&
\centered{$\frac{1}{|\Omega_{m}|^{0.6}}$\\}
&
\centered{$\frac{1}{|\Omega_{m}|^{0.6} }$ \\}
&
\centered{$0.6$ \\}
\\
\hline
\end{tabular}
\label{table_3}
\end{table}

In this paper,  we consider the set of $\gamma$-models with  $0<\gamma < 1$.
  The analysis for $\gamma \geq 1$ requires  separate consideration because of divergencies in the r.h.s. of both equations in (\ref{eq:gapeq}) (but not  in (\ref{ss_11})) and  will be presented in  subsequent papers.

Notice that for any $\gamma >0$ the pairing interaction $V (\Omega)$ decays at large frequencies as $1/|\Omega|^\gamma$.  A simple experimentation shows that the integrals in (\ref{eq:gapeq}) and (\ref{ss_11}) are then convergent in the  ultraviolet.  The only exceptions are the BCS case $\gamma =0$, when $V (\Omega)$ is just a constant,  and the case   $V (\Omega_m) \propto \log{|\omega_m|}$, i.e., $\gamma = 0_+$
(color superconductivity/pairing in 3D, Refs. \cite{son,*son2,max_last}). For these two cases one needs to set the upper cutoff of frequency integration at some $\Lambda$ to cut ultra-violet divergence.
 We discuss the limit $\gamma \to 0$ in Appendix \ref{app:gamma_to_0}.

The full set of Eliashberg equations for electron-mediated pairing contains also the equation describing the feedback from the pairing on the bosonic propagator. This feedback is small by $\delta_E$ in the case  of electron-phonon interaction, but
 is  generally not small when the pairing is mediated by a collective mode because the dispersion of a collective mode may change qualitatively below $T_p$. The most known example of this kind is the transformation of Landau overdamped
   spin collective mode in the normal state to a
  propagating mode (often called a resonance mode) below the onset of  $d-$wave pairing mediated by antiferromagnetic spin fluctuations (see, e.g., Refs.\cite{Millis_1996,Eschrig_2006,abanov-Norman,abanov-ARPES-2}).
 To avoid additional complications, we do not include this feedback explicitly into our consideration.  In general, the feedback from the pairing makes bosons less incoherent and can be modeled by assuming that
  $\gamma$ moves towards a larger value as $T$ decreases
  ( e.g., from $\gamma =1/2$ to $\gamma =1$ for the case of antiferromagnetic fluctuations and from $\gamma =1/3$ to $\gamma =2/3$ for the case of Ising-nematic fluctuations).

The coupled equations for $\Phi$ and ${\tilde \Sigma}$,   Eq.  (\ref{eq:gapeq}), describe the interplay between the two competing tendencies -- one towards superconductivity, specified by $\Phi$,  and the other towards incoherent non-Fermi liquid behavior, specified by ${\tilde \Sigma}$.  The competition between the two tendencies is encoded in the fact that ${\tilde \Sigma}$ appears in the denominator of the equation for $\Phi$ and $\Phi$ appears in the denominator of the equation for ${\tilde \Sigma}$.
 In more physical terms a self-energy ${\tilde \Sigma}$ is an obstacle to Cooper pairing, while if $\Phi$  is non-zero, it reduces the strength of the self-energy and moves the system back into a FL regime.

In Eq. (\ref{eq:gapeq}) the couplings in the particle-particle and particle-hole channels have the same magnitude ${\bar g}$.  To study the interplay, it is convenient to have a parameter, which would increase either the tendency towards NFL or towards pairing.
 With this in mind,  we multiply the coupling in the particle-particle channel by a factor $1/N$, i.e., set it to be ${\bar g}^\gamma/N$ instead of $\bar{g}^\gamma$, and keep the coupling in the particle-hole channel intact. For $N <1$, the tendency towards pairing is enhanced, for $N >1$ the tendency towards NFL effectively gets larger. We will treat $N$ as a free parameter, but use the extension as just a way to see the relevant physics more clearly, as our ultimate goal is to understand system behavior in the physical case of $N =1$.
    We note in passing that the  extension to integer $N >1$  can be formalized by extending the original model to matrix  $SU(N)$ model~\cite{raghu_15}.

The modified equations for $\Phi (\omega_m)$ and ${\tilde \Sigma} (\omega_m)$ are
 \bea
   && \Phi (\omega_m) =
    \frac{{\bar g}^\gamma}{2N}  \int d \omega'_m  \frac{\Phi (\omega'_{m})}{\sqrt{{\tilde \Sigma}^2 (\omega'_{m}) +\Phi^2 (\omega'_{m})}}
    ~\frac{1}{|\omega_m - \omega'_{m}|^\gamma}, \label{eq:gapeq_1} \\
   &&  {\tilde \Sigma} (\omega_m) = \omega_m
   +  \frac{{\bar g}^\gamma}{2} \int d \omega'_m
      \frac{{\tilde \Sigma} (\omega'_m)}{\sqrt{{\tilde \Sigma}^2 (\omega'_{m})  +\Phi^2 (\omega'_{m})}}
    ~\frac{1}{|\omega_m - \omega'_{m}|^\gamma}, \label{eq:gapeq_1_1}
\eea
 and the equation for $\Delta (\omega_m)$ becomes
 \beq
   \Delta (\omega_m) = \frac{ {\bar g}^\gamma }{2N} \int d \omega'_m  \frac{\Delta (\omega'_{m}) -N  \Delta (\omega_m) \frac{\omega'_{m}}{\omega_m}}{\sqrt{(\omega'_{m})^2 +\Delta^2 (\omega'_{m})}}
    ~\frac{1}{|\omega_m - \omega'_{m}|^\gamma}.
     \label{ss_111}
  \eeq
Below we will occasionally refer to the equation
  for
  $\Phi (\omega_m)$ as the gap equation, notwithstanding that the  gap equation is given by Eqs. (\ref{ss_111}).
    Indeed, once we know $\Phi (\omega_m)$ and ${\tilde \Sigma} (\omega_m)$, we also know $\Delta (\omega_m) = \Phi (\omega_m) \omega_m/{\tilde \Sigma} (\omega_m)$.

We will be searching for normalized solutions for $\Phi(\omega_m)$ and $\Delta (\omega_m)$.   Physically, the  normalizability of a solution  follows from the requirement that the Free energy of a superconductor should be free from divergencies.
 The Free energy of a superconductor, $F_{sc}$, can be obtained by either applying Hubbard-Stratonovich formalism~~\cite{Emil2020} or by using a generic Luttinger-Ward-Eliashberg expression~\cite{lw,eliashberg}, and explicitly integrating over momentum, approximating the density of states by its value at the Fermi level~\cite{haslinger,Wu_19_1}. The result is
    \begin{widetext}
 \beq
 \frac{F_{sc}}{N_{0}} = - \int d\omega_m \frac{\omega^2_m}{\sqrt{\omega^2_m + \Delta^2 (\omega_m) }}
- \frac{{\bar g}^\gamma}{4} \int d \omega_m d \omega'_m \frac{\omega_m \omega'_{m} + \frac{1}{N} \Delta (\omega_m) \Delta (\omega'_m)}{\sqrt{\omega^2_m + \Delta^2 (\omega_m) } \sqrt{(\omega'_{m})^2 + \Delta^2 (\omega'_m)}} \frac{1}{|\omega_m-\omega'_{m}|^\gamma}.
\label{a_5}
 \eeq
 \end{widetext}
The gap equation (\ref{ss_111}) is obtained from the condition $\delta F_{sc}/\delta \Delta(\omega_{m}) =0$.

The condensation energy $E_c$ is the difference between $F_{sc}$,  with $\Delta (\omega_m)$ satisfying the Eliashberg equation (\ref{ss_111}), and $F_n$ (the free energy for $\Delta =0$).
Using Eq. (\ref{a_5}) and the gap equation (\ref{ss_111}) we obtain~\cite{Wada,Bardeen,Emil2020}
\bea
&&\frac{E_c}{N_{0}} =  -\int d\omega_m \frac{|\omega_m|}{\sqrt{1 + D^2 (\omega_m)}} \left(\sqrt{1+ D^2 (\omega_m)}-1\right)^2
- \frac{{\bar g}^\gamma}{4}  \int \frac{d \omega_m d \omega'_m}{\sqrt{1 + D^2 (\omega_m)}\sqrt{1 + D^2 (\omega'_m)}}
\nonumber \\
&&
\times \left(\sqrt{1 + D^2 (\omega_m)}-\sqrt{1 + D^2 (\omega'_m)}\right)^2 \left(\frac{1}{|\omega_m - \omega'_m|^\gamma} - \frac{1}{|\omega_m + \omega'_m|^\gamma}\right) \label{a_51}
\eea
where $D (\omega_m) = \Delta (\omega_m)/\omega_m$. Both terms in the r.h.s. of (\ref{a_51}) are negative, hence, the condensation energy is negative for {\it any} solution of the gap equation $\Delta (\omega_m)$. Note that the factor  $N$ is not present in (\ref{a_51}).

To understand whether or not the ground state at $T=0$ is a NFL state or a superconducting state, we first consider infinitesimally small $\Phi (\omega_m)$.
 The corresponding equation is obtained by neglecting $\Phi$ in the denominator of Eq. (\ref{eq:gapeq_1}) and using (\ref{ss_111_a}) for ${\tilde \Sigma}$:
  \beq
       N \Phi (\omega_m) =\frac{1-\gamma}{2} \int d \omega'_m \frac{\Phi(\omega'_m)}{|\omega'_{m}| ^{1-\gamma} |\omega_m-\omega'_{m}|^\gamma}
 \frac{1}{1 + \left(\frac{|\omega'_m|}{\omega_0}\right)^\gamma}
  \label{eq:lineargap_1}
  \eeq
 This is an equation for an eigenfunction of a linear operator, in which $N$ plays a role of the eigenvalue.
  Observe that the fermion-boson coupling ${\bar g}$  appears only in the last term in the denominator, via
   $\omega_0 \propto {\bar g}$.  Without this term,  the r.h.s. of (\ref{eq:lineargap_1}) is marginal by power counting (the total exponent in the denominator is $1-\gamma + \gamma =1$).  Then,  once we rescale frequency to
 ${\bar \omega}_m = \omega_m/\omega_0$, the equation for $\Phi ({\bar \omega}_m)$ becomes fully universal:
   \beq
      \Phi ({\bar \omega}_m) =\frac{1-\gamma}{2N} \int d {\bar \omega}'_m \frac{\Phi({\bar \omega}'_m)}{|{\bar \omega}'_{m}| ^{1-\gamma} |{\bar \omega}_m-{\bar \omega}'_{m}|^\gamma}
 \frac{1}{1 + |{\bar \omega}'_m|^\gamma}
  \label{eq:lineargap_1_1}
  \eeq
 The same is true for the linearized equation for $\Delta (\omega_m)$:
\beq
   \Delta (\omega_m) = \frac{ {\bar g}^\gamma }{2N} \int d \omega'_m  \frac{\Delta (\omega'_{m}) -N  \Delta (\omega_m) \frac{\omega'_{m}}{\omega_m}}{|\omega'_{m}|}
    ~\frac{1}{|\omega_m - \omega'_{m}|^\gamma}.
     \label{ss_111_l}
  \eeq
Using ${\bar g}^\gamma = \omega^{\gamma}_0 (1-\gamma)$ and introducing again ${\bar \omega}_m = \omega_m/\omega_0$,  we re-express (\ref{ss_111_l}) as
\beq
   \Delta ({\bar \omega}_m) = \frac{1-\gamma}{2N} \int d {\bar \omega}'_m  \frac{\Delta ({\bar \omega}'_{m}) -N  \Delta ({\bar \omega}_m) \frac{{\bar \omega}'_{m}}{{\bar \omega}_m}}{|{\bar \omega}'_{m}|}
    ~\frac{1}{|{\bar \omega}_m - {\bar \omega}'_{m}|^\gamma}.
     \label{ss_111_l1}
  \eeq
For infinitesimal $\Phi$ and $\Delta$, $\Delta (\omega_m) = \Phi (\omega_m)/\left(1 + (\omega_0/|\omega_m|)^\gamma\right)$, or, equivalently, $\Delta ({\bar \omega}_m) = \Phi ({\bar \omega}_m)/(1 + |{\bar \omega}_m|^{-\gamma})$.

  A conventional wisdom, borrowed from BCS theory, would imply that Eqs. (\ref{eq:lineargap_1_1}) and (\ref{ss_111_l1}) must have solutions at a single critical $N_{cr}$, separating NFL and superconducting states, assuming that such critical $N_{cr}$ exists. By the same logic,  the full non-linear gap equation has no solution at $N > N_{cr}$, while at $N < N_{cr}$, the solution
   $\Phi (\omega_m)$ (or $\Delta (\omega_m)$) has a finite magnitude, which increases with  $N_{cr}-N$.

 We show below that the situation in the $\gamma$-model is different.  Specifically, we show that critical $N_{cr}$ exists, but the linearized gap equation has a solution not only at $N = N_{cr}$, but for all $N < N_{cr}$.  Furthermore, $N= N_{cr}$  turns out to be a multi-critical point of the $\gamma $-model (for $0<\gamma <1$), below which there exists an infinite number of solutions of the full non-linear
  gap equation, $\Delta_n (\omega_m)$. The magnitude of the $n$th solution initially increases as $e^{-A_n/\sqrt{N_{cr} -N}}$, where the factor $A_n$ depends on the number $n$ of the solution.

In the next four sections we discuss the linearized gap equation, Eqs.  (\ref{eq:lineargap_1_1}) and (\ref{ss_111_l1}). We return to the non-linear gap equation  in Sec. \ref{sec:nonlinear}.

To verify that a potential solution of the linearized gap equation is normalizable, we will need to analyze the Free energy (\ref{a_5}) order $\Delta^2$.  Expanding in (\ref{a_5})  we obtain
 \bea
 F_{sc} &=& F_{norm} + \frac{N_0}{2N}  J (\Delta, N) \label{a_5_1} \\
 J (\Delta, N) &=& N \int d {\bo}_m \frac{\Delta^2 (\bo_m) (1 + |\bo_m|^\gamma)}{|\bo_m|^{1+\gamma}} - \frac{1-\gamma}{2} \int d \omega_m d \omega'_m
 \frac{\Delta (\omega_m) \Delta (\omega'_m)}{|\bo_m| |\bo'_m|} \frac{1}{|\bo_m-\bo'_{m}|^\gamma}. \nonumber
\eea
The linearized gap equation (\ref{ss_111_l1}) is obtained by varying this $F_{sc}$ over $\Delta(\omega_{m})$.
   A solution is normalizable if the variation $\delta J(\Delta, N)/\delta N$ is finite, i.e., if
\beq
 \int d\bo_m \frac{\Delta^2 (\bo_m) (1 + |\bo_m|^\gamma)}{|\bo_m|^{1+\gamma}}
 \label{a_5_2}
 \eeq
 is non-divergent.  In terms of $\Phi (\bo_m)$,  the same integral is
 \beq
 \int d\bo_m \frac{\Phi^2 (\bo_m)}{|\bo_m|^{1-\gamma} (1 + |\bo_m|^\gamma)}
 \label{a_5_3}
 \eeq

\section{The limits of small and large $\omega_m$.}
\label{sec:truncated}

We begin with the analysis of the truncated gap equation at small and large frequencies.  The analysis can be done most straightforwardly for Eq. (\ref{eq:lineargap_1_1}) for the pairing vertex.  At large  $|\bo_m| \gg  1$,  we can pull out the external $\bo_m$ from the integral in the r.h.s. of (\ref{eq:lineargap_1}) and obtain
\beq
      \Phi ({\bar \omega}_m) =\frac{1 }{|\bo_m|^\gamma} \frac{1-\gamma}{N}  \int_0^{O(|\bo_m|)} d {\bar \omega}'_m \frac{\Phi({\bar \omega}'_m)}{|{\bar \omega}'_{m}| ^{1-\gamma}}
 \frac{1}{1 + |{\bar \omega}'_m|^\gamma}.
  \label{eq:lineargap_1_4}
\eeq
Substituting $\Phi({\bar \omega}'_m) \propto 1/ |\bo_m|^\gamma$ into the r.h.s of (\ref{eq:lineargap_1_4}) we find that the integral converges at $|{\bar \omega}'_{m}| = O(1)$.  This shows that pulling out $\bo_m$ from the integral is justified when $|\bo_m| \gg 1$.   The outcome is that at large frequencies
\beq
 \Phi ({\bar \omega}_m)  = \frac{C_\infty}{|\bo_m|^\gamma}
 \label{i_1}
 \eeq
In this limit, $\Phi ({\bar \omega}_m) = \Delta (\bo_m)$, hence, we also have
\beq
 \Delta ({\bar \omega}_m)  = \frac{C_\infty}{|\bo_m|^\gamma}
 \label{i_2}
 \eeq

In the opposite limit of small $\bo_m$ we assume and then verify that one can truncate (\ref{eq:lineargap_1_1})  by
 replacing $1/(1+ |\bo'_m|^\gamma)$ in the r.h.s. by the upper cutoff at $|{\bar \omega}'_m| = O(1)$. The precise value of the cutoff frequency will play no role, and we set the cutoff at $|{\bar \omega}'_m| = 1$.
 The equation for the pairing vertex  then becomes
 \beq
      \Phi (\bo_m) =\frac{1-\gamma}{2N} \int_{-1}^{1} d \bo'_m \frac{\Phi(\bo'_m)}{|\bo'_{m}| ^{1-\gamma} |\bo_m-\bo'_{m}|^\gamma}
        \label{eq:lineargap_1_3}
  \eeq
Below we analyze this equation for different $N$.

\subsection{Large $N$.}
\label{sec:critical_N_a}

We first consider the limit of large $N$.
 The effective coupling constant in (\ref{eq:lineargap_1_3}) scales as $1/N$, hence, the solution with a non-zero $\Phi (\omega_m)$ emerges only if the smallness of the coupling is compensated by a large value of the frequency integral in the r.h.s. of (\ref{eq:lineargap_1_3}).  This is what happens in a BCS superconductor (the case $\gamma =0$). There, the pairing kernel  scales as $1/|\bo_m|$,  $\Phi (\bo_m)= \Phi$ is independent of the running fermionic frequency,
  and the integral $\int_{-1}^{1} d \bo'_m  \Phi/|\bo'_m|$ is logarithmically singular.  The logarithm compensates for the smallness of the coupling $1/N$, and superconductivity emerges already for arbitrary weak attraction.  A way to see this is to compute the pairing vertex in the presence of an infinitesimally small initial $\Phi_0$.
   At $T=0$ this has to be done at a non-zero total incoming bosonic frequency $\Omega_{tot} = {\bar \Omega}_{tot}/\omega_0$ to avoid divergencies. The result for a BCS superconductor is well known: to logarithmic accuracy
\beq
\Phi ({\bar \Omega}_{tot}) = \Phi_0 \left(1 + \frac{1}{N} \log{\frac{1}{|{\bar \Omega}_{tot}|}} +  \frac{1}{N^2} \log^2{\frac{1}{|{\bar \Omega}_{tot}|}} + ...\right) = \frac{\Phi_0}{1 - \frac{1}{N} \log{\frac{1}{|{\bar \Omega}_{tot}|}}}
\label{su_1}
\eeq
The ratio $\Phi ({\bar \Omega}_{tot})/\Phi_0$ (the pairing susceptibility) diverges at $|\Omega_{tot}| = \omega_0 e^{-N}$ and becomes negative at smaller $|\Omega_{tot}|$, indicating that the normal state is unstable towards pairing. (In a more accurate description, the pole in $\Phi (\Omega_{tot})$ moves from the lower to the upper half-plane of  complex frequency~\cite{agd}).

 For a non-zero $\gamma$,  the pairing kernel is the function of both internal ${\bar \omega}'_m$ and external ${\bar \omega}_m$
 \beq
 K({\bar \omega},{\bar \omega}'_m) = \frac{1}{|\bo'_{m}|^{1-\gamma} |\bo_m-\bo'_{m}|^\gamma}
 \eeq
 If we set the external $\bo_m$ to zero, we find that $K (0,\bo'_m) = 1/|\bo'_{m}|$ is marginal,  like in BCS theory.  Then, if we add $\Phi_0$ and compute $\Phi (\bo_m)$ perturbatively, the series will be logarithmical, like in the BCS case.  In distinction to BCS, however, each logarithmical integral $\int d \bo'_m/|\bo'_m|$  runs between  the upper cutoff at $|\bo'_m|=1$  and
   the lower cutoff at $|\bo'_m| \sim |\bo_m|$. Because the lower cutoff is finite at a finite $\bo_m$, we can safely set $\Omega_{tot} =0$.
   Summing up logarithmical series we then obtain
\beq
\Phi (\bo_m) = \Phi_0 \left(1 + \frac{1-\gamma}{N}  \log{\frac{1}{|\bo_m|}} +  \frac{(1-\gamma)^2}{2N^2} \log^2{\frac{1}{|\bo_m|}}
 + ....\right)  = \Phi_0 \left(\frac{1}{|\bo_m|}\right)^{\frac{1-\gamma}{N}}
\label{su_2}
\eeq
We see that the pairing susceptibility remains finite and positive for all finite $\omega_m$, even when $\Omega_{tot} =0$.  Re-doing calculations at a finite $\Omega_{tot}$ we find the same result as in (\ref{su_2}), but with $|\bo_m|$ replaced by
 max$(|\bo_m|, {\bar \Omega}_{tot})$.  The implication is that at a finite $\gamma$, summing up logarithms does not give rise to the divergence of the pairing susceptibility, i.e.,  within a logarithmic approximation, the system at $T=0$ remains in the normal state.

 The difference between  logarithmic approximation at $\gamma =0$ and at a finite $\gamma$ can be also understood by expressing the flow of $\Phi (\bo_m, {\bar \Omega}_{tot})$ under external $\Phi_0$ in terms of RG-type differential equation.
   In BCS theory one can re-sum logarithmical ladder series by selecting a cross-section in the middle with the smallest running frequency $\bo_m$ and integrate over frequencies larger than $\bo_m$  in the cross-sections on both sides of the selected one. One such integration gives $\Phi (\bo_m)$, another gives the pairing susceptibility $\Phi (\bo_m)/\Phi_0$. Combining and differentiating over
   $L = \log{1/{\bar \Omega}_{tot}}$,
   we obtain the RG equation $d \Phi (L)/dL = \Phi^2 (L)/N \Phi_0$, with the boundary condition $\Phi (0) = \Phi_0$.   The solution of this equation is Eq. (\ref{su_1}).  For a non-zero $\gamma$, the logarithmical integral in a given cross-section is cut not by an external bosonic ${\bar \Omega}_{tot}$, but by the external fermionic frequency in the neighboring cross-section.  A simple experimentation shows that in this case the corresponding RG equation is $d \Phi (L)/dL = \Phi (L)/N$, where now $L = \log (1/|\bo_m|)$.  The solution of this equation is Eq. (\ref{su_2}).

We now go beyond perturbation theory and analyze Eq. (\ref{eq:lineargap_1_3}) without the $\Phi_0$ term.  Our first observation is that
 $\Phi (\bo_m) \propto \left(1/|\bo_m|\right)^{(1-\gamma)/N}$, which we found by summing up logarithms, does satisfy Eq.  (\ref{eq:lineargap_1_3}) at $|\bo_m| \ll 1$.
Indeed, substituting this form into  (\ref{eq:lineargap_1_3}) we find that it is satisfied, because to leading order in $1/N$,
\beq
 \int_0^\infty\frac{dx}{|x|^{(1-\gamma)(N+1)/N}} \frac{1}{|1-x|^\gamma} =\frac{N}{1-\gamma}
 \label{nn_1}
 \eeq
  We next observe that   there is another  possibility to compensate for the $1/N$ smallness of the coupling constant in (\ref{eq:lineargap_1_3}), by choosing
$\Phi (\bo_m) \propto \left(1/|\bo_m|\right)^{\gamma -(1-\gamma)/N}$, such that the integral over $\bo'_m$ in the r.h.s. of (\ref{eq:lineargap_1_3}) almost diverges at small $\bo'_m$. Indeed, substituting this form into (\ref{eq:lineargap_1_3}) we find that the equation is satisfied because to leading order in $1/N$,
\beq
\int_0^\infty  \frac{dx}{|x|^{1-(1-\gamma)/N}} \frac{1}{|1-x|^\gamma} = \frac{N}{1-\gamma}
 \label{nn_2}
 \eeq
Note that  the factor $N$  now comes from small $|x|\ll 1$.  This implies that this solution could not be obtained within a conventional logarithmic approximation, or, equivalently,  from the  RG equation, as the latter assumes that  the logarithms, which sum up into the anomalous power-law form,  come from internal frequencies, which are much larger than the external one.

The solution of (\ref{eq:lineargap_1_3}) at large $N$ is then the combination of the two power-law forms
\beq
\Phi (\bo_m) =  \frac{C_A}{|\bo_m|^{(1-\gamma)/N}} + \frac{C_B}{|\bo_m|^{\gamma-(1-\gamma)/N}}.
\label{nn_3}
\eeq
The overall factor doesn't matter because $\Phi (\omega_m)$ is defined up to a constant factor, but the ratio $C_A/C_B$ is a free parameter at this moment.

We now argue that while both terms in (\ref{nn_3}) satisfy (\ref{eq:lineargap_1_3}), only one component represents  the normalized  solution and should be  kept. Indeed,  substituting $\Phi (\bo_m)$  into (\ref{a_5_3}) and using the fact find that  for large $N$, $(1-\gamma)/N \ll \gamma/2$, we find that the integral in
 (\ref{a_5_3}) is finite
 for the first term in (\ref{nn_3}) but diverges for the second one.  We then have to drop this term and keep only the $C_A$ term in (\ref{nn_3}).

  The full gap equation (\ref{eq:lineargap_1_1}) has the solution if $\Phi (\bo_m) = C_A/|\bo_m|^{(1-\gamma)/N}$ at small $\bo_m$ and $\Phi (\bo_m) = C_\infty/|\bo_m|^{\gamma}$ at large $\bo_m$ can be smoothly connected (see Fig. \ref{fig5}).   We show below that these two limiting forms cannot be connected, i.e., the linearized  equation for the pairing vertex doesn't have a solution.

 \begin{figure}
	\begin{center}
		\includegraphics[width=0.45\columnwidth]{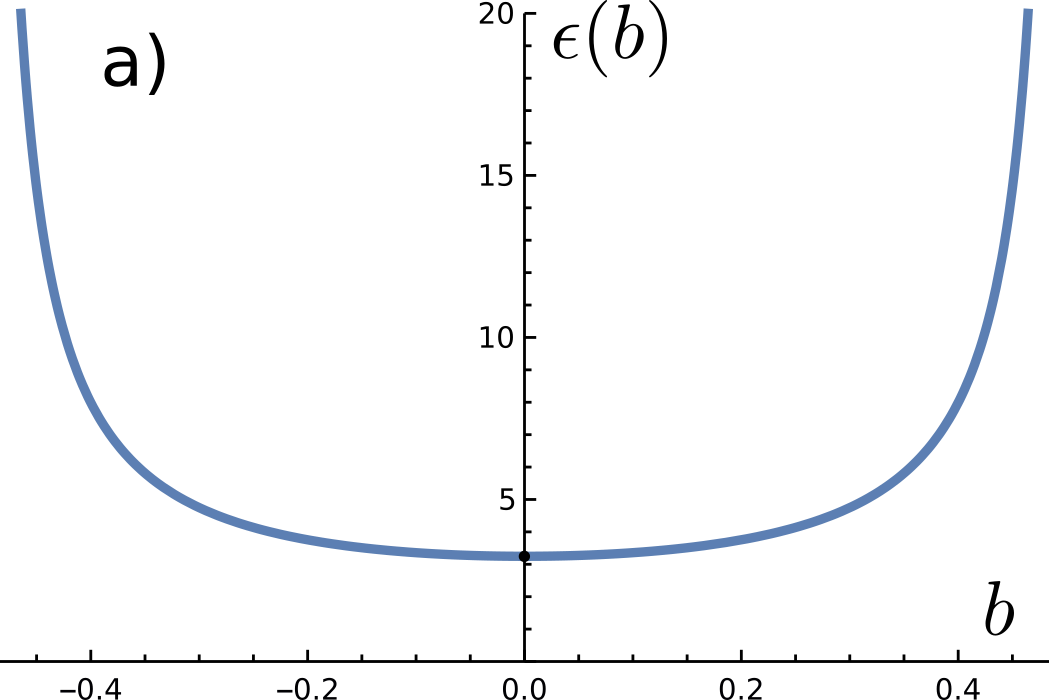}
		\includegraphics[width=0.45\columnwidth]{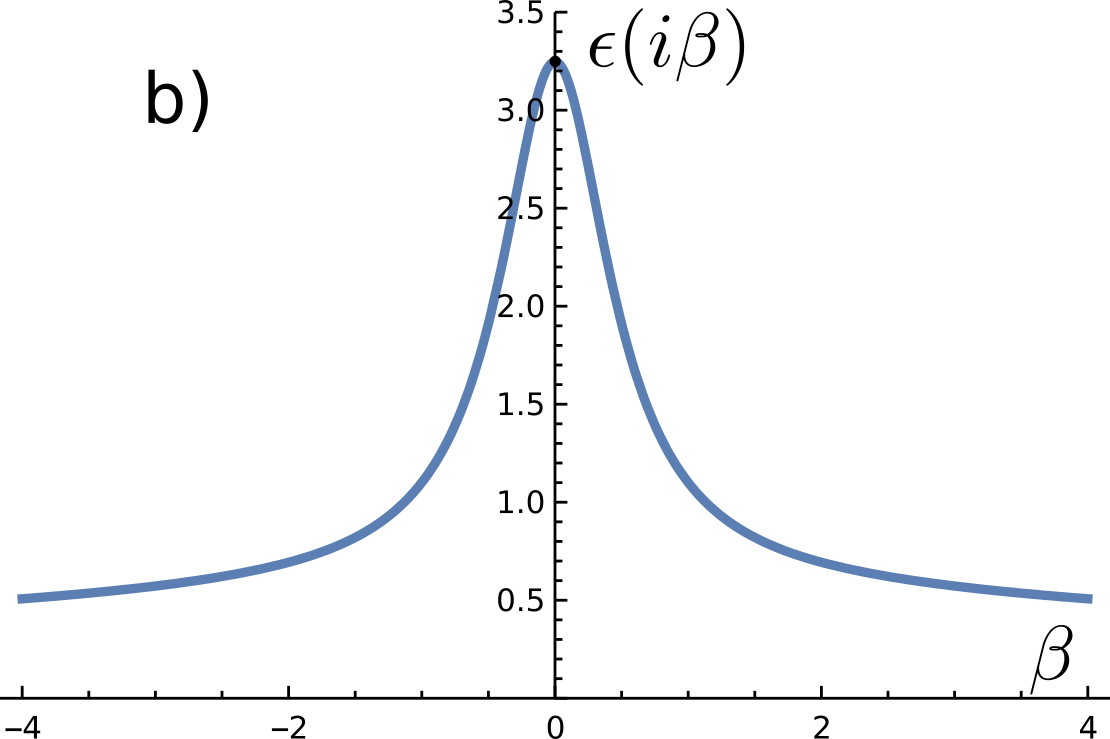}
		\caption{The function $\epsilon(b)$ from Eq. \ref{su_15} for real $b$ (left panel) and imaginary $b$ (right panel) for $\gamma =0.6$.  The solution of the truncated gap equation at small $\bo$ exists if the equation $\epsilon (b) = N$ has a solution.}
\label{fig6}
	\end{center}
\end{figure}

\subsection{Arbitrary $N$.}
\label{sec:critical_N_b}

 The analysis of the truncated equation for the pairing vertex can be straightforwardly extended to arbitrary $N$.  Like before, we use the fact that the kernel in (\ref{eq:lineargap_1_3}) is marginal and search for power-law  solution  of (\ref{eq:lineargap_1_3}) in the form $\Phi (\bo_m) \propto 1/|\bo_m|^{\gamma (1/2 + b)}$, where $b$ needs to be obtained self-consistently  (we pulled out $\gamma$ in the exponent for future convenience).  Substituting into (\ref{eq:lineargap_1_3}) and evaluating the integral we find that $b$ is the solution of
 \beq
 \epsilon_b = N
  \label{su_14}
 \eeq
 where
   \beq
    \epsilon_b   =\frac{1-\gamma}{2}\frac{\Gamma (\gamma/2  -\gamma b ) \Gamma (\gamma/2 +\gamma b)}{\Gamma (\gamma )}\left(1+\frac{\cos (\pi\gamma b)}{\cos (\pi \gamma /2)} \right)
\label{su_15}
\eeq
 The  integral in the r.h.s. of (\ref{eq:lineargap_1_3}) is evaluated using the identities
  \begin{eqnarray}
 \int_0^\infty \frac{dy}{y^{a} (1+y)^{\gamma}} &=& \frac{\Gamma (1-a) \Gamma (\gamma+a-1)}{\Gamma (\gamma)} \nonumber \\
\int_0^1 \frac{dy}{y^{a} (1-y)^{\gamma}}  &=& \frac{\Gamma (1-a) \Gamma (1 -\gamma)}{\Gamma (2-a- \gamma)} \nonumber \\
 \int_1^\infty
\frac{dy}{y^{a} (y-1)^{\gamma}}  &=& \frac{\Gamma (a+\gamma-1) \Gamma (1 -\gamma)}{\Gamma (a)}
\label{nn_2a}
\end{eqnarray}
We plot $\epsilon_b$ in Fig. \ref{fig6}a.
At small $\gamma $,
\beq
\epsilon_b  \approx \frac{4}{\gamma} \frac{1}{1-4 b^2}
 \label{new_a}
 \eeq
 We see that $ \epsilon_b$ is an even function of $b$, i.e., if $b$ is a solution, then $-b$ is also a solution. This implies that
\beq
\Phi (\bo_m) = \Phi (z) = \frac{C_A}{z^{1/2-b}} + \frac{C_B}{z^{1/2+b}},
\label{nn_3_d}
\eeq
where
\beq
z = |\bo_m|^\gamma = \left(\frac{|\omega_m|}{\omega_0}\right)^\gamma
\label{eeee}
\eeq
 At large $N$, $b \approx 1/2 - (1-\gamma)/(N\gamma)$ and (\ref{nn_3_d}) coincides with (\ref{nn_3}).  As  $N$ increases,  $b$ gets smaller.   As long as it remains positive,
  the second term in the r.h.s. of (\ref{nn_3_d}) has to be dropped
  as it does not  satisfy the normalization condition
  (the integral in (\ref{a_5_3})  diverges for $\Phi (\bo_m) \propto 1/(z^{1/2+b}$ ).  Then at small $z$,
  \beq
  \Phi (z) =  \frac{C}{z^{1/2 -b}}, ~~~  \Delta (z) =  C z^{1/2 +b}.
  \label{i_3}
  \eeq
    This is very similar to the case of large $N$.

\begin{figure}
	\begin{center}
		\includegraphics[width=0.6\columnwidth]{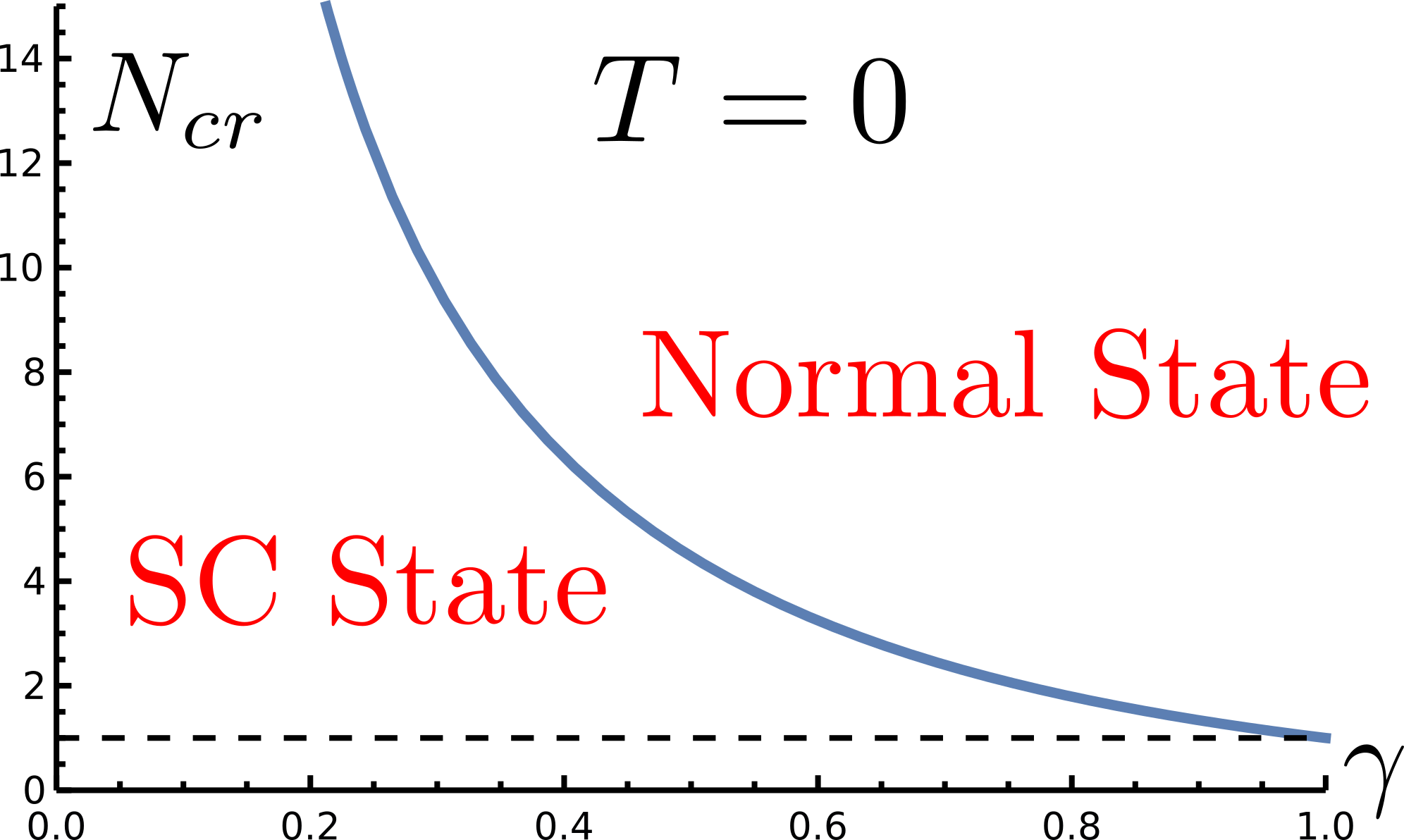}
		\caption{The critical $N_{cr}$ (for which Eq. (\ref{su_15}) has a solution for $b=0$) as a function of $\gamma$. At small $\gamma$, $N_{cr} \approx 4/\gamma$. At $\gamma \to 1$, $N_{cr} \to 1$.}
\label{fig7}
	\end{center}
\end{figure}

  We next observe that for any $0 <\gamma \leq 1$, there exists a critical $N_{cr}$, for which $b =0$. It is given by
 \bea
N_{cr} = \epsilon_0 &=& \frac{\frac{\pi}{2} (1-\gamma)}{\sin{\frac{\pi}{2} (1-\gamma)}} \frac{\pi}{\Gamma (\gamma)} \frac{\left(1 - \cos{\frac{\pi \gamma}{2}}\right)^{-1}}{\Gamma^2\left(1-\gamma/2\right)}
=\frac{1-\gamma}{2}\frac{\Gamma^{2} (\gamma /2)}{\Gamma (\gamma )}\left(1+\frac{1}{\cos (\pi \gamma /2)} \right),
\label{su_15_1}
\eea
We plot $N_{cr}$ vs $\gamma$ in Fig. \ref{fig7}

We see that  $N_{cr} >1$ for all $\gamma <1$.
  At small $\gamma$, we have from (\ref{new_a}),
    $N_{cr} \approx 4/\gamma$.
At $\gamma \to 1$, $N_{cr} \to 1$.

Right at $N = N_{cr}$, the two terms in (\ref{nn_3_d}) become equal, i.e., one solution of (\ref{eq:lineargap_1_3}) is
 $\Phi (z) \propto 1/\sqrt{z}$. On a more careful look, we find that
  $\Phi (z) \propto  \log {z}/\sqrt{z}$ is also the solution. Indeed, substituting this  form  into (\ref{eq:lineargap_1_3}) and using  $\int dy |y|^{\gamma/2-1} |y-1|^{-\gamma }\log |y| =0$ and
      $ ((1-\gamma)/2N_{cr})\int dy |y|^{\gamma/2-1} |y-1|^{-\gamma } =1$,  we find that this equation is satisfied, i.e.,
      \beq
  \frac{1-\gamma}{2N_{cr}} \int d \omega'_m \frac{\log {|\omega'_m|}}{|\omega'_{m}|^{1-\gamma/2} |\omega_m-\omega'_{m}|^\gamma} =  \frac{\log {|\omega_m|}}{|\omega_{m}|^{\gamma/2}}
 \label{su_19}
 \eeq
The full solution of (\ref{eq:lineargap_1_3}) at $N = N_{cr}$ is then
\beq
\Phi (z) =  \frac{1}{\sqrt{z}} \left( C_A \log{z} + C_B \right)
\label{nn_3_e}
\eeq
 or, equivalently,
 \beq
\Phi (z) =  \frac{C}{\sqrt{z}} \left(\log{z} + \phi \right),~~~ \Delta (z) = C \sqrt{z} \left(\log{z} + \phi \right).
\label{i_4}
\eeq
 This $\Phi (z)$ contains two free parameters: $C$ and $\phi$. The overall factor $C$ is irrelevant for the solution of the linearized equation, but  $\phi$ is a free parameter, which we can vary in a hope that at a particular value
  of $\phi$, $\Phi (z)$ and $\Delta (z)$ interpolate smoothly between
 small $z$ and large $z$ limits.  This gives an indication that $N = N_{cr}$ may be the onset for superconductivity.

We now move to $N < N_{cr}$. There is no solution of (\ref{su_14}) for real $b$, consistent with the conventional wisdom that the linearized gap equation has no solution inside the superconducting phase.
However, it turns out that the solution of (\ref{eq:lineargap_1_3}) still exists, but for imaginary  $b = i\beta_N$, i.e., with complex exponents $1/2 \pm i \beta_N$. Indeed, for a generic $b = i \beta$,
   \bea
    \epsilon_
    {i\beta}
      &=&
\frac{\frac{\pi}{2} (1-\gamma)}{\sin{\frac{\pi}{2} (1-\gamma)}}
\frac{\pi}{\Gamma (\gamma)}
\frac{\left(\cosh{\pi \gamma \beta}- \cos{\pi \gamma/2}\right)^{-1}}{\Gamma\left(1-\gamma/2(1+ 2i\beta)\right) \Gamma\left(1-
    \gamma/2(1 -2i\beta)\right)} \nonumber \\
    &=& \frac{1-\gamma}{2}\frac{|\Gamma (\gamma /2(1 +2i\beta) )|^{2}}{\Gamma (\gamma )}\left(1+\frac{\cosh (\pi \gamma \beta)}{\cos (\pi \gamma /2)} \right)
    \label{su_15_2}
\eea
is real, even, and monotonically decreases with increasing $\beta$  from its maximum value $\epsilon_{0}=N_{cr}>1$ to $0$, i.e.,
 Eq. (\ref{su_14}) has a solution for $N < N_{cr}$ at some non-zero $\beta =\beta_N$.  We plot $\epsilon_
 {i\beta}$ in Fig. \ref{fig6}b.
At small $\gamma$,
\beq
\epsilon_
{i\beta} \approx \frac{4}{\gamma} \frac{1}{1 + 4 \beta^2}
 \label{new_f}
 \eeq
and
\beq
\beta_N = \frac{1}{2} \sqrt{\frac{N_{cr}-N}{N}}
\label{aaaa}
\eeq
up to corrections of order $\gamma$.
Solutions with complex exponents have been recently found in several other physics problems~\cite{Liu_2011,Cortez_2011,Klebanov_19}. For superconductivity, the solution with complex exponents has been found in Ref. \cite{acf}  for the $\gamma$ model with $\gamma =1/2$.

At  $N$ slightly below $N_{cr}$,  $\beta_N \propto \sqrt{N_{cr}-N}$ for all $\gamma <1$.   For $N=1$,  $\beta_{N=1} \approx \gamma^{-1/2} (1-((15-\pi^2)/24)\gamma + O(\gamma^2))$ for small $\gamma$, $\beta_{N=1} \approx 1.268$ for $\gamma =1/2$, and $\beta_{N=1} \to 0.792$ for $\gamma \to 1$. For $N = O(1)$, but $N \neq 1$, the behavior at $\gamma <1$ is very similar, but for $\gamma \to 1$, $\beta_N$ diverges as  $\beta_N \approx 0.561 (1/N)^{1/(1-\gamma)}$.

Combining the contributions with $\beta_N$ and $-\beta_N$, we obtain
\beq
\Phi (z) =  \frac{C}{2} \frac{1}{\sqrt{z}} \left(\frac{e^{i\phi}}{z^{i\beta_N}} + \frac{e^{-i\phi}}{z^{-i \beta_N}}\right) =
\frac{C}{\sqrt{z}}  \cos{\left(\beta_N \log{z}+ \phi\right)}
\label{nn_3_b}
\eeq
where $\phi$ is a free parameter.
  We see that $\Phi (z)$ oscillates on a logarithmical scale down to the lowest  frequencies.
 Such an oscillating solution could not be obtained perturbatively, starting from
 $\Phi (z) = \Phi_0$, because  within perturbation theory $\Phi (z)$ remains of the same sign as
 $\Phi_0$.

 We also see that $ {\bar \Phi} = \Phi (z) \sqrt{z} \propto e^{i(\beta_N \log{z} + \phi)} + c.c$  has the form of a wave function of a free fermion if we associate $x=\log{z}$ with the coordinate.  In terms of $x$, the integral  in (\ref{a_5_3}) is  expressed as $\int dx {\bar \Phi} (x)^2$, and is nothing but the norm of the eigenfunction of continuum spectrum. The norm can be made finite by adding infinitesimal $e^{-\delta |x|}$ to the integral, after which it converges.   The addition of $e^{-\delta |x|}$ also makes the integral in (\ref{a_5_3}) convergent for $\Phi (z)$ at $N = N_{cr}$,  however for $N > N_{cr}$, the integral in (\ref{a_5_3}) still diverges for one term in (\ref{nn_3_d}).  We discuss the normalizability for $N > N_{cr}$ and $N \leq N_{cr}$ in more detail in  Appendix \ref{app:exact}.

The gap function $\Delta(z)$ behaves as
 \beq
\Delta (z) = C \sqrt{z} \cos{\left(\beta_N \log{z}+ \phi\right)}
\label{nn_3_c}
\eeq

We see that
 the solution of the truncated equation for the pairing vertex at small $z$ again has two free
 parameters. In Eq. (\ref{nn_3_c}) these are the overall factor $C$ and the phase $\phi$.
 The overall factor is irrelevant, but the phase $\phi$ is a free parameter, which we can vary to verify
  whether at some particular $\phi$, $\Delta (z)$ interpolates between (\ref{nn_3_c}) and (\ref{i_2}).

We emphasize that a free parameter for $\Delta (z)$ at small $z$ (or, equivalently, for $\Phi (z)$) exists both at $N = N_{cr}$ and $N < N_{cr}$. By conventional wisdom, one would expect that the linearized gap equation has the solution only  for one value of $N$, which in our case is expectedly $N = N_{cr}$ (see Fig. \ref{fig4}). However, the presence of a free parameter for all $N \leq N_{cr}$ hints that our case may go against the conventional wisdom.

 Matching the limiting forms of an integral equation is a non-trivial procedure as in general one should find a finite frequency range where the two forms coincide. This cannot be implemented in our case because the regions,
   where the integral equation can be truncated, do not overlap.
 Because of this, we will be searching for the solution of  (\ref{eq:lineargap_1_1}) without specifically looking at the limits of small and large $\bo_m$.

\section{The case $\gamma \ll 1$: The linearized gap equation as the differential equation.}
 \label{sec:diff}

  We first consider the limit $\gamma \ll 1$, when the pairing interaction $({\bar g}/|\Omega_m|)^\gamma$ is a shallow function of frequency, and reduce the linearized integral gap equation to the differential equation, for which we obtain the exact analytical solution,
   $\Delta_{{\text{diff}}}  (\omega_m)$, which for convenience of notations we  present below as a function of $z$,
   defined in \eqref{eeee}. We  show that a generic normalized solution  exists for $N \leq N_{cr}$. At small $\omega_m$, $\Delta_{{\text{diff}}}  (z)$ coincides with the solution of the truncated gap equation.  It oscillates as a function of $\beta_N \log {z} + \phi$.
    For arbitrary $\phi$, $\Delta_{{\text{diff}}}  (z)$ tends to a constant at large $z$,  instead of decaying as  $1/z$.  However,  for a certain value of $\phi$ the constant term cancels out, and the gap function has the correct
    asymptotic behavior at high frequencies.  In other words, by fixing the phase in the oscillations of $\Delta_{{\text{diff}}}  (z)$ at small $z$ one recovers the correct form of sign-preserving $\Delta_{{\text{diff}}}  (z)$ at large frequencies.

To obtain the differential equation, we return to the linearized equation for the pairing vertex $\Phi (\omega_m)$, Eq.  (\ref{eq:lineargap_1_1}),
and use the fact that for small $\gamma$,  the integral in the r.h.s. of (\ref{eq:lineargap_1}) comes from internal $\omega'_m$, which are either substantially  larger or substantially smaller than external  $\omega_m$. We then split the integral over $\omega'_m$ into two parts:  in one we approximate $|\omega_m - \omega'_m|$ by $|\omega'_m|$, in the other by $|\omega_m|$.
The equation for $\Phi (\omega_m) = \Phi (z)$ then simplifies to
    \beq
    \Phi (z) = \frac{1-\gamma}{N\gamma} \left[\int^{\infty}_{z} dy \frac{\Phi (y)}{y (1+y)} + \frac{1}{z} \int_0^z dy \frac{\Phi (y)}{1+y}\right].
   \label{su_6}
   \eeq
Differentiating this equation twice over $z$ and replacing $\Phi(z)$ by $\Delta (z) = \Phi (z) z/(1 +z)$, we
  obtain  second order differential gap equation
   \beq
   (\Delta_{{\text{diff}}} (z) (1+z))^{''}  = - \frac{N_{cr}}{4 N} \frac{\Delta_{{\text{diff}}} (z)}{z^2},
   \label{su_7}
   \eeq
   where  $(...)^{''} = d^2 (...)/dz^2$ and we  used the fact that for small $\gamma$, $N_{cr} \approx 4/\gamma$.
This $\Delta_{{\text{diff}}} (z)$ has to be real and satisfy the boundary conditions
 at small and large $z$. At large $z$ we have from (\ref{i_2}),
\begin{equation}\label{eq:boundary_1}
\qquad \lim\limits_{z\rightarrow \infty }\Delta_{{\text{diff}}} (z)
= C_{\infty }/z,
\end{equation}
At small $z$ we have  from (\ref{i_3}), (\ref{i_4}) and (\ref{nn_3_c})
\begin{equation}\label{eq:boundary_2}
 \lim\limits_{z\rightarrow 0}\Delta_{\text{diff}} (z) =
\left\{ \begin{array}{ll}
C_A z^{1/2+b},&\quad \mbox{for $N > N_{cr}$} \\
C z^{1/2} \left(\log{z} + \phi\right),&\quad \mbox{for $N = N_{cr}$} \\
C z^{1/2} \cos \left(\beta_{N}\log z+\phi \right),&\quad \mbox{for $N < N_{cr}$}
\end{array}
\right.
 \end{equation}
We recall that $\Delta_{{\text{diff}}} (z)$, which satisfies the boundary condition (\ref{eq:boundary_2}),
  is normalizable in the sense that the corresponding condensation energy is finite.

A  similar differential gap equation has been obtained  in Ref. \cite{Wang_H_17} for $\gamma \ll 1$ and $N \sim N_{cr}$. These authors, however, set by hand a UW cutoff at some $z = O(1)$.  In our case,
 the differential equation (\ref{su_7}) is valid for all $z$ and the solution must recover $1/z$ behavior at
  high frequencies, see (\ref{eq:boundary_1}).

Before we proceed with the analysis of Eqs. \eqref{su_7}, \eqref{eq:boundary_1}, and (\ref{eq:boundary_2}),
  a remark is in order.
  The integral equation \eqref{su_6} is
   ``non-local'' in the sense that
      the integrals in the r.h.s. are determined by
      all $y$, not only $y \approx z$.  The differential equation (\ref{su_7}), on the other hand, is local -- both r.h.s. and l.h.s  of (\ref{su_7}) contain
       $\Delta_{\text{diff}} (z) $  with the same $z$.
        However, the boundary conditions (\ref{eq:boundary_1}) and (\ref{eq:boundary_2}) imply that once we fix
          the asymptotic form of $\Delta_{\text{diff}} (z)$ at $z\rightarrow 0$,
          we also determine the constant $C_{\infty }$ in the asymptotic  form at $z\rightarrow \infty $.
          Let us take
            $z$ large enough such that
              the asymptotic form of $\Delta_{\text{diff}}(z) $
              holds.
                From Eq. \eqref{su_7}
                  we obtain  for such $z$,  $C_{\infty }= \frac{N_{cr}}{4N}\int_{0}^{\infty }\Delta_{\text{diff}}(y)dy/y$. One can verify
                   that the main contribution to this integral comes from $y\ll z$, i.e., the
                    prefactor in $\Delta_{\text{diff}}(z)$ for $z \to \infty$ is determined by
                    the form of $\Delta_{\text{diff}}(z)$ for much smaller $z$, roughly by $z \leq 1$, for which
                    $\Delta_{\text{diff}}(z)$ is close to its form at small $z$.
                   In this sense the non-locality of the initial integral equation \eqref{su_7} reflects itself  in the boundary conditions \eqref{eq:boundary_1}, \eqref{eq:boundary_2} for the local differential gap equation.
 We also emphasize that  \eqref{su_7} is a second order differential equation,  hence, $\Delta_{\text{diff}}(z)$  is fully determined by the two boundary conditions \eqref{eq:boundary_1} and \eqref{eq:boundary_2}

We now analyze
 Eqs. \eqref{su_7}, \eqref{eq:boundary_1}, \eqref{eq:boundary_2} separately for $N<N_{cr}$, $N=N_{cr}$, and $N>N_{cr}$. We will show that the solution  exists for all $N\leq N_{cr}$.

\subsection{${\bf {N < N_{cr}}}$}
\label{sec:small_N_diff}

We use Eq. (\ref{aaaa}) and re-express $N_{cr}/(4N)$ as $N_{cr}/(4N) =\beta^2_N +1/4$.
Eq. (\ref{su_7}) then becomes
    \beq
   (\Delta_{{\text{diff}}} (z) (1+z))^{''}  = - \left(\beta^2_N +\frac{1}{4} \right) \frac{\Delta_{{\text{diff}}} (z)}{z^2},
   \label{su_7_11}
   \eeq
At $z \ll 1$ (i.e., at $\omega \ll \omega_0$)  Eq. (\ref{su_7}) simplifies to
    \beq
   (\Delta_{{\text{diff}}} (z))^{''}  = - \left(\beta^2_N +\frac{1}{4} \right) \frac{\Delta_{{\text{diff}}} (z)}{z^2},
   \label{su_7_1}
   \eeq
  The solution of this equation is the combination of two power-law functions $\Delta_{{\text{diff}}} (z) \propto z^{-1/2 \pm i \beta_N}$.  Combining the two,  we obtain the expected result
  \beq
\Delta_{{\text{diff}}} (z) = C z^{1/2} \cos\left(\beta_N \log{z} + \phi\right)
\label{nn_3_c_1}
\eeq
 At $z \gg 1$, we have
   \beq
   (\Delta_{{\text{diff}}}  (z) z)^{''}  = -
   \left(\beta^2_N +\frac{1}{4} \right)
    \frac{\Delta_{{\text{diff}}} (z)}{z^2}
   \label{su_10}
   \eeq
The solution of (\ref{su_10})  is
\beq
\Delta_{{\text{diff}}} (z) =   \frac{1}{\sqrt{z}} \left[A_1  J_1 \left(\sqrt{\frac{4 \beta^2_N+1}{z}}\right) + A_2 Y_1 \left(\sqrt{\frac{4\beta^2_N +1}{z}}\right)\right],
 \label{su_11}
 \eeq
 where $J_1$ and $Y_1$ are Bessel and Neumann functions.
 At small value of the argument (or large $z$) $J_1 (p) \approx p$ and $Y_1 (p) \approx 1/p$.  Substituting into (\ref{su_11}) we find  that to satisfy the boundary condition (\ref{eq:boundary_1}) we must set $A_2=0$.  Then
 \beq
\Delta_{{\text{diff}}} (z) =   \frac{A_1}{\sqrt{z}} J_1 \left(\sqrt{\frac{4 \beta^2_N+1}{z}}\right).
 \label{su_11_1}
 \eeq
  A similar result has been obtained in Ref. \cite{Wang_H_17}.

\begin{figure}
	\begin{center}
		\includegraphics[width=0.9\columnwidth]{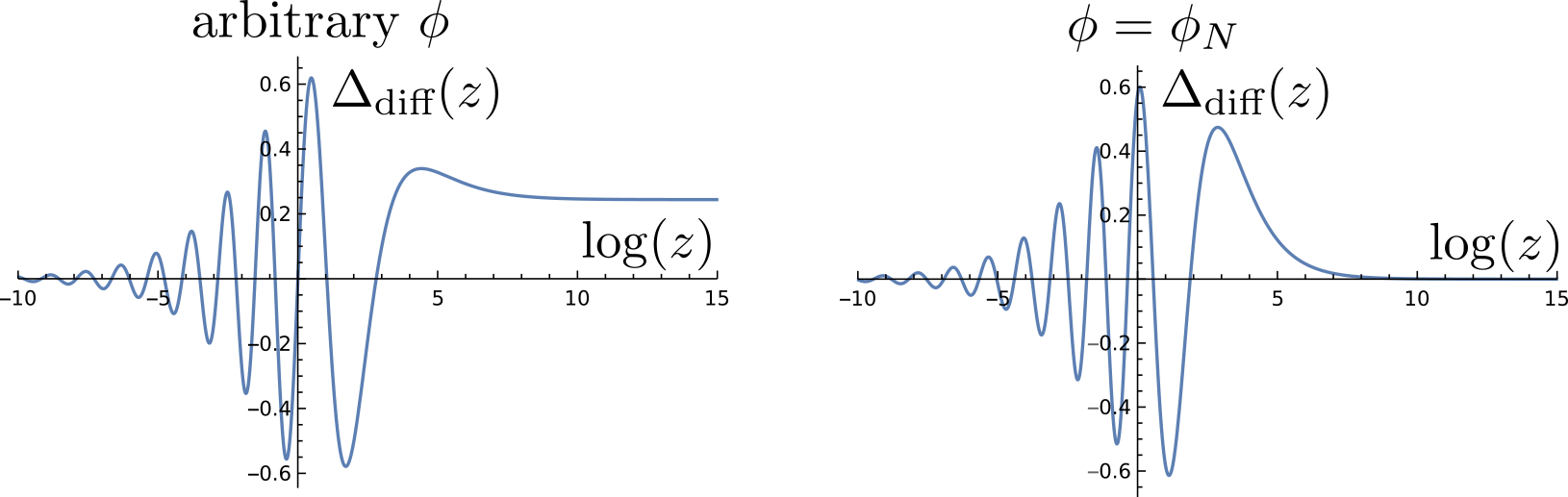}
		\caption{The solution of the differential gap equation, Eq. (\ref{su_17}) for arbitrary $\phi$ (left panel) and for particular $\phi = \phi_N$ given by Eq. (\ref{def_2}) (right panel).  In the last case $\Delta_{\text{diff}} (z)$ satisfies both boundary conditions at small and at large $z$. }
\label{fig8}
	\end{center}
\end{figure}

At  arbitrary $z$, the  solution of (\ref{su_7_11}) is expressed via hypergeometric function
\begin{widetext}
     \beq
 \Delta_{{\text{diff}}} (z) = C \sqrt{z} \times Re\left(e^{i \phi} z^{i \beta_N}
  {_2}F{_1} \left[\frac{1}{2} + i\beta_N,  \frac{3}{2} + i\beta_N, 1 + 2  i \beta_N, -z\right] \right)
 \label{su_17}
 \eeq
 \end{widetext}
At small $z$, ${_2}F{_1} \left[..., -z\right] \approx 1$, and $\Delta_{{\text{diff}}}  (z)$ is the same as in (\ref{nn_3_c_1}). At $z \gg 1$, we use large $z$ asymptotic of the
hypergeometric function
\bea
&&{_2}F{_1} \left[1/2+ i\beta_N,3/2 + i\beta_N, 1 +2\beta_N, -z\right] = \nonumber \\
&&z^{-1/2-i\beta_N} \left[\frac{\Gamma(1 + 2i\beta_N)}{\Gamma(3/2 + i \beta_N) \Gamma (1/2 + i \beta_N)}
 \left(1 + \left (\frac{1}{4} + \beta^2_N\right)\frac{\log{z}}{z} \right) + O\left(\frac{1}{z}\right)\right]
\label{def_1}
\eea
and obtain that for arbitrary $\phi$, $\Delta_{{\text{diff}}} (z)$ tends to a constant, which is inconsistent with
 the boundary condition (\ref{eq:boundary_1}) (see Fig. \ref{fig8} left panel). However, for a particular $\phi = \phi_N$, where
   \beq
   \phi_N =\arctan{\frac{\Re L}{\Im L}},~~\text {where}~ L = \frac{\Gamma(1+2i\beta_N)}{\Gamma(1/2+i \beta_N) \Gamma(3/2+i\beta_N)},
   \label{def_2}
   \eeq
 the constant term in $\Delta (z)$ vanishes (along with $\log{z} /z$ correction), and $\Delta_{{\text{diff}}} (z)$  does scale as $1/z$ at large $z$,  consistent with (\ref{eq:boundary_1})  (see Fig. \ref{fig8} right panel).

 \begin{figure}
	\begin{center}
\includegraphics[width=0.9\columnwidth]{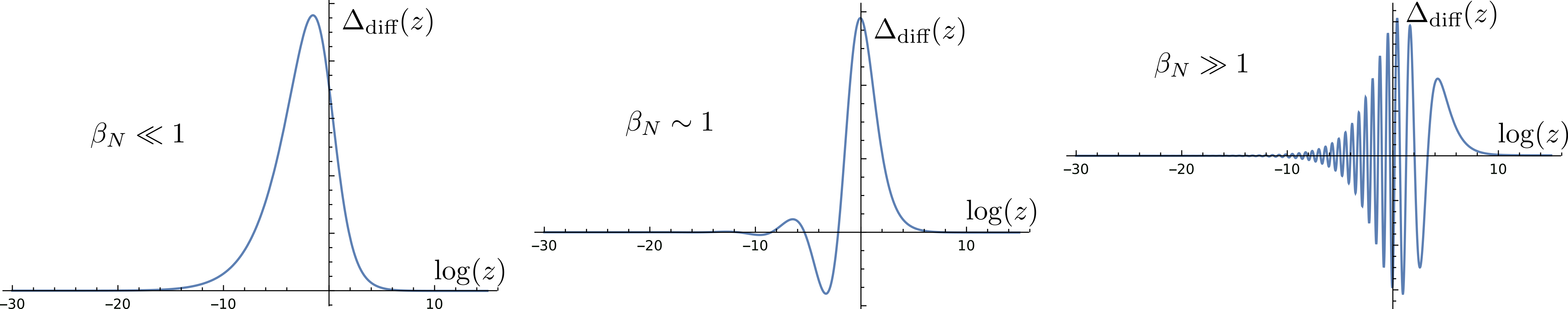}
\bigskip
\bigskip\\
\includegraphics[width=0.6\columnwidth]{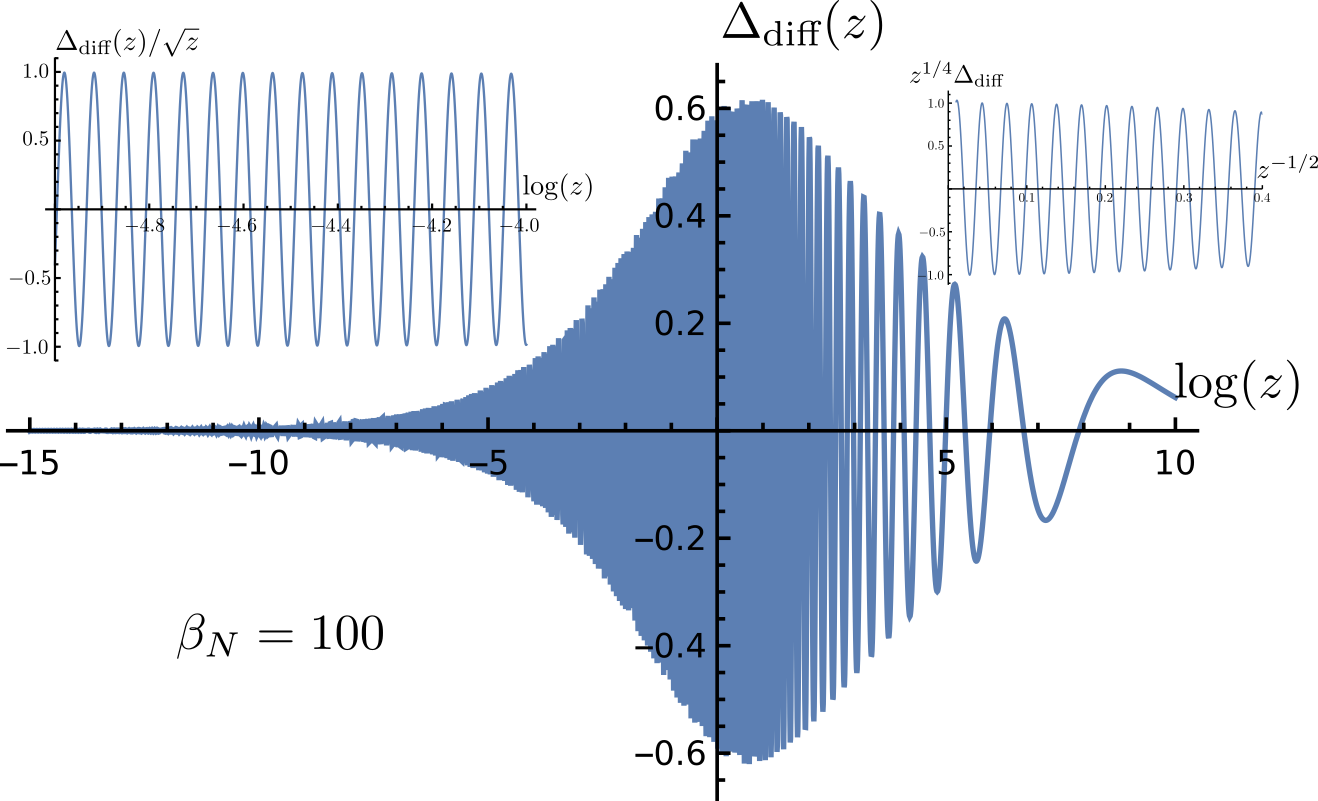}
		\caption{The solution of the differential gap equation, Eq. (\ref{su_17}) for various $\beta_N$.
  For $\beta_N \ll 1$ (panel a)  $\Delta_{{\text{diff}}} (z)$ oscillates as a function of $\log{z}$ up to $z \sim e^{-1/\beta_N}$ and decays as $1/z$ at
   $z >1$.
   For $\beta_N = O(1)$ (panel b) logarithmic oscillations extend to $z = O(1)$, and at larger $z$, $\Delta_{{\text{diff}}} (z) \propto 1/z$.  For $\beta_N \gg 1$ (panel c) logarithmic oscillations persist up to $z =O(1)$, and at larger $z$
 $\Delta_{{\text{diff}}} (z)$ again oscillates  as $\Delta_{{\text{diff}}} (z) \propto \cos(3\pi/4 -2\beta_N/\sqrt{z})/z^{1/4}$.
These oscillations  persist up to $z \sim (\beta_N)^2 \gg 1$.  At larger $z$, $\Delta_{{\text{diff}}} (z) \propto 1/z$. Panel d -- A closer look at oscillations at $z \ll 1$ and $1\ll z \ll \beta^2_N$ for large $\beta_N$ (small $\gamma$).}
\label{fig9}
	\end{center}
\end{figure}

   At small $\beta_N$, the solution of (\ref{def_2}) is $\phi_N = \pi/2 -0.77259 \beta_N$. At $\beta_N =1$,  $\phi_N = \pi/2 -0.93251$, and  at
 large $\beta_N$, $\phi_N \approx 3\pi/4 - \beta_N \log{4}$.
 Once $\phi = \phi_N$ is fixed, one can also
  express the prefactor $A_1$ in (\ref{su_11}) in terms of $C$ in (\ref{nn_3_c_1}), e.g.,
   $A_1 = C (\pi \beta_N)^{1/2}$ for $\beta_N \gg 1$.

Analyzing  Eq. (\ref{su_17}) further,  we note that both the location and the width of the crossover between small $z$ and large $z$ behavior of $\Delta_{{\text{diff}}} (z)$ depend on $N$ (Fig. \ref{fig9}). For $N \leq N_{cr}$, $\beta_N \propto (N_{cr}-N)^{1/2}$ is small. In this situation $\Delta_{{\text{diff}}} (z)$ oscillates as a function of $\log{z}$ up to $z \sim e^{-1/\beta_N}$, then monotonically increases up to $z \sim 0.2$, passes through a maximum, and decays as $1/z$ at larger frequencies (Fig. \ref{fig9}a).  For $N$ such that $\beta_N = O(1)$, logarithmic oscillations extend to $z = O(1)$, and at larger $z$, $\Delta_{{\text{diff}}} (z) \propto 1/z$ (Fig. \ref{fig9}b).  For $N = O(1)$, $\beta_N  \approx 0.5(N_{cr}/N)^{1/2}$ is large because $N_{cr} \approx 4/\gamma \gg 1$. In this situation, logarithmic oscillations of $\Delta (z) = C \sqrt{z} \cos(\beta_N \log{z/4} + 3\pi/4)$, persist up to $z =O(1)$, and at larger $z$ there is a range $1 < z < \beta^2_N$, where
 $\Delta_{{\text{diff}}} (z)$ again oscillates  as $\Delta_{{\text{diff}}} (z) \approx C \cos(3\pi/4 -2\beta_N/\sqrt{z})/z^{1/4}$.
These oscillations  persist up to $z \sim (\beta_N)^2 \sim {1/\gamma}$, i.e., up to  $\omega = \omega_{max} \sim {\bar g}  e^{|\log \gamma|/\gamma}$. At larger $z$, $\Delta_{{\text{diff}}} (z) \propto 1/z$ (Fig.\ref{fig9}c).
Oscillations at small $z$ are best seen if we use the  logarithmical variable
 \beq
  x = \log{z} = \log{\left(\frac{|\omega_m|}{{\bar g}}\right)^\gamma}
  \label{ll_2}
  \eeq
 while oscillations at  $z>1$ are best seen if we plot $\Delta_{\text{diff}} (z)$ as a function of $1/\sqrt{z}$. We present both plots in Fig. \ref{fig9}(d).

The existence of a large intermediate range $1< z < \beta^2_N$ for $\beta^2_N \gg 1$  can be inferred already from Eq. (\ref{su_11}).  Indeed,
  in this range the argument of the Bessel  function $y = 2\beta_N/\sqrt{z}$ is  large,  $J_1 (y) \propto y^{-1/2} \cos(y-3\pi/4)$,  and
   $\Delta (z)$ displays an oscillating behavior with the period set by  $1/\sqrt{z}$ rather than by $\log{z}$.

 The scale $z \sim \beta^2_N $ also shows up in the expansion of $\Delta_{{\text{diff}}} (z)$ in both
$1/z$ and $z$. Expanding in $1/z$, we find
 \beq
 \Delta_{{\text{diff}}} (z) \propto  \frac{1}{z} \left(1 - \frac{\beta^2_N}{2 z} + O\left(\frac{1}{z^2}\right)\right)
  \label{bbbb}
  \eeq
The expansion in $z$ yields
    \bea
    &&\Delta_{{\text{diff}}}(z) \approx C \frac{\sqrt{z}}{(1+z)^{3/4}}
\cos{\left(\frac{3 \pi}{4} + \beta_N Q(z)\right)} f \left(\frac{z}{\beta^2_N}\right),
 \label{dd_1}
 \eea
  where $f(0) =1$ and
 \beq
  Q (z) = \log{\frac{z}{4}} -\frac{z}{2} +\frac{3z^2}{16} - \frac{5 z^3}{48} +\frac{19 z^4}{256} +  ... ,
  \label{dd_1a}
  \eeq
We see that oscillations extend to $z > 1$.  At $1 \ll z \ll \beta_N^{2}$, the form of $Q(z)$ is determined by
by comparison with (\ref{su_11}). We have in this range $Q (z) \approx -2/\sqrt{z}$. Then $\beta_N Q(z)$ becomes of order one at $z \sim \beta^2_N$, where
 the argument of $f (z/\beta^2_N)$ also becomes of order one. This sets $z \sim \beta^2_N$ as the scale where
  Eqs. (\ref{dd_1}) and (\ref{bbbb}) match.

 The scale $z \sim \beta^2_N$ (or, equivalently, $\omega_m \sim \omega_{max} \sim g (c/\gamma)^{1/\gamma}$, $c = O(1)$) is also the scale at which the boundary condition at $z \to \infty$ becomes relevant and the phase $\phi$ gets locked at $\phi = \phi_N$.  To see this,  we substituted  $\Phi_{\text{diff}} (z) = \Delta_{\text{diff}} (z) (1+z)/z$ back into the r.h.s. of (\ref{su_6}) and evaluated the integrals.  We found that
 $\Phi_{\text{diff}} (z) \propto  z^{-1/2} \cos{(\beta_N \log{z} + \phi)} $  at $z <1$ and   $\Phi_{\text{diff}} (z) \propto z^{-1/4} \cos{(2\beta_N/\sqrt{z} + \phi)}$ at $1< z < \beta^2_N$ are reproduced for any $\phi$, and the corresponding integrals are confined to internal $y \sim z$ in (\ref{su_6}) (for $1< z < \beta^2_N$ the integrals are expressed via Fresnel C-function, which have to be expanded to appropriate order).
 However, for $z > \beta^2_N$, Eq. (\ref{su_6}) is reproduced only if we set $\phi = \phi_N$.

 \begin{figure}
	\begin{center}
		\includegraphics[width=0.6\columnwidth]{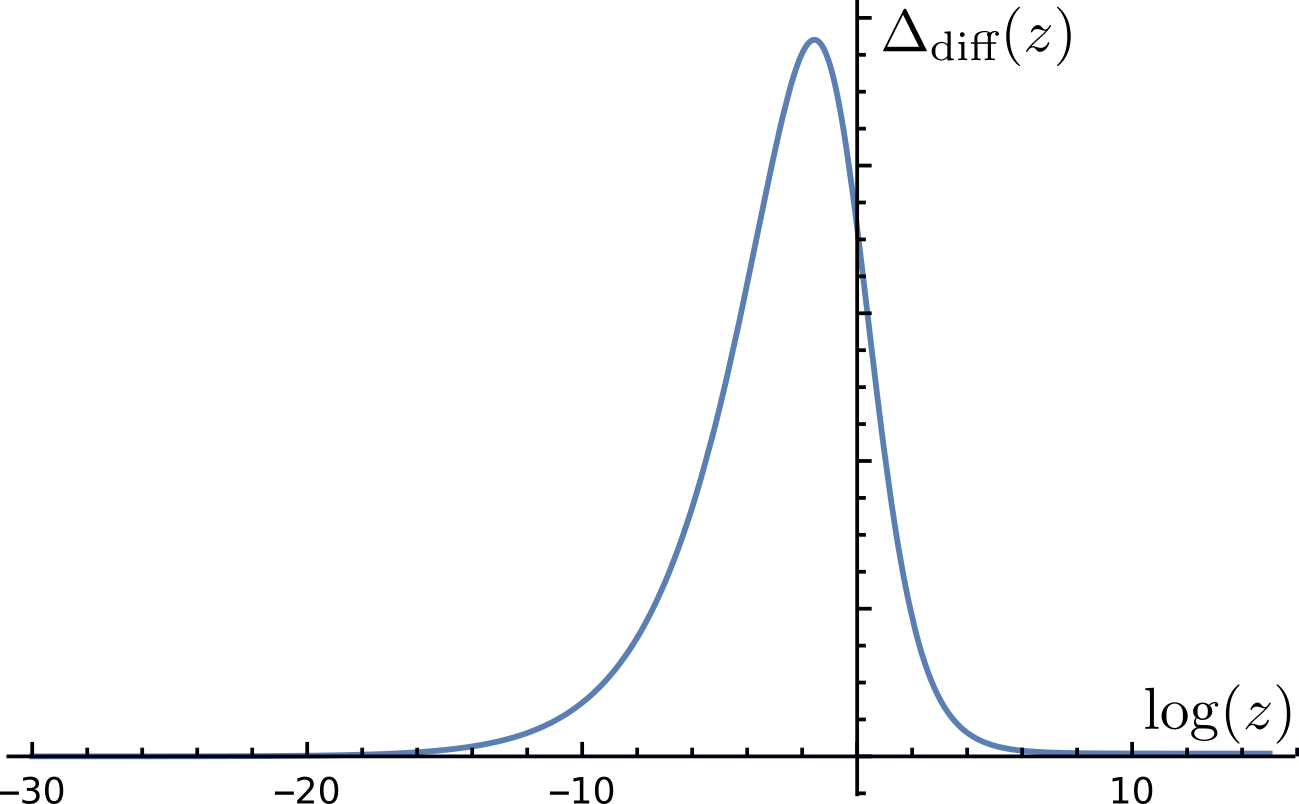}
        \caption{The gap function $\Delta_{\text{diff}} (z)$ for $N = N_{cr} -0$ ($\beta_N \to 0$). The function is sign-preserving. It scales as $\log{z}
        \sqrt{z}$ at small $z$, passes through a maximum at $z \sim 0.2$, and decreases as $1/z$ at $z >1$.}
\label{fig10}
	\end{center}
\end{figure}

\subsection{${\bf{ N=N_{cr}}}$}
\label{sec:critical_N_d}

 We now analyze the form of $\Delta_{\text{diff}}  (z)$ at $N= N_{cr}$, where, we expect, the normalized solution for $\Delta_{\text{diff}}$ first to appear.
 This analysis requires the same extra care as we exercised in Sec. \ref{sec:critical_N_b},
 when we analyzed the truncated gap equation. Namely, we need to take the limit
   $\beta_N \to 0$ rather than just set $\beta_N =0$.
   To take the limit properly, we re-express Eq. (\ref{su_17}) as
  \beq
  \Delta_{{\text{diff}}} (z) = C_1 S(\beta_N,z) + C_2 S(-\beta_N,z)
  \label{cccc}
  \eeq
   where
   \beq
   S(\beta_N,z) = z^{i\beta_N-1/2} {_2}F{_1} [1/2+i\beta_N, 3/2+i\beta_N, 1+2i\beta_N, -z],
   \eeq
    and $C_1$ and $C_2$ are two independent free parameters.  Expanding to linear order in $\beta_N$, we obtain
  \beq
  \Delta_{{\text{diff}}} (z) = C^*_1 S(0,z) + C^*_2 S'(0,z)
  \label{cccc_1}
  \eeq
 where the derivative is with respect to $i\beta_N$. As long as $\beta_N$ is non-zero,  $C^*_1$ and $C^*_2$ are  two independent parameters.
 The functions $S(0,z)$ and $S' (0,z)$ are expressed in terms of Meijer G-function and tend to finite values at large $z$, with subleading term of order $1/z$. Then  $ \Delta_{{\text{diff}}} (z) =a + b/z$ at large $z$.
   A constant  $a$ vanishes once we choose  $C^*_2/C^*_1 = 1.294$. For this particular $C^*_2/C^*_1$, $\Delta_{{\text{diff}}} (z)$ from (\ref{cccc}) satisfies the boundary condition at $z \to \infty$.  This implies that the  solution does indeed exist at $N = N_{cr} -0$.
   At small $z$,  $S(0,z) \approx
   \sqrt{z}$, $S'(0,z)\approx \log{z}
   \sqrt{z}$, hence, $\Delta (z) \propto \log{z}
    \sqrt{z}$, consistent with (\ref{eq:boundary_2}).    Computing the subleading terms and using  $C^*_2/C^*_1 =1.294$,
    we obtain at $z < 1$,
    \beq
\Delta_{{\text{diff}}} (z) = C \sqrt{z} \left(1 - \frac{3z}{4}\right)  \left(\log{\frac{z}{2.165}} - \frac{z}{4} + O(z^2)\right).
\label{su_20}
 \eeq
  In Fig. \ref{fig10} we show $\Delta_{{\text{diff}}} (z)$ over the whole range of $z$.  We see that $\Delta_{{\text{diff}}} (z)$ does not oscillate. It monotonically increases with $z$ at small $z$,  passes through a maximum at $z \sim 0.2$,  and then decreases as $1/z$.
We will study the consequences of this behavior  in Sec. \ref{sec:nonlinear}, where we analyze the non-linear gap equation.
 A similar behavior has been obtained in Ref. \cite{Wang_H_17}. These authors, however,  put a hard UV cutoff $\Delta' (z) =0$ at the maximum (at $z =0.2$ in our case),  and only analyzed the range  $z <0.2$.

\subsection{${\bf {N > N_{cr}}}$}

At $N > N_{cr}$, the exponent $b_N = ((N-N_{cr})/4N)^{1/2}$ is a real number, and the solution of (\ref{su_10}) is
\begin{widetext}
     \beq
     \Delta_{{\text{diff}}} (z) = C_1 S[b_N,z] + C_2 S[-b_N,z]
     \eeq
     where
     \beq
     S[b_B,z]= x^{b_N+1/2} {_2}F{_1} \left[\frac{1}{2} +b_N, \frac{3}{2} + b_N, 1 +2 b_N, -z\right].
 \label{su_9_3}
 \eeq
 \end{widetext}
 The boundary conditions at $z \gg 1$ and $z \ll 1$, Eqs. (\ref{eq:boundary_1}) and (\ref{eq:boundary_2}),
 set two conditions on the prefactors:
   \beq
    C_2 =0 ~~{\text {and}} ~~
\frac{C_2}{C_1} = - \frac{\Gamma(3/2 - b_N) \Gamma(1/2-b_N) \Gamma(1 + 2b_N)}{\Gamma(3/2+b_N)\Gamma(1/2+b_N) \Gamma(1-2b_N)}.
\eeq
These two conditions are incompatible for any $b_N$, hence, there is no normalized solution of the gap equation for $N > N_{cr}$.

\subsection{Larger $\gamma$}

At larger $\gamma$, the approximations used to reduce the integral equation for the pairing vertex to Eq. (\ref{su_6}) are not justified.  If we formally extend (\ref{su_6}) to $\gamma \leq 1$, we still obtain Eq. (\ref{su_7}), but with  $N^{{\text{diff}}}_{cr} = 4(1-\gamma)/\gamma$ instead of the actual $N_{cr}$ given by (\ref{su_15_1}).   At small $\gamma$,  $N^{{\text{diff}}}_{cr} \approx N_{cr} \approx 4/\gamma$, but for larger $\gamma$, they differ. The difference becomes crucial for $\gamma \to 1$, where $N^{{\text{diff}}}_{cr}$ tends to zero, while the actual $N_{cr}$  approaches $1$.  This can be partly corrected  by (i) expressing the differential equation in terms of  $\beta_N$, as in (\ref{su_7_11}), and using the correct expression for $\beta_N$ in terms of $\gamma$ and $N$, and (ii)  modifying the derivation of the differential equation by adding the contribution from internal $\omega' \approx \omega$,  as we show in Appendix (\ref{app:largergamma}).  With this additional contribution, the differential equation for $N \leq N_{cr}$  remains the same as Eq. (\ref{su_7_11}), but $z$ is rescaled to
${\bar z} = z/d$, where  $d = 4(1-\gamma)/(N\gamma (4 \beta^2_N +1))$.  For small $\gamma$, $d \approx 1$, but for $\gamma \to 1$ and $N \to 1$, $d = (1-\gamma)/(\beta^2_N +1/4)$. In this limit $z = (|\omega_m|/g) (1-\gamma)$ vanishes, but ${\bar z} \approx (|\omega_m|/g) (\beta^2_N +1/4)$ remains finite.

Still, the assumption that the approximation of the actual
 integral gap equation by the differential one can be rigorously justified only for small  $\gamma$, when the pairing interaction is a weak function of frequency.  At larger $\gamma = O(1)$ one must analyze the actual, integral gap equation.  This is what we will do next.

 \section{The exact solution of the linearized gap equation}
\label{sec:exact_solution}

In this section we prove that for any $\gamma$ in the interval  $0<\gamma < 1$, the
 linearized gap equation  has the solution for any $N \leq N_{cr}$.
  We obtain the exact analytical solution that satisfies Eq. \eqref{eq:lineargap_1_1} and
 the  normalization condition, Eq. (\ref{a_5_2}).  The analysis is somewhat involved, so we discuss the details
in Appendix \ref{app:exact} and here list the computational steps and present the final result.

We start by  re-writing Eq.  \eqref{eq:lineargap_1_1}  as  eigen-value/eigen-function equation of a linear integral operator
\begin{equation}\label{eq:eq1}
N\Phi (\bo_m)=\frac{1-\gamma}{2}\int_{-\infty }^{\infty } \frac{\Phi (\bo'_m )d\bo ' }{|\bo_m -\bo'_m |^{\gamma }|\bo'_m|^{1-\gamma }}\frac{1}{1+|\bo'_m|^{\gamma }}
\end{equation}
We then  introduce a set of functions
\begin{equation}\label{eq:basisN}
\Phi_{\beta }(\Omega_m )\equiv \frac{|\Omega_m |^{i\beta \gamma +\delta_{\Omega_m }}}{|\Omega_m |^{\gamma /2}},
\end{equation}
  which obey the orthogonality condition:
  \beq
  \int \frac{d\Omega_m}{|\Omega_m|^{1-\gamma}} \Phi^*_\beta \Phi_{\beta'} = \int dy e^{-i \gamma(\beta-\beta') y } = \frac{2\pi}{\gamma} \delta (\beta -\beta'),
  \label{ll_1}
  \eeq
 and define the function $b_{\beta}$ as
\begin{equation}\label{eq:abetaN}
b_{\beta}=  \frac{1}{2\epsilon_{\beta}}\int \frac{d\Omega_m }{|\Omega_m |^{1-\gamma }}\Phi_{\beta}(\Omega_m )\Phi (\Omega_m ).
\end{equation}

\begin{figure}
\includegraphics[width=.9\columnwidth]{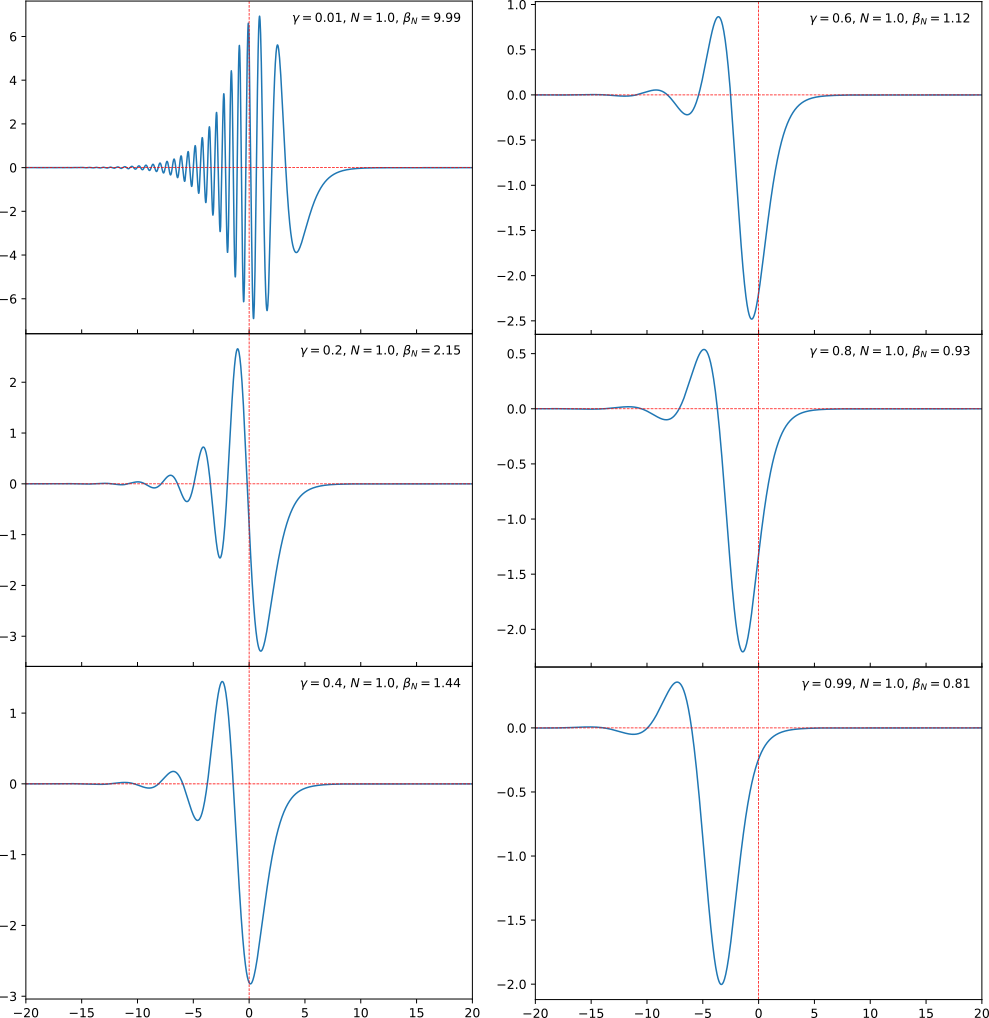}
\caption{\label{fig:fN}
The exact solution of the linearized gap equation, $\Delta_{\text{ex}} (x)$, as a function of $x = \log{(|\omega_m|/\omega_0)^{\gamma}}$ for $N=1$ and various $\gamma $ from the interval $0 <\gamma <1$.  For all $\gamma$, $\Delta_{\text{ex}} (x)$ oscillates at small frequencies with the period set by $x$, and decays as
$(\omega_0/|\omega_m|)^\gamma = e^{-x}$ at large $x$.  }
\end{figure}

\begin{figure}
\includegraphics[width=3in]{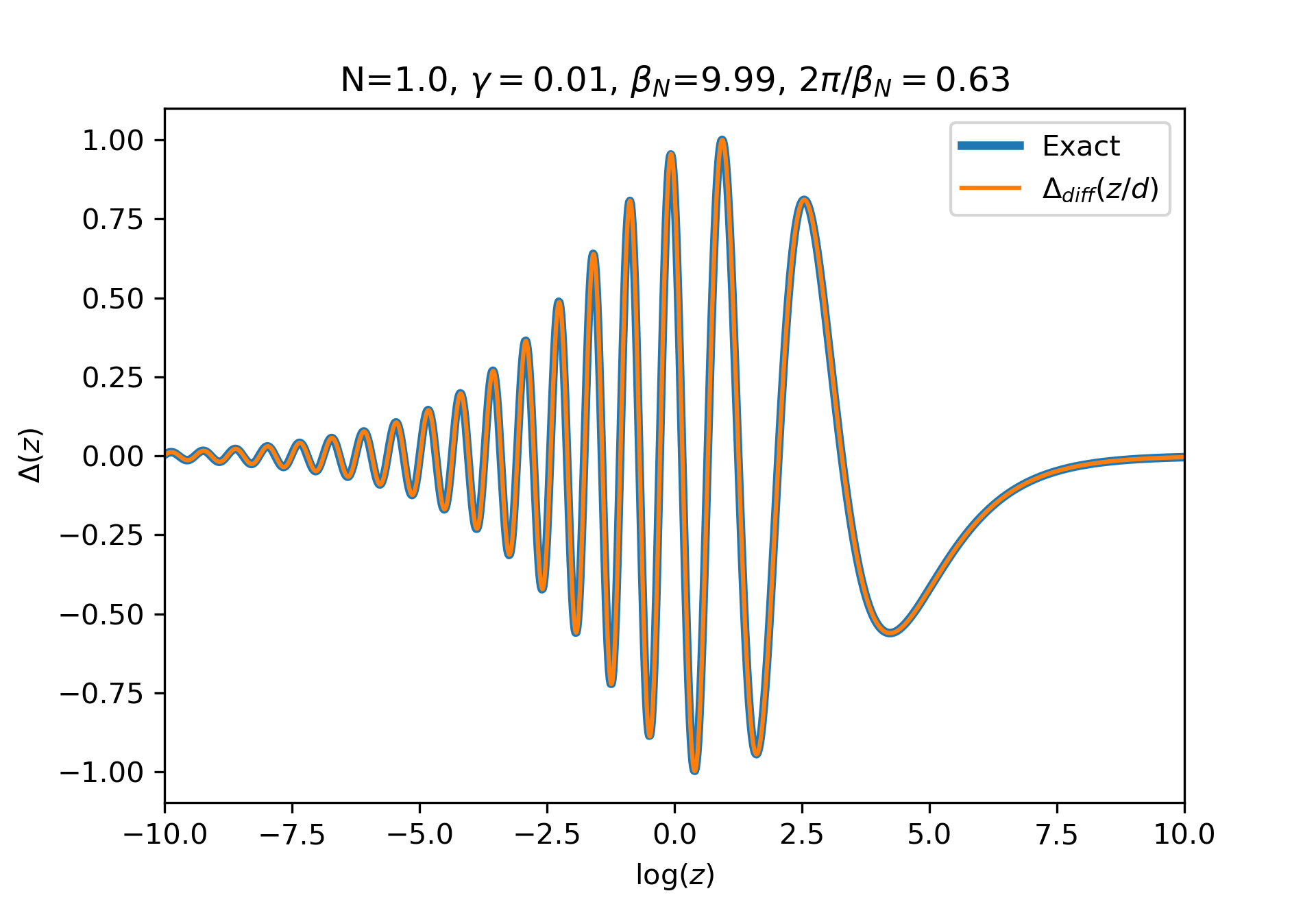}
\includegraphics[width=3in]{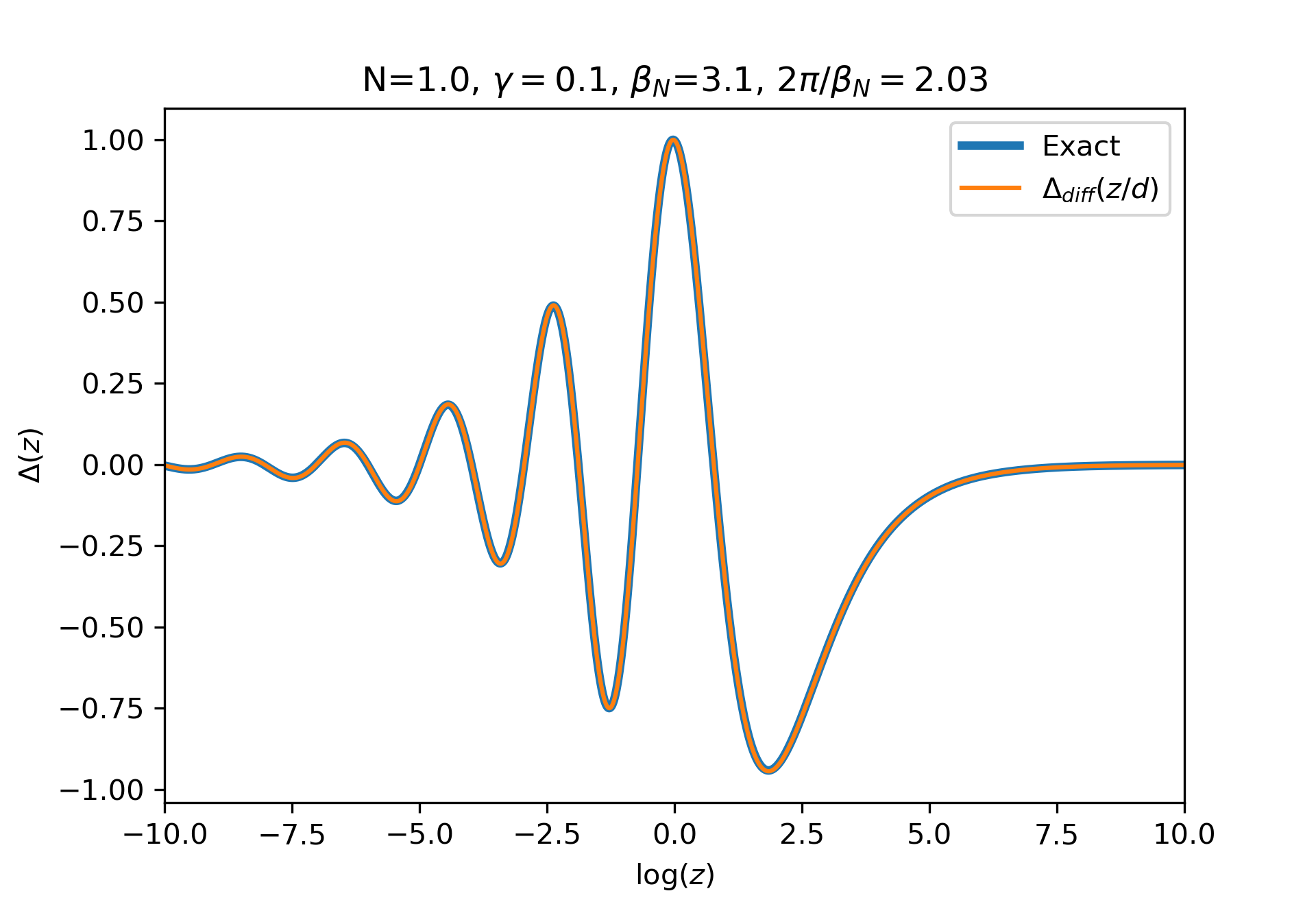}
\includegraphics[width=3in]{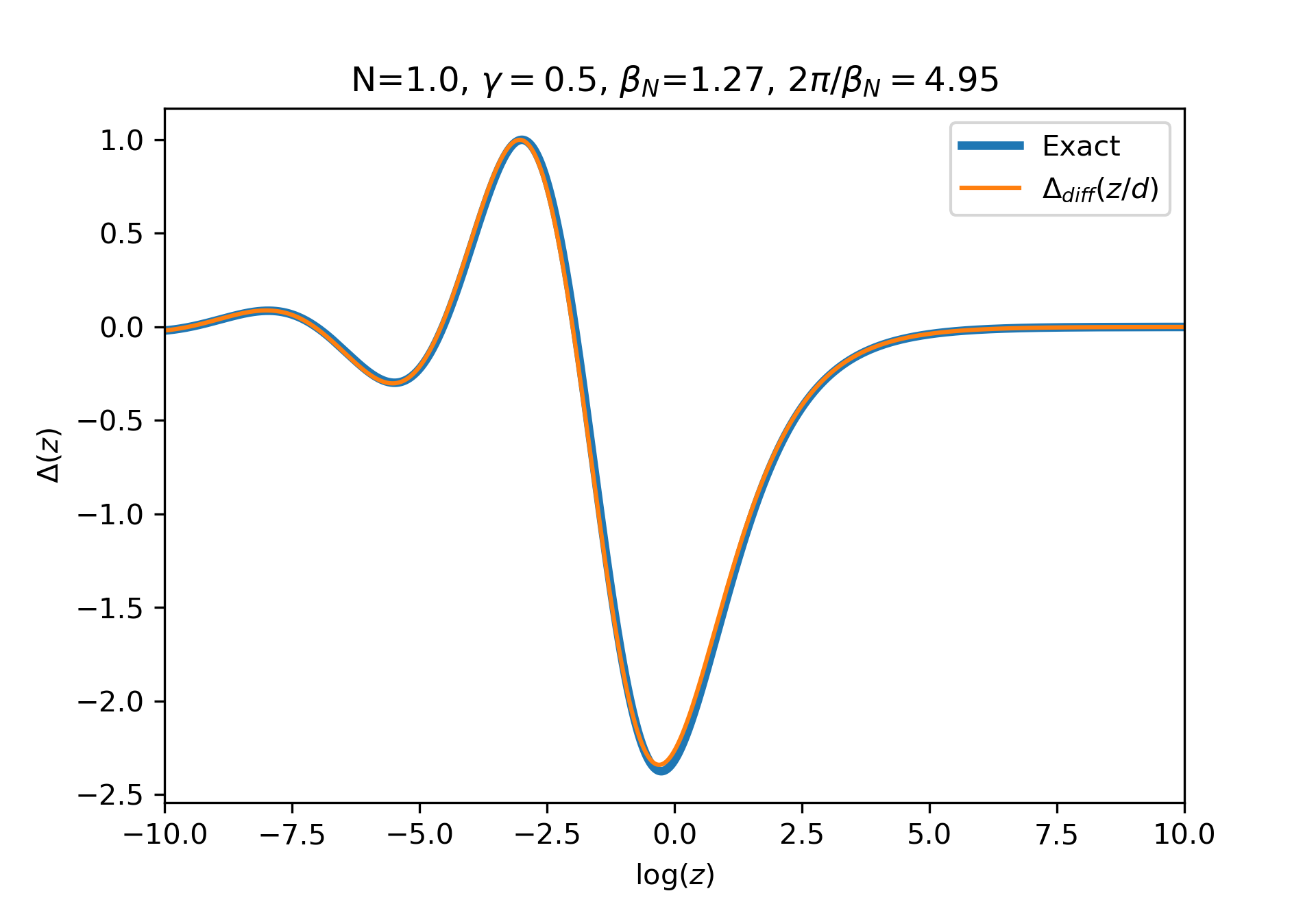}
\includegraphics[width=3in]{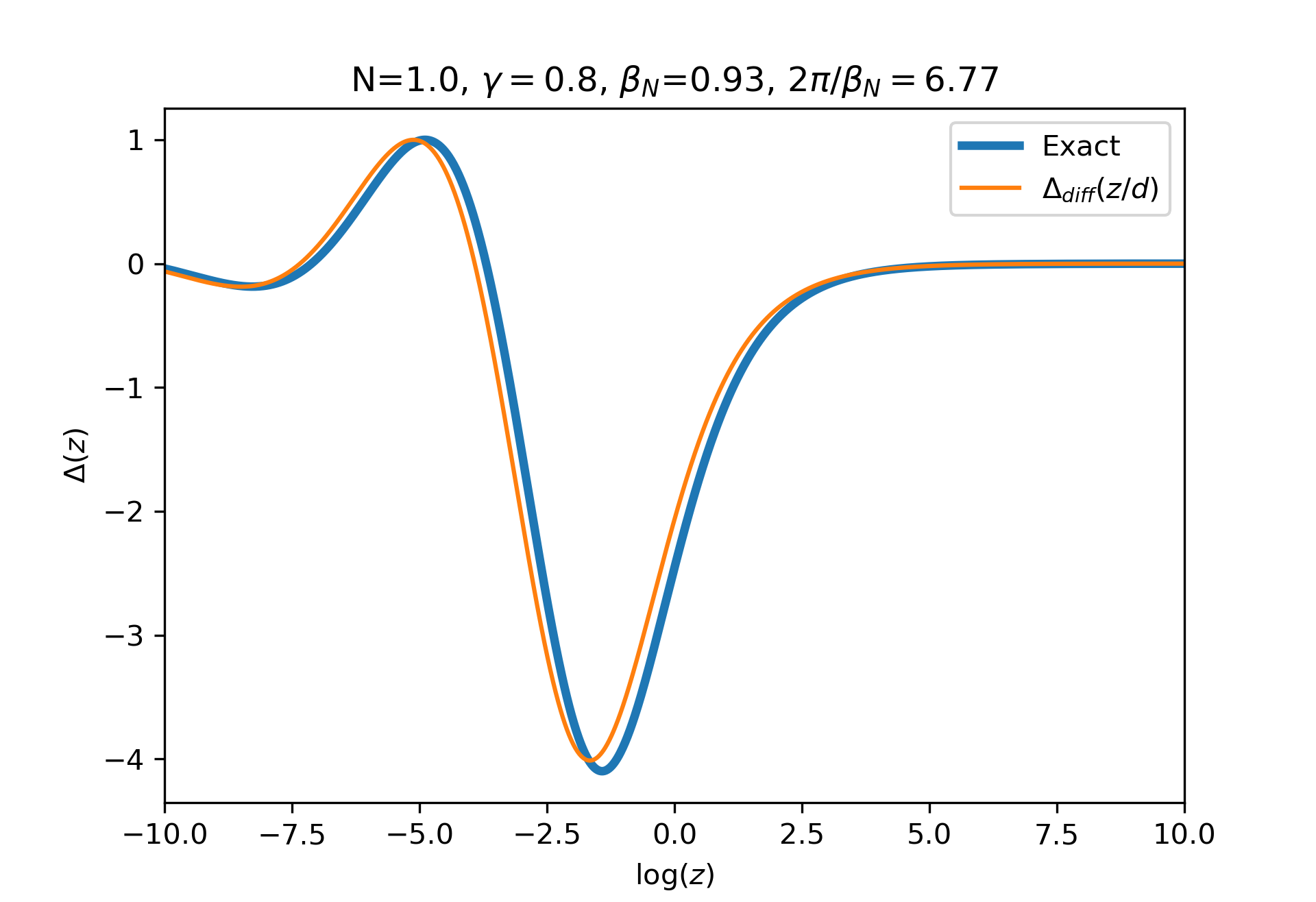}
\includegraphics[width=3in]{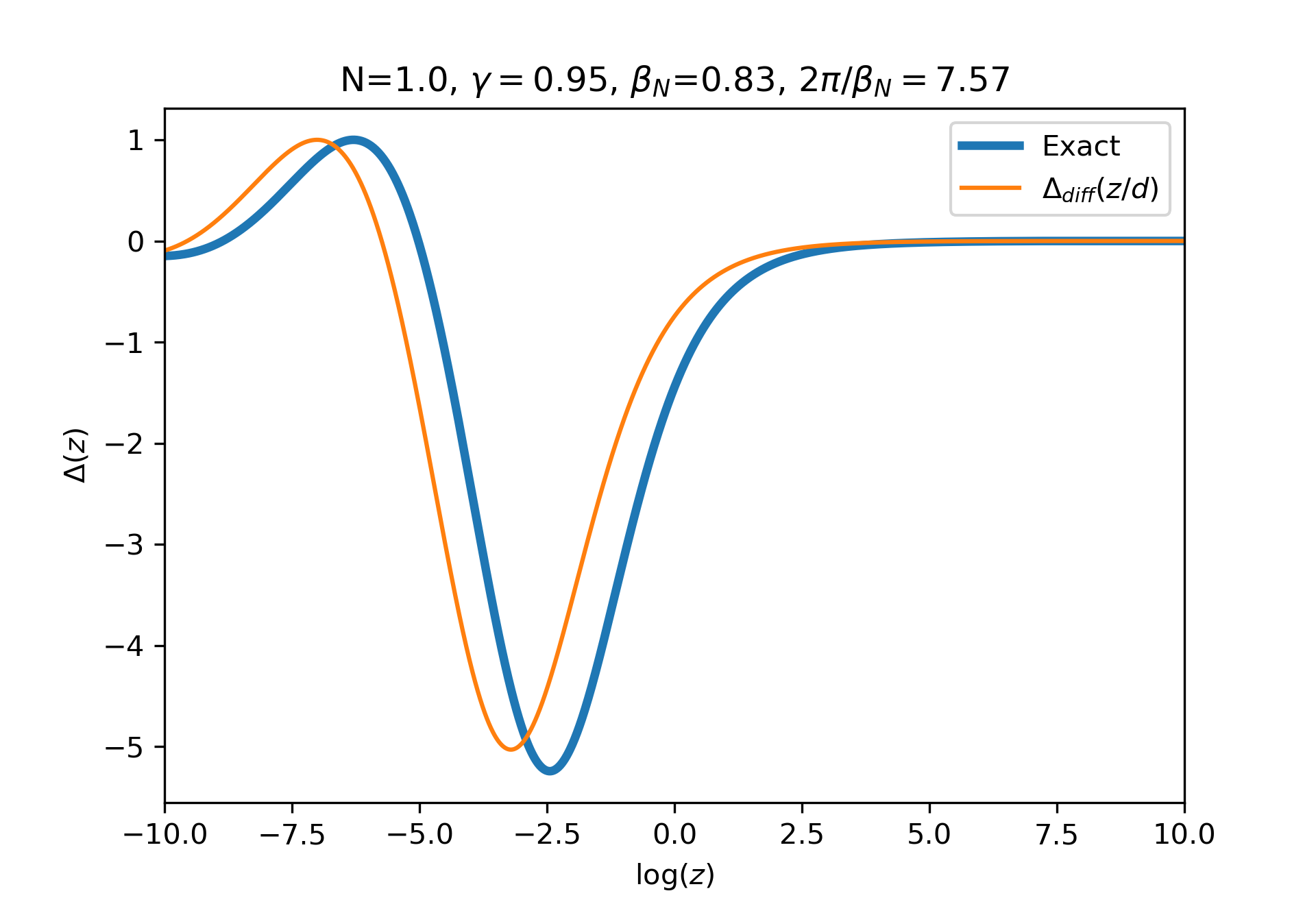}
\caption{\label{fig_11}
The exact  $\Delta_{\text{ex}} (x)$ and the solution of the differential gap equation $\Delta_{\text{diff}} (x)$
(Eq. (\ref{su_17})  as functions of $\log z = \log((|\omega_m|/\omega_0)^\gamma)$.  For $\Delta_{\text{diff}}$, we  rescaled $z$  to
${\bar z} = z/d$, where  $d = 4(1-\gamma)/(N\gamma (4 \beta^2_N +1))$, see  Appendix \ref{app:largergamma}.}
\end{figure}
 Multiplying both sides of Eq. \eqref{eq:eq1} by $\Phi_\beta (\Omega_m)$ and  integrating over $d\Omega_m/|\Omega_m|^{1-\gamma}$, we obtain the equation
  for
  $b_\beta$:
\begin{equation}\label{eq:mainN}
 Nb_{\beta}= i \int_{-\infty }^{\infty }  \frac{\epsilon_{
 i\beta'} }{\sinh \left(\pi (\beta' -\beta +i0) \right)}b_{\beta'} d\beta'
\end{equation}
 where $\epsilon_
 {i\beta}$ is defined in (\ref{su_15_2}).
 The normalizability condition for
 $\Phi (\bo_m)$ for given $N$ and $\gamma$ is expressed in terms of $b_\beta$ as  the  non-divergence of
 the integral
 \begin{equation}\label{eq:normN}
\int_{-\infty }^{\infty } \left[ 1-\frac{1}{N}\epsilon _{\beta'}\right] \epsilon _{\beta'}  b_{\beta'} b^{\ast}_{\beta'} d\beta' .
\end{equation}
 We solve Eq. (\ref{eq:mainN})
 subject to (\ref{eq:normN})  in  Appendix \ref{app:exact} and use (\ref{eq:abetaN}) and (\ref{ll_1})  to extract $\Phi (\bo_m)$ and $\Delta (\bo_m) = \Phi (\bo_m) /\left(1 + (\bo_m)^{-\gamma}\right)$.
  We find that there is no normalizable solution for $N > N_{cr}$, but for
  $N \leq N_{cr}$, the  normalizable solution does exist.
  The solution $\Delta_{\text{ex}} (z)$ is expressed as
\begin{equation}\label{eq:fTildeBN}
 \Delta_{\text{ex}} (z) =\int_{-\infty }^{\infty }\frac{\sinh (\pi \beta_N)e^{-i\beta {\log z}- \frac{i}{2} I (\beta)}d \beta}{\sqrt{\cosh (\pi (\beta-\beta_N)\cosh ( \pi (\beta+\beta_N))}}.
\end{equation}
  Here $\beta_N$ is the same as before (the solution of $\epsilon_{\beta_N} =N$),
  and the function $I(\beta)$ is given by
\begin{equation}\label{eq:IEN}
I (\beta)= \int_{-\infty }^{\infty } d \beta^{'} \log \left|1-\frac{1}{N}\epsilon_{\beta^{'}}\right|\left[ \tanh \left(\pi (\beta^{'}-\beta)\right)-\tanh \left(\pi \beta^{'} \right)\right].
\end{equation}
The gap function $\Delta_{\text{ex}} (z)$ likely cannot be represented by a simple analytical formula, but can be straightforwardly computed numerically.
   We plot $\Delta_{\text{ex}} (z)$ for different $\gamma$ and $N=1$ in Fig. \ref{fig:fN}.  To show both the oscillations at small $z$ and the power-law behavior at larger $z$, we use  the logarithmical variable
 $ x = \log (\bo_m)^\gamma = \log{z}$, defined in (\ref{ll_2}).

 \begin{figure}
\includegraphics[width=0.5\columnwidth]{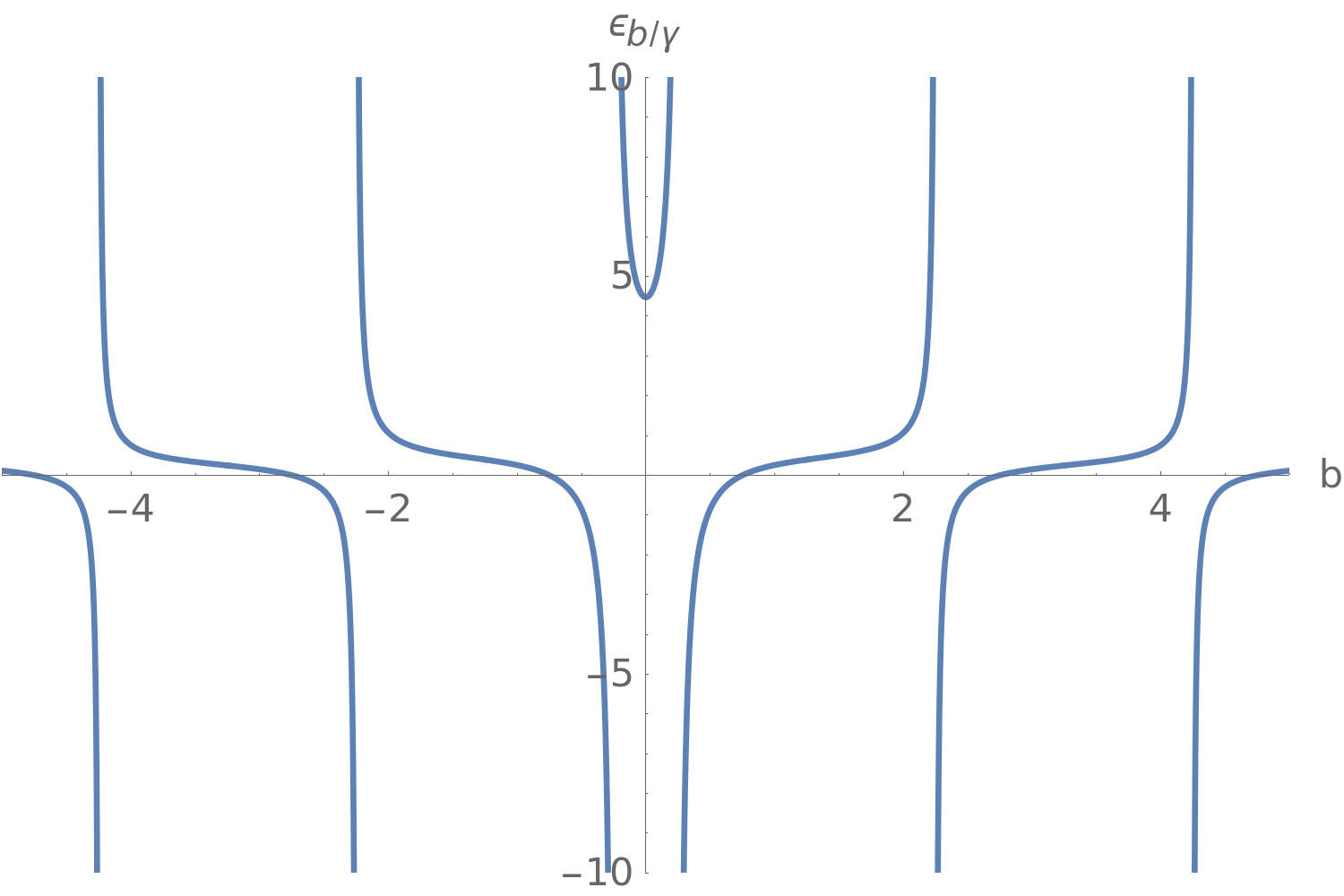}
\caption{\label{fig:epsilon}
 The function $\epsilon_{b/\gamma}$  from Eq. (\ref{su_15}) for $\gamma =0.5$. }
\end{figure}

\subsection{The exact $\Delta_{\text{ex}} (z)$ vs $\Delta_{\text{diff}} (z)$}
\label{sec:comparison}

In Fig. \ref{fig_11} we compare the exact $\Delta_{\text{ex}} (z)$ for $N=1$ and various $\gamma$ with the solution of the differential equation $\Delta_{\text{diff}} (z)$ (both are expressed in terms of $x = \log{z}$).  We see that the two essentially coincide for small $\gamma$ and remain close to each other for $\gamma = O(1)$, but the discrepancy increases with increasing $\gamma$. To understand where the discrepancy comes from, we expanded the exact $\Delta_{\text{ex}} (z)$ in powers of $z$ at small $z$ and in powers of $1/z$ at large $z$ and compared with the expansion of $\Delta_{\text{diff}}$. We present the details of the calculations in Appendix \ref{app:asymptotic} and here list the results.  The expansion of $\Delta_{ex} (z)$  in $z$ holds in
\begin{equation}\label{eq:x<}
 \Delta_{\text{ex}} (z)= \frac{1}{1+z} {\text Re}~ \sum_{n=0}^{\infty } e^{(i \beta_N \log{z} + \phi_N)} C^{<}_n~ z^{n+1/2}
+  \sum_{n,m=0}^{\infty }D^{<}_{n,m} z^{(n + b_m/\gamma+1/2)}.
\end{equation}
 The expansion of $ \Delta_{\text{ex}} (z)$ in $1/z$ holds in
 \begin{equation}\label{eq:x>}
 \Delta_{\text{ex}} (z)= \sum_{n=0}^\infty C^{>}_n~ z^{-(n+1)} +
 \sum_{n,m=0}^{\infty }D^{>}_{n,m} z^{-(n + 2(m+1)/\gamma+1)}
\end{equation}
 The real $b_m$ in (\ref{eq:x<}) satisfy $\epsilon_{b_m/\gamma} = N$. There are no solutions for $|b_m| < \gamma/2$, but there are solutions for larger $b$, see Fig. \ref{fig:epsilon}. A given $b_m$ is in the interval $\gamma/2 + 2m < b_m < \gamma/2 + 2(m+1)$.   The crossover between the two regimes is at $z = z_{max} = (\omega_{max}/\omega_0)^\gamma$. This is the scale at which oscillating behavior of $\Delta_{\text{ex}} (z)$ crosses over to $1/z$  behavior.

 The expansion of $\Delta_{\text {diff}} (z)$ yields
 \begin{equation}\label{eq:x<_d}
 \Delta_{\text{diff}} (z)= \frac{1}{1+z}
 {\text Re}~ \sum_{n=0}^{\infty } e^{(i \beta_N \log{z} + \phi_N)} {\tilde C}^{<}_n~z^{n+1/2}
 \end{equation}
 for the expansion in $z$, and
 \begin{equation}\label{eq:x>_d}
 \Delta_{\text{diff}} (z)= \sum_{n=0}^\infty {\tilde C}^{>}_n~ z^{-(n+1)}
\end{equation}
  for the expansion in $1/z$.

We see that the expansion of $\Delta_{\text {diff}} (z)$ is a regular expansion in powers of either $z$ or $1/z$.
 The coefficients ${\tilde C}^{<}_n$ and ${\tilde C}^{>}_n$ can be easily obtained by expanding the hypergeometric function either in $z$ or in $1/z$, see Eqs. (\ref{dd_1}) and (\ref{dd_1a}).

The expansion of $\Delta_{\text {ex}} (z)$ is more complex and contains  two types of terms: the $C_n$ series and $D_{n,m}$ series.
 The $C^{>}_n$ series in (\ref{eq:x<}) are  local terms in the sense that in the direct
    perturbative expansion  of the gap equation (\ref{ss_111})
     they come
  from $\omega'_m \sim \omega_m$ in the integral in (\ref{ss_111}).
 The $D^{<}_{n,m}$ series describe non-local corrections, for which the integral over $\omega'$ is  determined by $\omega_{max} \gg  \omega_m$.
Examining perturbation series, we find that the coefficients $C^{<}_n$ can be obtained analytically and are given by
 \beq
  C^{<}_n = C^{<}_0 ~ \left[I_n  \displaystyle\prod_{m=1}^{n} \frac{1}{I_m-1}\right]
  \label{dd_10}
  \eeq
  where
 \begin{widetext}
 \bea
&&I_m = \frac{1-\gamma}{2N} Q(m, \gamma, \beta_N),  \nonumber \\
&&  Q(m, \gamma, \beta_N)  =  \frac{\Gamma((m+1/2)\gamma +i{\beta_N} \gamma) \Gamma((1/2-m)\gamma -i{\beta_N} \gamma)}{\Gamma(\gamma)} +  \nonumber \\
&&\Gamma(1-\gamma) \left(\frac{\Gamma((m+1/2)\gamma +i{\beta_N} \gamma)}{\Gamma(1-(1/2-m)\gamma +i{\beta_N} \gamma)} + \frac{\Gamma((1/2-m)\gamma -i{\beta_N} \gamma)}{\Gamma(1-(m+1/2)\gamma -i{\beta_N} \gamma)}\right)
\label{dd_7a}
\eea
\end{widetext}
The series do not converge absolutely because at large $n$, $ C^{<}_n \propto 1/n$, but each term in the series can be easily calculated.
At small $\gamma$ and $N = 1$, when $\beta_N \approx (N_{cr}/4N) \gg 1$, the $C^{<}_n$ series in
(\ref{eq:x<}) reproduce, order by order, the expansion of $\Delta_{\text{diff}} (z)$,  Eqs. (\ref{dd_1}) and (\ref{dd_1a}).
The same holds for the expansion in $1/z$ -- the $C^{>}_n$ series in (\ref{eq:x>}) reproduce the expansion in (\ref{bbbb}).
  The $D$ terms are small  for $\gamma \ll 1$ by a combination of two factors:  (i) the prefactors are relatively small, e.g., $D^{>}_{0,0} \approx - C^{>}_0~ \gamma/4$, and (ii)  $b_m \approx 2(m+1)$, hence, the non-local terms contain additional small factors $z^{2m/\gamma}$ in the expansion in $z$ and
  and $(1/z)^{-2m/\gamma}$ in the expansion in $1/z$.   As the consequence, for small $\gamma$, $\Delta_{\text{ex}} (z)$  and $\Delta_{\text {diff}} (z)$ coincide, up to small corrections, as Fig. \ref{fig_11}(a) demonstrates.

  At larger $\gamma = O(1)$, the non-local terms become relevant, particularly near $z_{max}$, where $\Delta_{\text{ex}} (z)$ crosses over  to $1/z$ behavior. Numbers-wise, $\Delta_{\text{diff}} (z)$ still agrees reasonably well with  $\Delta_{\text{ex}} (z)$, as Fig. \ref{fig_11} shows, but qualitatively, the exact $\Delta_{\text{ex}} (z)$ is different from $\Delta_{\text{diff}} (z)$.

\section{Non-linear gap equation}
\label{sec:nonlinear}

We now analyze the full non-linear  gap equation, Eq. (\ref{ss_111}).  We argue that  it
  has an infinite, discrete set of solutions, which can be specified by topological index $n$, and the limit $n \to \infty$ corresponds to the solution of the linearized gap equation.

\subsection{Qualitative reasoning}

We begin with qualitative reasoning. We assume that for any finite $n$, the gap function $\Delta_n (\omega_m)$ tends to a finite value at $\omega_m =0$, while at $\omega_m > \omega^*_n$, $\Delta_n (\omega_m)$ is small and has the same dependence on $\omega_m$ as $\Delta_{\infty} (\omega_m) = \Delta_{\text{ex}} (\omega_m)$.  We further assume that $\omega^*_n$ is the only relevant scale for $\Delta_n (\omega_m)$, i.e., $\Delta_n (\omega_m)$  saturates at $\omega_m <\omega^*_n$.  This reasoning has two implications. First, $\Delta_n (\omega^*_n) \sim \omega^*_n$. Second, $\Delta_n (\omega_m)$  must satisfy the modified linearized gap equation, in which the lower limit of integration over positive and negative $\omega'_m$ is set at a frequency of order $\omega^*_n$.  Because $\Delta_{\text{ex}} (\omega_m)$ is the solution for $\omega^*_n =0$, the modified linearized gap equation has a solution if
   \beq
   \int_0^{\omega^*_n} d \omega_m' \frac{\Delta_{ex} (\omega'_m)}{\omega'_m} =0.
   \label{sat_11}
   \eeq
  These two conditions determine $\omega^*_n$ and $\Delta_n (\omega^*_n)$.
 Below we use the  variable $z$ instead of $\omega_m$ and define $z^*_n  = (\omega^*_n/\omega_0)^\gamma$.
 In these variables, $\Delta_n (0) \sim \Delta_n (z^*_n) \sim \omega_0 (z_n)^{1/\gamma}$.

 For $N \leq N_{cr}$,  $\Delta_{\text{ex}} (z)$  behaves at $z < 1$ as  $C \sqrt{z} \cos(\beta_N \log{z} + \phi_N)$, where $\beta_N \propto (N_{cr}-N)^{1/2}$ is small and
 $\phi_N = \pi/2 - 0.773 \beta_N$. Evaluating the integral in (\ref{sat_11}) and re-expressing the result in terms of $z$, we obtain
     \beq
       \sin\left(\beta_N \left(\log{z^*} - 2.77 \right)\right) =0,
       \label{sat_12a}
       \eeq
  This equation  has a discrete set of solutions for $z^* \ll 1$:  $\log{z^*} = 2.77 - \pi (n+1)/\beta_N$, where $n =0, 1, 2...$, i.e,
      \beq
       z^*_n \propto e^{-\pi (n+1)/\beta_N},
       \label{sat_12_1}
       \eeq
        All $z^*_n$, including $z^*_0$,  are exponentially small at small $\beta_N$.
    The gap functions $\Delta_n (0)
     \sim
      \omega_0 (z^*_n)^{1/\gamma}$ are also exponentially small:
     \beq
     \Delta_n (0) \propto \omega_0 e^{-\pi (n+1)/(\gamma \beta_N)}.
     \label{sat_12_2}
     \eeq
       One can easily verify that $\Delta_n (z)$ changes sign $n$ times between $z=0$ and $z = \infty$.

  The result $z^*_n \propto  e^{-\pi n/\beta_N}$ at $n \gg 1$  also follows from a generic  reasoning
  that $\Delta_{\text{ex}} (z)$
 should have an extremum at $z \sim z^*_n$ to match with a constant $\Delta_n (z) \approx \Delta_n (0)$ for smaller $z$.  We note, however, that the presence of an extremum in $\Delta_{\text{ex}} (z)$  by itself does not guarantee that the solution of the non-linear gap equation exists. Indeed, $\Delta_{\text{ex}} (z)$ for $N = N_{cr}$ has a maximum at $z = O(1)$, but there is no solution with a finite $\Delta (z)$ for this $N$.

   When  $\beta_N \gg 1$ (e.g., for $\gamma \ll 1$ and $N = O(1)$),  we can still apply the same reasoning
    for large enough $n$, when  $z^*_n$ is exponentially small in $n$. However, for $n = O(1)$,
     $z^*_n \geq 1$, and we should use appropriate expression for $\Delta_{\text{ex}} (z)$ in Eq. (\ref{sat_11}).  Replacing $\Delta_{\text{ex}} (z)$ by $\Delta_{{\text{diff}}} (z)$ as the two are very close, and using the asymptotic form of $\Delta_{{\text{diff}}} (z)$ for $z >1$: $\Delta_{{\text{diff}}} (z)  \sim (1/\sqrt{z}) J_1 (2 \beta_N/\sqrt{z})$, we obtain the condition on $z^*_n$ in the form
    \beq
    J_0 \left(\frac{2 \beta_N}{\sqrt{z^*_n}} \right) = O\left(1\right).
    \label{sat_12_3}
    \eeq
     This equation has a discrete set of solutions $ z^*_n = 4 \beta^2_N s_n$, where
      where $s_n$ is a decreasing function of $n$.
        The  corresponding
       $\Delta_n (z^*_n ) \sim \Delta_n (0) \sim \omega_0 (\beta^2_N)^{1/\gamma}$.  For small $\gamma$, $\beta^2_N \approx 1/\gamma$, hence, $\Delta_n (0) \sim \omega_0 (1/\gamma)^{1/\gamma}$
         (Refs. \cite{max_last,Wang2016,Wu_19,Emil2020}).  This analysis holds up to $n =n_{max}$, for which $z^*_{n_{max}} = O(1)$ For larger $n$,
        $z^*_n$ are given by Eq. (\ref{sat_12_1}).  Still, $\Delta_n (z)$ changes sign  $n$ times between $z=0$ and $z = \infty$.

        Note that at small $\gamma$, $\Delta_n (0)  \propto \omega_0 (z^*_n)^{1/\gamma}$ rapidly evolves  from $\Delta_n (0) \ll \omega_0$ when $z^*_n <1$ to $\Delta_n (0)  \gg \omega_0$ when $z^*_n >1$.
        In Ref. \cite{Wu_19} we solved the non-linear gap equation at small $\gamma $ and $N=1$ for sign-preserving $\Delta_0 (z)$. We obtained $\Delta_0 (0) =
         0.885 {\bar g} (1/1.4458 \gamma)^{1/\gamma}
         \equiv 0.326 \omega_0 (1/1.4458 \gamma)^{1/\gamma}$, consistent with our estimate.   The number $1.4458$
          comes from the fact that at small $\gamma$, the more explicit form of (\ref{sat_12_3}) is
          $J_0 (2\beta_N/\sqrt{z^*_0})  = O(\gamma)$ (Refs. \cite{max_last,Wang2016,Wu_19,Emil2020}).

\subsection{Non-linear differential equation}

We now corroborate qualitative  reasoning by more direct analysis in which we derive and solve the non-linear  differential  gap equation.   To derive this equation, we again depart from the original integral equations for the pairing vertex and the self-energy, Eqs.  (\ref{eq:gapeq_1}) and (\ref{eq:gapeq_1_1}), or, equivalently, from the non-linear gap equation (\ref{ss_111}) and
 approximate the integral over
  $\omega_m'$ in the l.h.s. of  (\ref{eq:gapeq_1}) by the integrals over $\omega'_m \gg \omega_m$ and $\omega'_m <\omega_m$, each time neglecting either $\omega'_m$ or $\omega_m$ in the pairing interaction.
   However, for a finite $\Delta (\omega_m)$ we have to take into account the fact
   that the self-energy has a more complex form in the presence of $\Delta$. In particular, one can easily verify that if
$\Delta (\omega_m)$ tends to a finite value at $\omega_m =0$, as we assume to be the case, $\Sigma (\omega_m)$ acquires a Fermi liquid form $\Sigma (\omega_m) \sim \omega_m (\omega_0/\Delta (0))^{\gamma}$ at the smallest $\omega_m$.
 The non-Fermi liquid
$\Sigma (\omega_m) = |\omega_m|^{1-\gamma} \omega^{\gamma}_0 {\text{sign} \omega_m} = \omega_m/z$ is  recovered at frequencies for which $|\omega_m| \gg \Delta (\omega_m)$.  These two limiting forms of the
 self-energy can be combined into the interpolation formula
 \beq
 \Sigma (\omega_m) = \omega_m \frac{\omega^\gamma_0}{(\omega^2_m + \Delta^2 (\omega_m))^{\gamma/2}}
 \label{last_1}
 \eeq
 If we include this self-energy and then  follow the same  derivation for the linearized differential gap equation, we would obtain
 \beq
 \left[{\bar \Delta} (z) \left(z + \frac{1}{(1+ D^2 (z))^{\gamma/2}}\right)\right]^{''} = - \frac{N_{cr}}{4N z^2} \frac{{\bar \Delta} (z)}{\sqrt{1+ D^2 (z)}}
 \label{last_1_1}
 \eeq
 where ${\bar \Delta}(z)= \Delta (z)/\omega_0$ and $D(z)= \Delta(z)/z^{1/\gamma}$.
 There is a second complication, however. One can verify that  a at non-zero $\Delta (0)$, the r.h.s of the actual, non-linear integral gap equation (\ref{ss_111}) (with the self-energy contribution kept in the l.h.s) has a regular expansion in powers of $\omega^2_m$. Indeed, a direct expansion in $\omega_m$  yields
 \beq
 \frac{ {\bar g}^\gamma }{2N} \int d \omega'_m  \frac{\Delta (\omega'_{m})}{\sqrt{(\omega'_{m})^2 +\Delta^2 (\omega'_{m})}}
    ~\frac{1}{|\omega_m - \omega'_{m}|^\gamma}  =  A + B \omega^2_m
  \label{last_2}
  \eeq
  where $A$ and $B$ are given by convergent integrals:
  \bea
  A &=& \frac{ {\bar g}^\gamma }{2N} \int d \omega'_m \frac{\Delta (\omega'_{m})}{\sqrt{(\omega'_{m})^2 +\Delta^2 (\omega'_{m})}} \frac{1}{|\omega'_m|^{\gamma}} \nonumber \\
   B &=& -\frac{\gamma (\gamma+1)}{2} \frac{ {\bar g}^\gamma }{2N}
    \int \frac{d \omega'_m}{\sqrt{(\omega'_{m})^2 +\Delta^2 (\omega'_{m})} \left(\Delta (\omega_m) + \sqrt{(\omega'_{m})^2 +\Delta^2 (\omega'_{m})}\right)} \frac{1}{|\omega'_m|^{\gamma}}.
\label{last_3}
 \eea
The combination of Eqs. (\ref{last_1}) and (\ref{last_2}) then implies that at small $\omega_m$, $\Delta (\omega_m)$ is a regular function of frequency $\Delta (\omega_m) = \Delta_0 + C \omega^2_m$, as is expected in a Fermi liquid regime.  Equivalently, ${\bar \Delta} (z) = {\bar \Delta}_0 + {\bar C}z^{2/\gamma}$.
  This regular expansion of ${\bar \Delta} (z)$ is not reproduced by Eq. (\ref{last_1_1}). The reason can be easily understood by analyzing how the $\omega^2_m$ term appears in the r.h.s. of (\ref{last_2}):
    to get it one has to keep both $\omega'_m$ and $\omega_m$ in the interaction, i.e., go beyond the approximation used to derive (\ref{last_1_1}).  A simple experimentation shows that to reproduce regular
     behavior of the gap function at small frequencies, one has to multiply the r.h.s. of (\ref{last_1_1}) by the factor $\left(\sqrt{1+D^2 (z)} - |D(z)|\right)^{1+\gamma}$.
      This factor reduces to $1$ when $D(z)$ is small, but gives additional  $z^{(1+\gamma)/\gamma}$ when $D(z)$ is large. With this extra factor, the non-linear differential gap equation takes the form
       \beq
 \left[{\bar \Delta} (z) \left(z + \frac{1}{(1+ D^2 (z))^{\gamma/2}}\right)\right]^{''} = - \frac{N_{cr}}{4N z^2} \frac{{\bar \Delta} (z)}{\sqrt{1+ D^2 (z)}} \left(\sqrt{1+D^2 (z)} - |D(z)|\right)^{1+\gamma}
 \label{last_4}
 \eeq
 One can  verify that at small $z$, the solution of this equation, if it exists, has an asymptotic form ${\bar \Delta} (z) =
 {\bar \Delta} (0) + {\bar C} z^{2/\gamma}$, consistent with asymptotic form of the solution of the actual integral equation.

The boundary conditions for (\ref{last_4}) are a finite $\Delta (0)$ and $\Delta (z) \propto 1/z$ at $z \to \infty$. The last condition is the same as for the linearized gap equation.  According to our qualitative reasoning, this equation should have an infinite, discrete set of solutions $\Delta_n (z)$.

 \begin{figure}
\includegraphics[width=0.9\columnwidth]{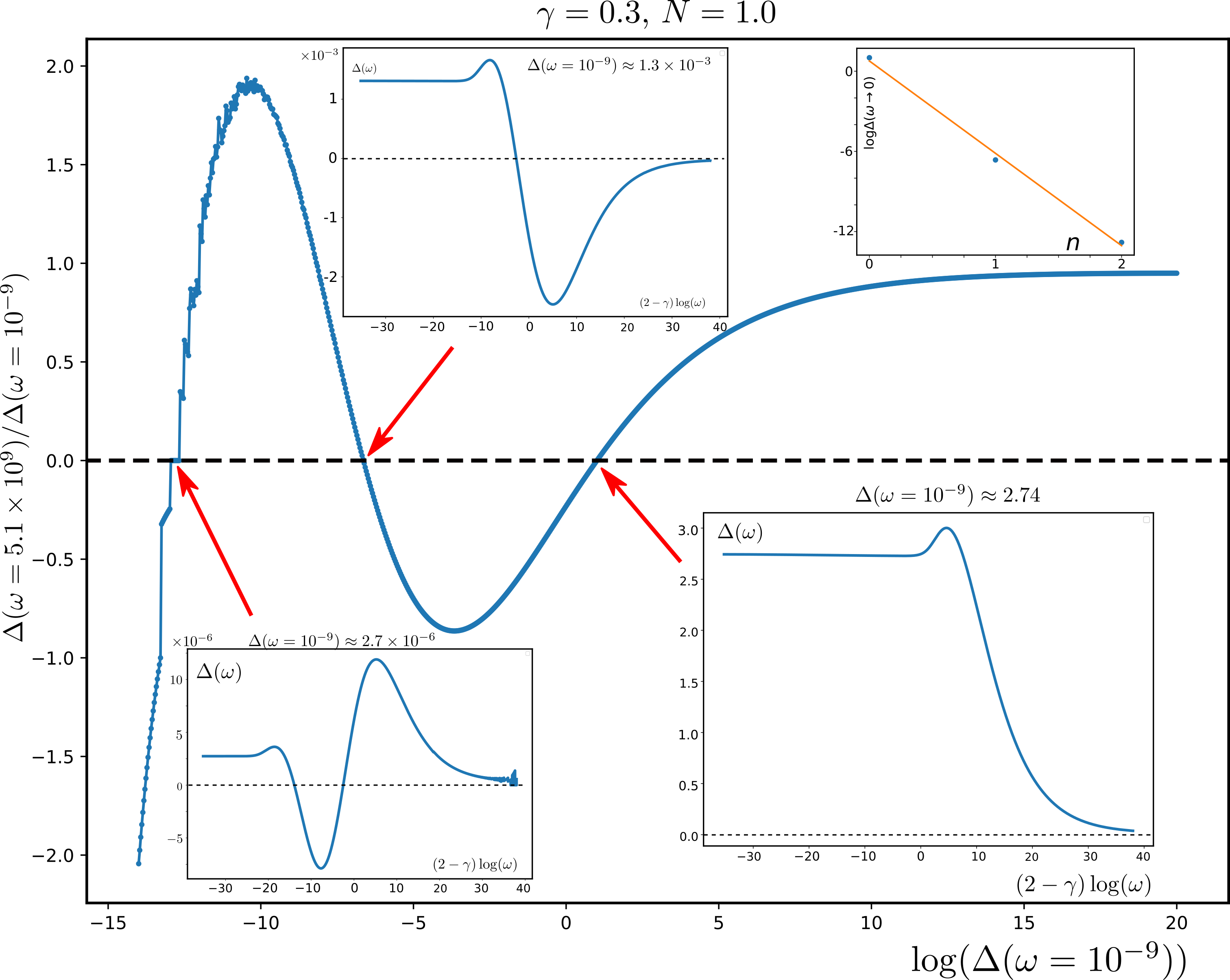}
\caption{\label{fig:NonLinearDiffN1g03}
The ratio of $\Delta (\omega  \to \infty)/\Delta (\omega \rightarrow 0)$ vs $\log \Delta (\omega \rightarrow 0)$.}
\end{figure}

We solve Eq. (\ref{last_4}) numerically.  We set $\Delta (0)$ to some value, which we vary. For a generic
$\Delta (0)$,  $\Delta (z)$ evolves with $z$ towards some constant at $z \to \infty$.  We verify whether for some particular  $\Delta (0)$, a constant vanishes and $\Delta (z)$ decays as $1/z$.
We show the result in Fig. (\ref{fig:NonLinearDiffN1g03}) for $\gamma =0.3$ and $N =1$ ($N_{cr}/N \approx 13$, $\beta_n \approx 1.75$).  We plot the ratio of $\Delta (z \to \infty)/\Delta (0)$ vs $\Delta (0)$. We see that for a generic $\Delta (0)$, $\Delta (\infty)$ remains finite, but there is a discrete set of $\Delta_n (0)$, for which $\Delta (\infty)$ vanishes.  With our numerical accuracy we can clearly  identify three such $\Delta_n (0)$.   In three inserts in Fig. (\ref{fig:NonLinearDiffN1g03}) we show $\Delta (z)$ for these three $\Delta_n (0)$. We see that the gap function changes sign $n$ times as a function of $z$ ($n =0,1,2$), precisely as we anticipated.

In the top right insert in Fig. (\ref{fig:NonLinearDiffN1g03}) we plot $\Delta_n (0)$ vs $n$ for $N=1$ and $\gamma =0.3$.
 We clearly see that $\Delta_n (0)$ has exponential dependence on $n$.  We fitted the results to Eq.  (\ref{sat_12_1}). The best fit yields $\beta_N \approx 1.51$. This is  quite consistent with the actual value $\beta_N \approx 1.75$ (we note that Eq. (\ref{sat_12_1}) is by itself only valid for $\beta_N <1$).

 \begin{figure}
\includegraphics[width=0.45\columnwidth]{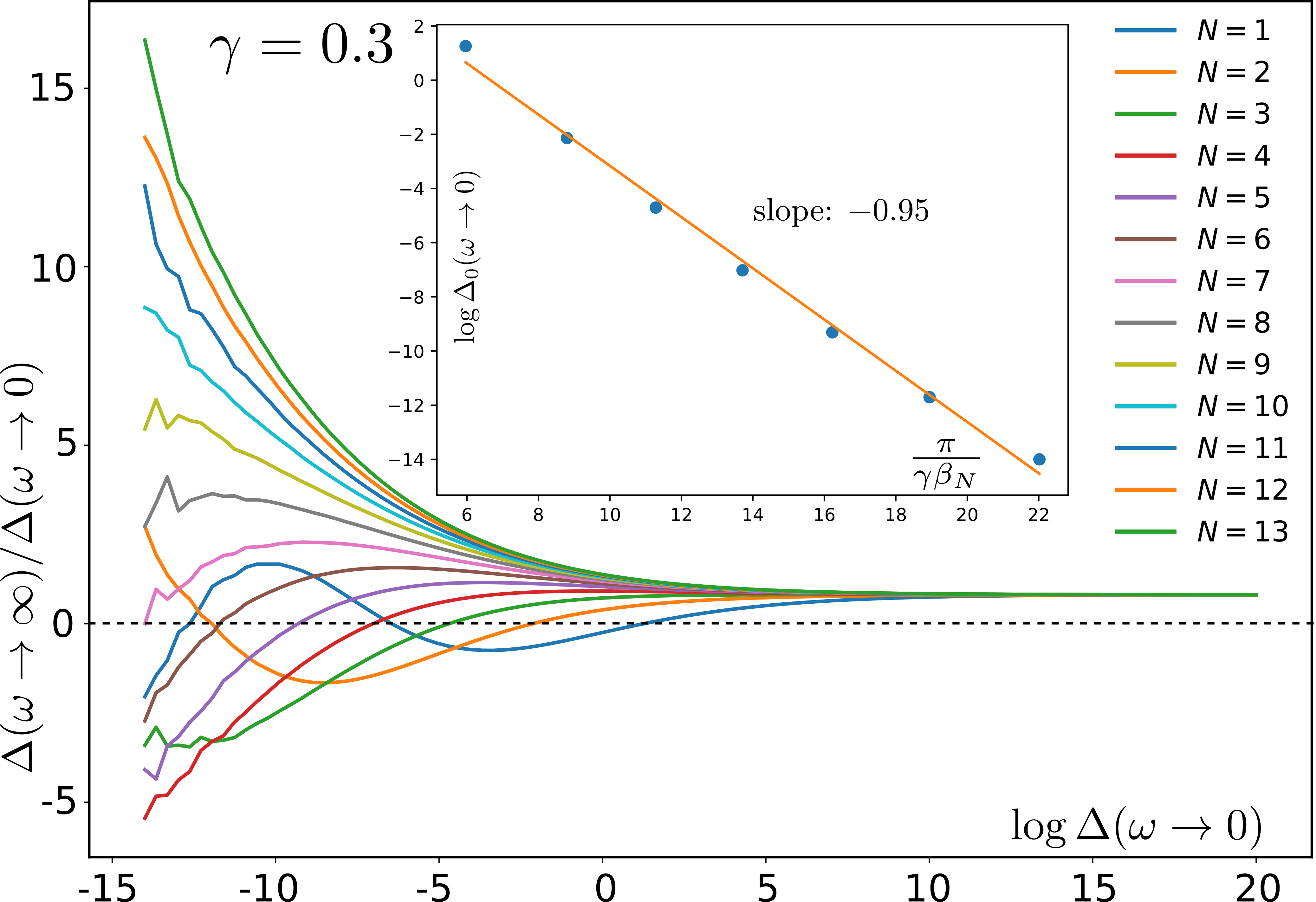}
\includegraphics[width=0.45\columnwidth]{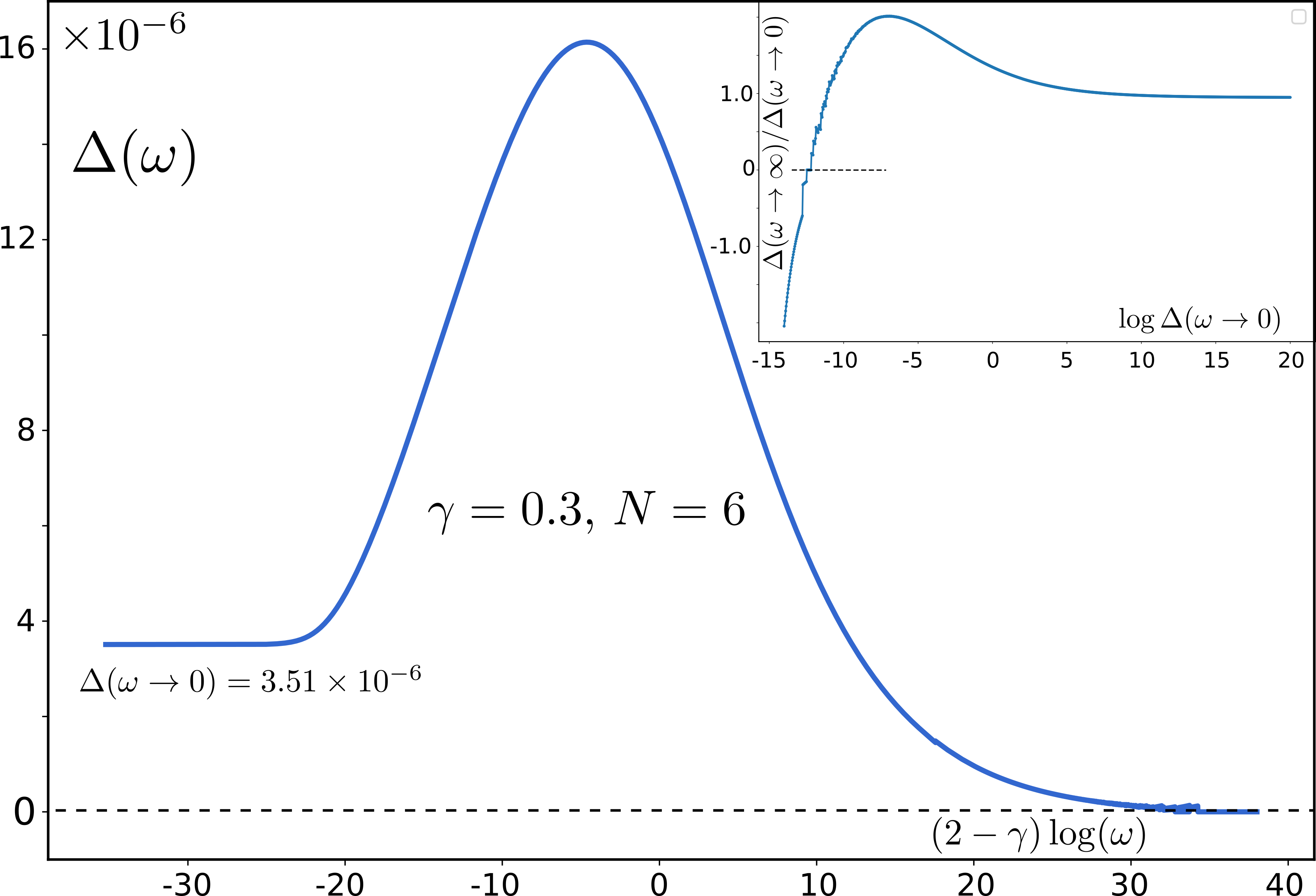}
\caption{\label{fig:NonLinearDiffFindInitialManyN}
Left: the ratio of $\Delta (\omega  \to \infty)/\Delta (\omega \rightarrow 0)$ vs $\log \Delta (\omega \rightarrow 0)$. for different $N$. The inset shows $\Delta_{0}$ vs $\frac{\pi }{\gamma \beta_{N}}$, the best fit has the slope of $-0.95$. Right $\Delta_{0}$ vs $z$ for $\gamma =0.3$ and $N=6\approx N_{cr}/2$.
}
\end{figure}

In the left panel of Fig. (\ref{fig:NonLinearDiffFindInitialManyN}) we collect the results for $\gamma =0.3$ and different $N$ between $N=1$ and $N_{cr}$,
the inset shows a comparison of $\log \Delta_0$ and
$\frac{\pi }{\gamma \beta_{n}}$,
where $\beta_N = 0.5 (N_{cr}/N-1)^{1/2}$.
  We see that the agreement is
  quite good: the theoretical slope is $-1$ (see Eq. (\ref{sat_12_2})) and the fitting of numerical data yields the slope $-0.95$.

 Finally, in the right panel of Fig. (\ref{fig:NonLinearDiffFindInitialManyN})  we plot $\Delta_0 $ vs $z$ for $N \approx N_{cr}/2$, where $\Delta_0 (0)$ is already very small. We see that the gap function flattens at $z^*  \sim (\Delta_0 (0))^{\gamma}$, which is much smaller than the scale at which
  $\Delta_0 (z)$ has a broad maximum.  This is again consistent with our qualitative reasoning.

 In the subsequent publication (Paper II, Ref.\cite{Paper_2}) we  collaborate the $T=0$ analysis  with the analysis of the linearized gap equation at a finite $T$. We show that
 for $N < N_{cr}$, there exists a discrete set of onset pairing temperatures $T_{p,n}$, and  the eigenfunction $\Delta_n (\omega_m)$  changes sign $n$ times as a function of  Matsubara frequency.  We argue that  these $\Delta_n (\omega_m)$ grow in magnitude below $T_{p,n}$, but preserve the number of sign changes,  and at $T=0$ coincide with $\Delta_n (\omega_m)$, which we found here.

We note in passing that $\Delta_n (\omega_m)$  has a maximum at $\omega \sim \omega_0 \sim {\bar g}$. The reason for the maximum is that at $T=0$, a non-zero $\Delta_n (\omega_m)$   emerges  only at $N < N_{cr}$, and $\Delta_n (0)$, including $n=0$, scale with  $N_{cr} -N$.  Meanwhile, the onset temperature for the pairing into the $n=0$ state, $T_{p,0}$,  is of order $\omega_0$. The gap  structure reflects this discrepancy:  $\Delta_n  (\omega_m)$ at the lowest frequencies scales with $N_{cr} -N$, but the gap at
$\omega_m \sim \omega_0$  scales with $\omega_0$.    This gives rise to a   maximum  at $\omega_m \sim \omega_0$.

\section{Conclusions}
\label{sec:summary}

In this paper we analyzed the competition between two opposing trends in the behavior of interacting fermions near a quantum critical point in a metal: non-Fermi-liquid physics and pairing.
Both trends are captured within the model of fermions interacting by exchanging soft bosonic fluctuations of an order parameter, which condenses at the critical point.
  The non-Fermi liquid behavior is the result of strong, non-analytic self-energy  due to boson-mediated scattering near the Fermi surface,  the pairing is due to strong attraction in at least one
   pairing channel, provided by the same soft boson exchange. We considered a class of quantum-critical models with an effective dynamical electron-electron interaction
 $V(\Omega_m) \propto 1/|\Omega_m|^\gamma$ (the $\gamma$-model). In this paper, the first in the series, we considered the case  $0<\gamma <1$ and restricted the analysis to $T=0$.
  The limit $\gamma = 0$ corresponds to BCS theory without the upper cutoff, but for all finite $\gamma$ the interaction drops off at large $\Omega$ and the pairing problem is ultra-violet convergent.
   To parametrically separate the tendencies towards non-Fermi liquid and pairing, we  extended the model and introduced the parameter $1/N$ as additional overall factor in the pairing interaction only.
   At large $N$, the tendency towards non-Fermi liquid normal state at $T=0$ is  stronger by $N$, at small $N$ the tendency towards pairing is stronger  by $1/N$.

 Our analysis brings two conclusions. First, we found that  there indeed exists a critical  $N = N_{cr}$,  separating non-Fermi liquid normal state at larger $N$ and superconducting state at smaller $N$.
    The critical $N_{cr} >1$ for all $\gamma <1$, such that the original $N=1$ model is superconducting at $T=0$.
  Second, we found that the system behavior  for $N < N_{cr}$  is rather unconventional in the sense that there exists an infinite set of solutions of the non-linear gap equation,
        $\Delta_n  (\omega_m)$, $n=0, 1,2...$, all within the same pairing symmetry. The solutions are topologically distinct:  $\Delta_n (\omega_m)$ changes sign $n$ times as a function of Matsubara frequency $\omega_m$.
         All $\Delta_n$ emerge at $N = N_{cr}$,
            and for $N \leq N_{cr}$
             their magnitudes scale as
           $\Delta_{n} \propto e^{-\pi (n+1)/\gamma \beta_N}$, where $\beta_N \propto  (N_{cr}-N)^{1/2}$.
            The end point of the set, $\Delta_{\infty} (\omega_m)$, is the  solution of the linearized gap equation at $T=0$. We obtained the exact analytical solution of the linearized gap equation at $T=0$
          and also a highly accurate simple approximate solution for all $N \leq N_{cr}$.
           In the subsequent paper we
      analyze the linearized gap equation at a finite $T$ and  show that there is an infinite set of the onset temperatures for the pairing, $T_{p,n}$, and the corresponding eigenfunctions change sign $n$ times as functions of $\omega_m$. We argue that these topologically distinct gap functions  grow in magnitude below the corresponding $T_{p,n}$, but largely preserve the functional forms  and at $T=0$ become $\Delta_n$ that we found in this work.

     The sign-preserving solution with $n=0$  has the largest condensation energy and is  the global minimum of the condensation energy $E_c$. All other solutions are local minima. Still, the presence of the infinite set of $\Delta_n (\omega_m)$, including the solution of the linear gap equation for all $N \leq N_{cr}$  is a highly non-trivial feature of the pairing at a QCP.
     In subsequent publications   we extend the $T=0$ analysis  to $\gamma >1$ and argue that  the local minima become more closely spaced with increasing $\gamma$  form a continuous set for $\gamma =2$. We argue that in this case all solutions have equal condensation energy, and superconducting order gets destroyed  already at $T=0$.   As the consequence, the true superconducting $T_c$ terminates at $\gamma =2$,  while the onset temperature for the pairing $T_{p,0} = T_p$ remains finite (see Fig. \ref{fig1a}). The two temperatures
      then must be different for all $0<\gamma \leq 2$, including $0<\gamma <1$, which we considered here. In between $T_c$ and $T_{p}$ the system displays pseudogap behavior. We emphasize again that the root to this highly unusual behavior is the existence of the solution of the linearized gap equation for all $N$, rather than only at $N = N_{cr}$.

  \acknowledgements
  We thank   I. Aleiner, B. Altshuler, E. Berg, D. Chowdhury, L. Classen,  K. Efetov, R. Fernandes,  A. Finkelstein, E. Fradkin, A. Georges, S. Hartnol, S. Karchu, S. Kivelson, I. Klebanov, A. Klein, R. Laughlin, S-S. Lee, G. Lonzarich, D. Maslov, F. Marsiglio, M. Metlitski, W. Metzner, A. Millis, D. Mozyrsky, C. Pepan, V. Pokrovsky,  N. Prokofiev,  S. Raghu,  S. Sachdev,  T. Senthil, D. Scalapino, Y. Schattner, J. Schmalian, D. Son, G. Tarnopolsky, A-M Tremblay, A. Tsvelik,  G. Torroba,  E. Yuzbashyan,  J. Zaanen,
   and particularly R. Combescot,  Y. Wang  and Y. Wu  for useful discussions.   The work by  AVC was supported by the NSF DMR-1834856.

 \appendix

\section{The case of  small $\gamma$ and a finite frequency cutoff}
\label{app:gamma_to_0}

The limit $\gamma \to 0$ of the $\gamma$ model attracted a lot of attention from various sub-communities in physics\cite{raghu_15,Wu_19,Wang_H_17,Wang_H_18,Fitzpatrick_15,max_last,senthil,son,*son2} and has been  analyzed in both Eliashberg-type and renormalization group approaches.  This  limit
 requires extra care because  strictly  at $\gamma =0$, the interaction $V(\Omega) \propto |\Omega|^{-\gamma}$ becomes frequency independent, and the pairing problem becomes equivalent to BCS, but without the upper cutoff for the interaction.  Meanwhile for any non-zero $\gamma$, the gap equation
 converges  at $\omega>\omega_{max} \propto g e^{|\log \gamma|/\gamma}$.
 More explicit calculation yields $\omega_{max} \sim g (1/1.446)^{1/\gamma} e^{|\log \gamma|/\gamma}$  (Refs. \cite{max_last,Wang2016,Wu_19,Emil2020}). This $\omega_{max}$ remains finite at any non-zero $\gamma$, but grows with decreasing $\gamma$ faster than the exponent.

 In this Appendix we consider how the gap equation gets modified if we introduce the upper cutoff for $V(\Omega_m)$ at $\Omega_m = \Lambda$,
  i.e.,
  modify $V(\Omega_m) = ({\bar g}/|\Omega_m|)^{\gamma}$ to
  \beq
  V(\Omega_m) = \left(\frac{{\bar g}}{|\Omega_m|}\right)^{\gamma} \left[1- \left(\frac{|\Omega_m|}{\Lambda}\right)^{\gamma}\right]
  \label{last_1a}
  \eeq
  for $|\Omega_m| < \Lambda$, and $V(\Omega_m) =0$ for $|\Omega_m| > \Lambda$.
   We  consider $\gamma$ for which $\Lambda \ll \omega_{max}$.
 In this case we can expand  $V(\Omega)$  to first order in $\gamma$ and express it for $\Omega_m < \Lambda$ as  $V(\Omega)= \gamma \log{(\Lambda/|\Omega|)}$.
 The normal state self-energy is
  $\Sigma (\omega_m) = \gamma \omega_m \log{\Lambda/|\omega_m|}$.
 Substituting these $V(\Omega)$ and $\Sigma (\omega_m)$ into the equation for the pairing vertex, we obtain
 \beq
 \Phi(\omega_m) = \frac{\gamma}{2N} \int \frac{d \omega'_m  \log{\frac{\Lambda}{|\omega_m - \omega'_m|}}}{|\omega'_m| (1 + \gamma \log{\frac{\Lambda}{|\omega'_m|}})}
 \label{app1_2}
 \eeq
 Introducting logarithmic variables $x = \log{\Lambda/|\omega_m|}$, $x' = \log{\Lambda/|\omega'_m|}$ and  restricting with the contributions from $\omega'_m \gg \omega_m$ and $\omega'_m \ll \omega_m$, as we did in the derivation of (\ref{su_6}), we reduce (\ref{app1_2}) to
 \beq
 \Phi (x) = \frac{\gamma}{N} \left[\int_0^x  dx' \frac{ x' \Phi(x')}{1+ \gamma x'} + \frac{x}{1+\gamma x} \int_x^\infty dx' \Phi (x') \right]
  \label{app1_2a}
 \eeq
 Differentiating twice over $x$ and introducing $y = \gamma x$, we obtain
 \beq
 \left(\Phi' (y) (1+y)^2 \right)' = -\frac{\Phi (y)}{N\gamma}
   \label{app1_3}
 \eeq
 The boundary conditions are $\Phi (y=0) =0$ ($\Phi (\omega_m)$ must vanish at $\omega_m = \Lambda$) and
 $\Phi (y = \infty) =0$, i.e., $\Phi (\omega_m =0) =0$. The last condition is set to  avoid the divergence in (\ref{app1_2}) at vanishing $\omega'_m$.
 The solution of (\ref{app1_3}) is
 \beq
 \Phi (y) = \frac{C}{\sqrt{1+y}} \cos \left(\phi + \sqrt{{\frac{1}{N\gamma} -\frac{1}{4}}} \log{(1 +y)}\right)
   \label{app1_4}
 \eeq
 This $\Phi (y)$ satisfies  the boundary condition $\Phi (y=\infty) =0$ provided $1/N\gamma > 1/4$, i.e., $N < N_{cr} = 4/\gamma$. To satisfy the other boundary condition we choose $\phi = \pi/2$.

\begin{figure}
\includegraphics[width=0.6\columnwidth]{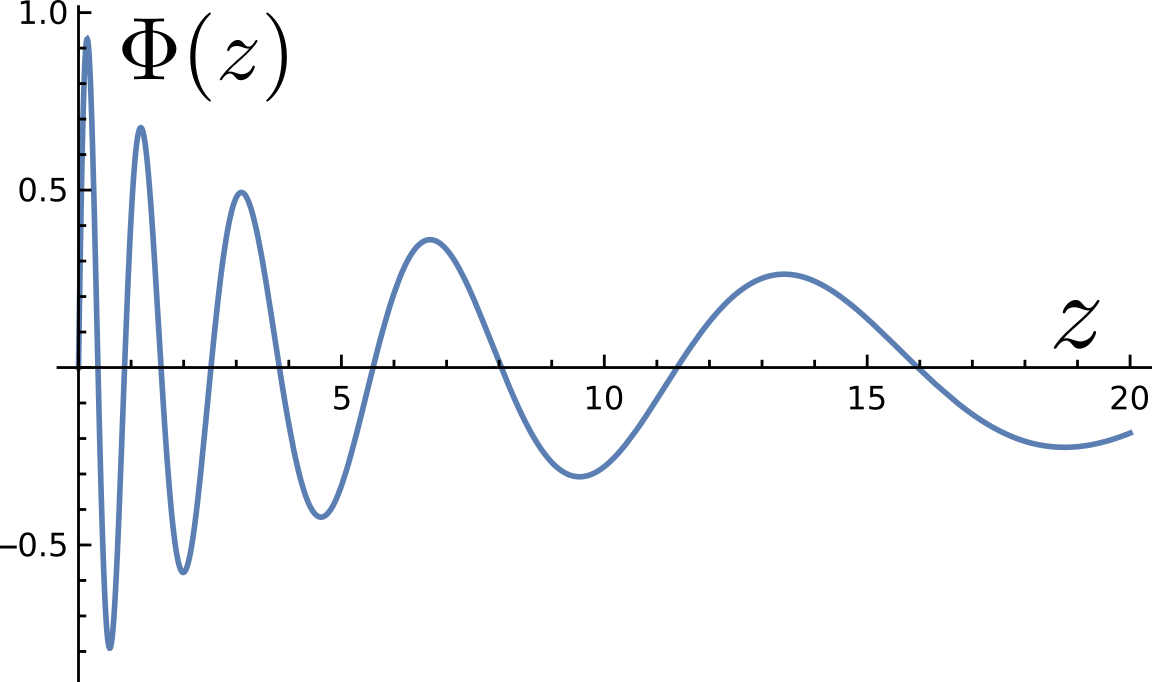}
\caption{\label{fig:phi_z_app_1}
 The function $\Phi (y)$ from Eq. (\ref{app1_4}).}
\end{figure}

 We plot $\Phi (y)$ in Fig.\ref{fig:phi_z_app_1} for $N=1$ and $\gamma = 0.01$.  We see that $\Phi (y)$ is an oscillating function of $y$ for $y \ll 1$ and an oscillating function of $\log{y}$ for $y \gg 1$.
 For $y \ll 1$ we have, restoring the units of frequency,
  \beq
 \Phi (\omega_m) = -C \sin \left(\gamma \sqrt{{\frac{1}{N \gamma} -\frac{1}{4}}} \log{\frac{\Lambda}{|\omega_m|}} \right)
   \label{app1_3_1}
 \eeq

  The largest frequency for oscillations  is  $\omega^*_0 \sim \Lambda e^{-\gamma \sqrt{1/(N\gamma) -1/4}}$.
This scale determines the magnitude of the gap
$\Delta_0 (0)$.
 For $N\gamma \ll 1$,  $\omega^*_0 \sim \Lambda e^{-\frac{\pi}{2}\sqrt{N/\gamma}}$.
The ratio $\gamma/N$ plays the role of a dimensionless coupling $\lambda$, and $\Delta_0$ can be cast into $\Delta_0 (0) \sim \Lambda e^{-\frac{\pi}{2\sqrt{\lambda}}}$. This agrees with Refs.
\cite{son,*son2,max_last}.
Comparing $\Delta_0 (0) $ obtained with and without the cutoff at $\Lambda$ we  see that the
 analysis of the $\gamma$ model without the upper cutoff is valid as long as $g (1/1.446\gamma)^{1/\gamma} \ll \Lambda e^{-\gamma\sqrt{1/N\gamma -1/4}}$.
 At smaller $\Lambda$, $\Delta_0 (0)  \sim \Lambda e^{-\gamma \sqrt{1/N\gamma -1/4}}$.   We emphasize that even in this case there are an infinite number of solutions
 \beq
\Delta_n (0) \sim \Lambda e^{-\frac{\pi}{2\sqrt{\lambda}}} e^{-\frac{n\pi}{\sqrt{\lambda}}}.
 \label{last_11}
 \eeq
This is valid up to $n \sim 1/\sqrt{N \gamma}$.  For larger $n$, relevant $y$ in (\ref{app1_3}) become larger than one, and the formula  for $\Delta_n (0)$ changes.

At small $\gamma$ and $N = O(1)$, the ratio $\Sigma (\omega)/\omega$ is small at $\omega \sim \omega^*_0$, both when $\Lambda$ is finite and when it is set to infinity. In the first case, the ratio is of order $\sqrt{\gamma}$, in the second it is of order $\gamma$.  This implies that  for the computations $\Delta_0$, the self-energy can be safely neglected.  Once self-energy is neglected, one can easily obtain the differential gap equation for the full $V(\Omega_m)$ in (\ref{last_1a}), without expanding it to first order in $\gamma$. This allows us to study the crossover between the behavior at finite and infinite $\Lambda$.

  Introducing, as before, $z = (|\omega_m|/\omega_0)^\gamma$ and  anticipating that relevant $\omega_m \gg \omega_0$, we find that
 the  equation for $\Delta_{{\text diff}} (z)$ for $z \gg 1$  and $V(\Omega_m)$ given by (\ref{last_1a}),
  is the same as when $\Lambda = \infty$, i.e.,
\beq
\left(\Delta_{{\text diff}} (z) z \right)'' = - \frac{1}{N \gamma} \frac{\Delta_{{\text diff}}}{z^2},
\label{last_2a}
\eeq
  However, now $\Delta_{{\text diff}} (z)$ has to satisfy the boundary condition  $\Delta_{{\text diff}} (\Lambda^*) =0$, where $\Lambda^* = (\Lambda/\omega_0)^\gamma$.  The proper solution of (\ref{last_2a}) is
\bea
&&\Delta_{{\text diff}} (z) \propto \frac{1}{\sqrt{z}} \times  \nonumber \\
&& \left[J_1 \left(\frac{2}{(N \gamma z)^{1/2}}\right) Y_1 \left(\frac{2}{(N \gamma \Lambda^*)^{1/2}}\right)
- Y_1 \left(\frac{2}{(N \gamma z)^{1/2}}\right) J_1 \left(\frac{2}{(N \gamma \Lambda^*)^{1/2}}\right)\right].
\label{last_3a}
\eea
In the limit $N\gamma \Lambda^* \ll 1$,  we use
\bea
&& J_1 (x) \approx \sqrt{\frac{2}{\pi x}} \cos{(x -3\pi/4)}, \nonumber \\
&& Y_1 (x) \approx \sqrt{\frac{2}{\pi x}} \sin{(x -3\pi/4)}
\label{last_4a}
\eea
and obtain
\beq
\Delta_{{\text diff}} (z) \propto \frac{1}{z^{1/4}} \sin{\left(\frac{2}{(N \gamma \Lambda^*)^{1/2} - \frac{2}{(N \gamma z)^{1/2}}}\right)}
\label{last_5a}
\eeq
In original variables, $\omega_m$ and $\Lambda$, this reduces, to the leading order in $\gamma$, to
\beq
\Delta_{{\text diff}} (\omega_m) \propto \frac{C}{|\omega_m|^{\gamma/4}} \sin{\left(\sqrt{\frac{\gamma}{N}} \log{\frac{|\omega_m|}{\Lambda}}\right)}.
\label{last_6a}
\eeq
This coincides with (\ref{app1_3_1}), up to subleading terms.

 In the opposite limit, $N \gamma \Lambda^* \gg 1$, we use $J_1 (x \ll 1) \sim x$, $Y_1 (x \ll 1) \sim 1/x$, where
 $x = 2/\sqrt{N \gamma \Lambda^*}$.  Keeping only $Y_1 (x)$, we obtain from (\ref{last_3a}) that the dependence on $\Lambda$ disappears, and
 \beq
\Delta_{{\text diff}} (z) \propto \frac{1}{\sqrt{z}} J_1 \left(\frac{2}{(N \gamma z)^{1/2}}\right)
\label{last_7}
\eeq
This coincides with Eq. (\ref{su_11_1}).

The same result can be obtained  from the RG equations~\cite{max_last,Wang_H_17}. One has to write a set of
 two RG equations for the two-fermion pairing vertex $g$ and the effective 4-fermion pairing interaction $u$. Both depend on the logarithmic scale  $L_n  = \log{\Lambda}/|\omega_n|$.  To leading order in $\gamma$, the equations for the running $g (L)$ and $u(L)$ are
 \beq
 g' = \gamma g,~~ u' = u^2 + g
 \label{last_8}
 \eeq
The initial condition is
\beq
g(0) = \left(\frac{\bar g}{\Lambda}\right)^\gamma \frac{\gamma}{N}, ~~ u(0) =0.
\eeq
The solution of (\ref{last_8}) is $g(L) = g(0) e^{\gamma L}$ and
\beq
u(L) = g^{1/2} (0)  e^{\gamma L/2} \frac{J_1\left(\frac{2 g^{1/2} (0)}{\gamma} e{\gamma L/2}\right) Y_1 \left(\frac{2 g^{1/2} (0)}{\gamma}\right) - Y_1\left(\frac{2 g^{1/2} (0)}{\gamma} e{\gamma L/2}\right) J_1 \left(\frac{2 g^{1/2} (0)}{\gamma}\right)}{J_0\left(\frac{2 g^{1/2} (0)}{\gamma} e{\gamma L/2}\right) Y_1 \left(\frac{2 g^{1/2} (0)}{\gamma}\right) - Y_0\left(\frac{2 g^{1/2} (0)}{\gamma} e{\gamma L/2}\right) J_1 \left(\frac{2 g^{1/2} (0)}{\gamma}\right)}
\label{last_9}
\eeq
For $g (0)/\gamma^2 = ({\bar g}/\Lambda)^\gamma (1/N\gamma) \gg 1$ (i.e., for small $\gamma$ and finite $\Lambda$),  relevant values of the arguments of Bessel and Neumann functions are large, and using
 (\ref{last_4a}) we find that $\gamma$ cancels out, and
 \beq
 u (L) =  g^{1/2} (0) \tan{(g^{1/2} (0) L)}
 \label{last_10}
 \eeq
 The 4-fermion vertex formally diverges at the set of $\omega^*_n = \Lambda e^{-\pi/(2g^{1/2} (0))} e^{-n \pi/g^{1/2} (0)}$.  Using $g (0) \approx \lambda$, valid when $\Lambda$ is finite and $\gamma \to 0$, and associating the corresponding $\omega^*_n$ with $\Delta_n (0)$, we reproduce (\ref{last_11}).

 In the opposite limit,  $g (0)/\gamma^2 \ll 1$, valid when $\gamma$ is kept finite and $\Lambda$ is set to be sufficiently large, we again use that at small $x$, $Y_1 (x) \gg J_1(x)$. Keeping only $Y_1 (2 g^{1/2}_0/\gamma)$ in (\ref{last_9}), we obtain
  \beq
 u (L) =  g^{1/2} (0) e^{\gamma L/2} \frac{J_1\left(\frac{2 g^{1/2} (0)}{\gamma} e{\gamma L/2}\right)}{J_0\left(\frac{2 g^{1/2} (0)}{\gamma} e{\gamma L/2}\right)}
\label{last_12}
 \eeq
 The 4-fermion vertex now diverges at the set of zeros of $J_0 (p)$, where
 \beq
 p = \frac{2 g^{1/2} (0)}{\gamma} \left(\frac{\Lambda}{|\omega_m|}\right)^{\gamma/2} = \frac{2}{(N\gamma z)^{1/2}}
 \eeq
  where, as before, $z = (\omega_0/|\omega_m|)^\gamma$  (we used that for small $\gamma$, ${\bar g}^\gamma = \omega^\gamma_0 (1-\gamma) \approx \omega^\gamma_0$.  We see that now $\Lambda$ cancels out, and the condition $J_0 [2/(N \gamma z)^{1/2}] =0$  gives the same set of $z^*_n$ as for the case when $\Lambda$ is infinite.

\section{ Differential gap equation for arbitrary $\gamma$.}
\label{app:largergamma}

The derivation  of the differential gap equation, Eq. (\ref{su_6}), from the original integral equation is only justified for small $\gamma$. In this appendix we modify the derivation
 and show that for $N=1$  the modified gap function $\Delta_{\text{diff}} (\omega_m)$ shows qualitatively the same behavior as the exact
$\Delta (\omega_m)$ even for $\gamma \leq 1$.

   Specifically, we  add to the r.h.s of (\ref{su_6}) the additional contribution from $\omega' \sim \omega$, where the interaction is strongly peaked at $\gamma \to 1$.   There is some uncertainty with this contribution as we have to specify the range of integration around $\omega' =\omega$. We choose this range to be
  $|\omega' - \omega| \leq a_\gamma \omega$ and keep  $a_\gamma$  as a  parameter.
   We assume that for small $\gamma$, $a_\gamma \to 0$, but for $\gamma \leq 1$, $a_\gamma = O(1)$.
   Adding this contribution to the r.h.s. of  Eq. (\ref{su_6}) we find that it becomes
 \beq
    \Phi (z) \frac{d +z}{1+z} = \frac{1-\gamma}{N\gamma} \left[\int^{\infty}_{z} dy \frac{\Phi (y)}{y (1+y)} + \frac{1}{z} \int_0^z dy \frac{\Phi (y)}{1+y} \right]
   \label{su_6_2}
   \eeq
    where $d = 1- a^{1-\gamma}_\gamma/N$.
  Introducing $\Delta$ instead of $\Phi$, rescaling $z$ to ${\bar z} = z/d$ and differentiating twice over ${\bar z}$ we obtain differential gap equation in the same form as in (\ref{su_7_11})
  \beq
   (\Delta_{{\text{diff}}} ({\bar z}) (1+{\bar z}))^{''}  = - \left(\beta^2_N + \frac{1}{4}\right)
    \frac{\Delta_{{\text{diff}}} ({\bar z})}{{\bar z}^2},
   \label{su_6_3}
   \eeq
 once we identify
 \beq
  \frac{1-\gamma}{N\gamma-a_\gamma^{1-\gamma}} = \beta^2_N +\frac{1}{4}
 \label{nnnn_1}
  \eeq
  Hence, the solution is given by Eq. (\ref{su_17}) with ${\bar z}$ instead of $z$.

  Eq. (\ref{nnnn_1}) determines $a_\gamma$. At small $\gamma$, $\beta^2_N +1/4 \approx 1/(N \gamma)$, hence, $a$ is small and ${\bar z} \approx z$.  However, for larger $\gamma$, $a_\gamma = O(1)$, and the rescaling $z \to {\bar z}$ is essential. Fig. \ref{fig_11} shows that it brings $\Delta_{{\text{diff}}} (z)$ closer to the exact $\Delta_{ex} (z)$.

 We note in this regard that for $N =1$ and $\gamma \to 1$,
  $a^{1-\gamma}_\gamma/N  \approx 1 + (1-\gamma) \log{a_\gamma}$, and the r.h.s. of (\ref{nnnn_1}) becomes of order one if we choose $a = O(1)$.  This is consistent with the fact that $\beta_{N=1}$ tends to a finite value $\beta_{N=1} \approx 0.79$ at $\gamma \to 1$. Also, the  rescaling factor $d = 1-a^{1-\gamma}/N \approx -(1-\gamma) \log{a_\gamma}$, hence, for $a = O(1)$,
       ${\bar z} = z/d$ becomes of order $|\omega_m|/g$.  As the consequence, the modified  $\Delta_{\text{diff}} (\omega_m)$ evolves at a non-singular $\omega_m \sim g$. This is consistent with the behavior of $\Delta_{\text{ex}} (\omega_m)$ at $\gamma \to 1$ and $N =1$ (see Fig. \ref{fig_11}(e)).

 \section{The exact solution of the linearized gap equation -- the computational details.}
\label{app:exact}

In this appendix we present the
 full details of the computations, leading to the formula for the exact solution of the equation \eqref{eq:lineargap_1_1}.  The presentation  is self-contained in the sense that it does not use specific notations, introduced in the main text.

\subsection{Reformulation of Eq. (\ref{eq:lineargap_1_1}).}\label{sec:new}

We write Eq. \eqref{eq:lineargap_1_1} as
\begin{equation}\label{eq:eqAlpha}
 E\Phi (\Omega )=\frac{1-\gamma}{2}\int_{-\infty }^{\infty } \frac{\Phi (\omega )d\omega }{|\omega -\Omega |^{\gamma }|\omega|^{1-\gamma }}\frac{1}{1+|\omega|^{
 \gamma }}
\end{equation}
 and use the latter $E$ instead of $N$ to emphasize that this is a continuous variable.\footnote{The method, presented in this appendix, will also work for a generalized version of Eq. \eqref{eq:eqAlpha} with $\frac{1}{1+|\omega|^{\alpha }}$ instead of $\frac{1}{1+|\omega|^{ \gamma }}$.}

We
 introduce a set of functions
\begin{equation}\label{eq:basis}
\Phi_{\beta }(\Omega )=\frac{|\Omega |^{2i\beta +\delta_{\Omega }}}{|\Omega |^{\gamma /2}}
\end{equation}
where $\beta $ changes from $-\infty $ to $+\infty $ and $\delta_{\Omega }=+0\mbox{sign}(1-|\Omega |)$ is a convergence factor.
It is clear that
$$
\int \Phi_{\beta }(\omega )\Phi_{\beta }^{\ast}(\Omega )\frac{d\beta}{2\pi }=\frac{1}{2}|\Omega |^{1-\gamma }\delta (|\Omega| -|\omega|)  .
$$
We now denote
\begin{equation}\label{eq:abeta}
a_{\beta }=  \frac{1}{2}\int \frac{d\Omega }{|\Omega |^{1-\gamma }}\Phi_{\beta }(\Omega )\Phi (\Omega ),
\end{equation}
multiply \eqref{eq:eqAlpha} by $\Phi_{\beta }(\Omega )$, and integrate it over $\frac{d\Omega }{|\Omega |^{1-\gamma }}$. We obtain
$$
Ea_{\beta }=\frac{1-\gamma }{2}\frac{1}{2}\int d\omega  \frac{\phi (\omega )}{|\omega |^{1-\gamma }}\frac{1}{1+|\omega|^{
\gamma }} \int \frac{\Phi_{\beta }(\Omega )  d\Omega }{|\Omega |^{1-\gamma }|\Omega -\omega |^{1-\gamma }}=
\frac{\epsilon_{\beta }}{2}\int d\omega  \frac{\Phi (\omega )\Phi_{\beta }(\omega )}{|\omega |^{1-\gamma }}\frac{1}{1+|\omega|^{
\gamma  }},
$$
where
\begin{equation}\label{eq:epsilon}
 \epsilon_{\beta }=\frac{1-\gamma}{2}\frac{|\Gamma (\gamma /2+2i\beta )|^{2}}{\Gamma (\gamma )}\left(1+\frac{\cosh (2\pi \beta )}{\cos (\pi \gamma /2)} \right).
\end{equation}

From the definition \eqref{eq:abeta} (we also use the fact that $\Phi (\omega )$ is
taken to be an even function of frequency)
we find
$$
 \int a_{\beta }\Phi_{\beta }^{\ast}(\omega )\frac{d\beta }{2\pi }=  \frac{1}{2}\int \frac{d\Omega }{|\Omega |^{1-\gamma }}\Phi (\Omega ) \int \Phi_{\beta }(\Omega )\Phi_{\beta }^{\ast}(\omega )\frac{d\beta}{(2\pi )}=\frac{1}{2}\Phi (\omega ).
$$
We then have
$$
 Ea_{\beta }= \epsilon_{\beta }\int \frac{d\beta'}{2\pi }a_{\beta'} \int d\omega  \frac{\Phi_{\beta' }^{\ast}(\omega )\Phi_{\beta }(\omega )}{|\omega |^{1-\gamma }}\frac{1}{1+|\omega|^{
 \gamma }}.
$$
We now
 compute the integral
$$
 \int d\omega  \frac{\Phi_{\beta' }^{\ast}(\omega )\Phi_{\beta }(\omega )}{|\omega |^{1-\gamma }}\frac{1}{1+|\omega|^{\alpha  }}=2  \int_{0}^{\infty } \frac{d\omega }{\omega }\frac{\omega^{2i(\beta -\beta ')+\delta_{\omega }}}{1+\omega^{\gamma
  }}=
\frac{2}{
\gamma }\int_{-\infty }^{\infty }dx\frac{e^{ix\frac{2}{
\gamma}(\beta -\beta ')+x\delta_{x}}}{1+e^{x}}
$$
where $\delta_{x}=-0\mbox{sign}(x)$. At $x\rightarrow \infty $ the integral perfectly converges with or without the convergence factor. At $x\rightarrow -\infty $ we do need the convergence factor, and
 we assume that $\delta_{x}$ is
 a small positive number.
  Evaluating the integral we then obtain
\begin{equation}\label{eq:help}
 \frac{2}{
 \gamma
  }\int_{-\infty }^{\infty }dx\frac{e^{ix\frac{2}{
  \gamma}(\beta -\beta '-i0)}}{1+e^{x}}
=-\frac{2}{
\gamma
}\frac{i\pi }{\sinh \left(\frac{2\pi }{
\gamma
}(\beta -\beta '-i0) \right)}   ,
\end{equation}
and
$$
 Ea_{\beta }= -i\epsilon_{\beta }\frac{1}{
  \gamma}\int_{-\infty }^{\infty }  \frac{a_{\beta'} d\beta ' }{\sinh \left(\frac{2\pi }{
\gamma
}(\beta -\beta '-i0) \right)} .
$$
 We now
  substitute $2\beta /
  \gamma =k$, and use $a_{k}$ instead of $a_{\beta }$. Then,
\begin{equation}\label{eq:a}
 Ea_{k}= -\frac{i}{2}\epsilon_{
\gamma
k/2 }\int_{-\infty }^{\infty }  \frac{a_{k'} dk ' }{\sinh \left(\pi (k -k '-i0) \right)}
\end{equation}
Introducing the new function $b_k$ via
\begin{equation}\label{eq:b}
a_{k}=b_{k}\epsilon_{\gamma k/2} ,
\end{equation}
we
 obtain from (\ref{eq:a}) the equation for $b_k$ in the form
\begin{equation}\label{eq:main}
 Eb_{k}= \frac{i}{2}\int_{-\infty }^{\infty }  \frac{\epsilon_{\gamma  k'/2 } }{\sinh \left(\pi (k' -k +i0) \right)}b_{k'} dk '
\end{equation}

\subsection{The functional $F[\phi]$.}\label{sec:functional}
For an infinitesimally small $\Phi (\omega)$, the free energy difference between the
 states with $\Phi (\omega) =0$ and $\Phi (\omega) \neq 0$    is given by
\begin{equation}\label{eq:F}
   \frac{F[\Phi]}{N_{0}\omega_{0}}=\frac{1}{2}\int\frac{\Phi_{\omega }^{2}}{|\omega |^{1-\gamma }\left(1+|\omega |^{\gamma } \right)}d\omega -\frac{1-\gamma }{4
    E}\int \frac{\Phi_{\omega }\Phi_{\omega '}d\omega d\omega '}{|\omega |^{1-\gamma }\left(1+|\omega |^{\gamma } \right)|\omega -\omega '|^{\gamma }|\omega' |^{1-\gamma }\left(1+|\omega' |^{\gamma } \right)},
\end{equation}
where we defined $\Phi (\omega_m) \equiv \Phi_\omega$ to shorten the notations.
Taking the first variation of this functional we obtain
equation \eqref{eq:eqAlpha}
 This functional defines our space of functions, namely we must only consider the functions for which  $F[\phi]$ is finite.

\subsection{Normalizability condition.}\label{sec:FonS}

Now, let us take $\Phi_{\omega }^{E}$ to be the solution of \eqref{eq:eqAlpha} with $E\not=1$.
Then
\beq
   F[\Phi _{\omega }^{\epsilon }]=N_{0}\bar{g}^{2}(1-\gamma )^{-2/\gamma }\frac{1-E}{2}\int_{-\infty}^{\infty }\frac{\Phi_{\omega }^{2}}{|\omega |^{1-\gamma }\left(1+|\omega |^{\gamma } \right)}d\omega.
\label{app_C1}
\eeq
We use
$$
\Phi_{\omega }=2\int a_{\beta }\Phi^{\ast}_{\beta }(\omega )\frac{d\beta}{2\pi },
$$
where $\Phi_{\beta }(\omega )$ is defined in \eqref{eq:basis}. According to \eqref{eq:help}, we have
$$
\int \frac{\Phi_{\omega}^{2}d\omega }{|\omega |^{1-\gamma }\left(1+|\omega |^{\gamma } \right)}=-\frac{4i}{\pi \gamma }\int \frac{a_{\beta }a_{\beta '}^{\ast}d\beta d\beta '}{\sinh \left[\frac{2\pi}{\gamma}(\beta -\beta '-i0) \right]}.
$$
Substituting into (\ref{app_C1}) and using
$$
k=2\beta /\gamma,
$$
we obtain
\beq
  \frac{F[a _{k }^{E }]}{N_{0}\omega_{0}}=-\gamma (1-E)\frac{i}{2\pi }\int \frac{a_{k }a_{k '}^{\ast}dk dk '}{\sinh \left[\pi(k -k'-i0) \right]}  .
\label{app_C2}
\eeq
Using
\eqref{eq:b}, we rewrite (\ref{app_C2})
in terms of $b_{k}$ as
$$
 \frac{F[b _{k }^{E }]}{N_{0}\omega_{0}}=-\gamma (1-E)\frac{i}{\pi }\int \frac{\epsilon _{\gamma k/2 }\epsilon _{\gamma k '/2}}{\sinh \left[\pi(k -k'-i0) \right]} b_{k}b_{k'}^{\ast}  dk dk '.
$$
It can be  re-expressed as
$$
\frac{F[b _{k }^{E }]}{N_{0}\omega_{0}}=- \gamma (1-E)\frac{i}{\pi }\int \epsilon _{\gamma k/2 }\epsilon _{\gamma k '/2}\left[ \frac{1}{\sinh \left[\pi(k -k'+i0)\right]}+2i\delta (k-k') \right] b_{k}b_{k'}^{\ast}  dk dk '.
$$
Comparing to \eqref{eq:main},  we finally obtain
\begin{equation}\label{eq:norm}
\frac{F[b _{k }^{E }]}{N_{0}\bar{g}^{2}(1-\gamma )^{-2/\gamma }\gamma }=-\frac{E(1-E)}{\pi }\int \left[ 1-\frac{1}{E}\epsilon _{\gamma k'/2 }\right] \epsilon _{\gamma k '/2}  b_{k'}b_{k'}^{\ast} dk '
\end{equation}

 Eq. \eqref{eq:norm} defines a norm for our solutions. If we manage to solve Eq. \eqref{eq:main} and find $b_k$
    for a given $E$, we will need to verify whether for these $b_k$, $F[b _{k }^{E }]$ is finite.
This will give us the spectrum.

\subsection{Solution of Eq. \eqref{eq:main}}\label{sec:sol}

We now show that the normalizable solution of Eq. \eqref{eq:main} exists for all $E<\epsilon_{0}$. We obtain $b_k$ and use them to obtain  the pairing vertex $\Phi (\omega_m)$ and the gap function $\Delta (\omega_m)$.
For this purpose we need to first  find
 symmetry properties of
$b_k$
in \eqref{eq:main}.

\subsubsection{$b_k$ vs $b_{-k}$.}\label{sec:reality}

The equation \eqref{eq:eqAlpha} is real. So,
(i) if an eigenfunction is complex, its complex conjugated must also be an eigenfunction with the same eigenvalue. The eigen value then is double degenerate; (ii)
 If an eigenvalue is non-degenerate, then the eigenfunction must be real (more precisely, it may have a trivial $\omega $ independent phase.)
If the eigenfunction of Eq.
\eqref{eq:eqAlpha} is real, then, according to
 Eqs. \eqref{eq:basis} and \eqref{eq:abeta}, we have
$$
  a_{
  k}=a^{\ast}_{-
  k}
$$
As the function $\epsilon_{\gamma k/2}$ is real and symmetric under the change
$k \rightarrow -k$
we have
\begin{equation}\label{eq:reality}
  b_{k}
=b^{\ast}_{-k}
\end{equation}

\subsubsection{Periodicity.}\label{sec:periodicity}

Equation \eqref{eq:main} has an obvious, but important property. If we change $k\rightarrow k+i $ the r.h.s will turn into itself, but with the opposite sign. So we must have
\begin{equation}\label{eq:periodicity}
  b_{k+in}=(-1)^{n}b_{k}.
\end{equation}
It means, in particular, that a solution of \eqref{eq:main} is periodic with the period $2i$.

\subsubsection{Analytical properties.}\label{sec:analytical}

Let us consider Eq.  \eqref{eq:main} for arbitrary complex $k$, keeping $k'$ to be
real. This way we define the analytical function $b_{k}$.
 The integral perfectly converges as long as $\Im k\not=in$ for integer $n$. So the function $b_{k}$ can only have singularities on the lines $\Im k=in$. It cannot have poles on these lines, as in this case the integral
  in the r.h.s. of  Eq. \eqref{eq:main}
 would not converge.

Let us return to Eq. \eqref{eq:main}  and  use it to define
an analytic function $B_{k}$:
$$
B_{k}= \frac{i}{2E}\int_{-\infty }^{\infty }  \frac{\epsilon_{\gamma  k'/2 } }{\sinh \left(\pi (k' -k) \right)}b_{k'} dk '
=\frac{i}{2\pi E}\sum_{n=-\infty }^{\infty }(-1)^{n}\int_{-\infty }^{\infty }  \frac{\epsilon_{\gamma  k'/2 } b_{k'}}{k' -k+in} dk '
$$
 (we assume $E\not=0$).
The function $b_k$ is related to $B_k$ as
\begin{equation}\label{eq:bB}
b_{k}=\lim_{\delta \rightarrow 0}B_{k-i\delta }  ,\qquad \Im k=0,\quad \delta >0.
\end{equation}
  Using the fact that both $\epsilon_{\gamma k/2}$ and $1/\sinh(\pi k)$ are analytic in narrow strip below the real axis, we can express $B_k$ as
\beq
B_{k}= \frac{i}{2E}\int_{-\infty-i\tilde{\delta } }^{\infty-i\tilde{\delta } }  \frac{\epsilon_{\gamma  k'/2 } }{\sinh \left(\pi (k' -k) \right)}B_{k'} dk '   ,\qquad \tilde{\delta}>0
\label{app_C4}
\eeq
This formula must be understood in the sense that to compute  $b_k
 =\lim_{\delta \rightarrow 0}B_{k-i\delta }$, for real $k$, we must first take the limit $\tilde{\delta }\rightarrow 0$ and  then $\delta \rightarrow 0$, in which case
  $k$ in the r.h.s. of (\ref{app_C4}) is below the integration path.
The function $B_{k}$ obviously
 satisfies the periodicity condition, Eq.
 \eqref{eq:periodicity}. It is also clear from (\ref{app_C4})  that $B_{k}$ has branch cuts along the lines $\Im k=in$.

Let us now compute $\lim_{\delta  \rightarrow 0}(B_{k+in+i\delta }-B_{k+in-i\delta })$, for real $k$ and positive $\delta $. By  Sokhotski-Plemelj theorem: $\frac{1}{x-x_{0}-i0}=\mbox{P}\frac{1}{x-x_{0}}+i\pi \delta (x-x_{0})$. Hence
$$
\lim_{\delta  \rightarrow 0}(B_{k+in+i\delta }-B_{k+in-i\delta })=-(-1)^{n}\frac{1}{E}\epsilon_{\lambda k/2}b_{k} =-\frac{1}{E}\epsilon_{\lambda k/2}B_{k+in-i\delta },\qquad \delta \rightarrow 0,\quad \delta >0
$$
which means that
\begin{equation}\label{eq:RH}
B_{k+in+i0 }=\left( 1-\frac{1}{E}\epsilon_{ \gamma k/2}\right)B_{k+in-i0} .
\end{equation}
This equation is a particular case of  the  Riemann-Hilbert problem.

Here, one has to distinguish between the following two
cases
\begin{itemize}
\item (i)] $E>\mbox{max}\epsilon_{\gamma k/2}=\epsilon_{0}$ -- in this case the expression $ 1-\frac{1}{E}\epsilon_{ \gamma k/2}$ does not change sign, and
\item (ii)] $0<E<\epsilon_{0}$  -- in this case the expression $ 1-\frac{1}{E}\epsilon_{ \gamma k/2}$ changes sign twice at $k=\pm k_{E}$, where
\end{itemize}
\begin{equation}\label{eq:kE}
 \epsilon_{\gamma k_{E}/2}=E
\end{equation}
We consider the two cases separately.

\noindent {\em Case (i). $E>\epsilon_{0}$:}
In this case we take the logarithms of the both sides of the equation \eqref{eq:RH}, we then get
$$
  \log B_{k+in+i0 }-\log B_{k+in-i0} =\log \left( 1-\frac{1}{E}\epsilon_{ \gamma k/2}\right)
$$
So the analytic function $\Phi_{k}=\log B_{k}$ has branch cuts along the lines $\Im k=in$ with the jumps given by above formula. According to Sokhotski-Plemelj theorem we have
$$
\log B_{k}
 =\frac{1}{2\pi i}\sum_{n=-\infty }^{\infty }\int_{-\infty }^{\infty } \frac{\log (1-\frac{1}{E}\epsilon_{ \gamma k'/2})}{k'-k+in}dk'
+F_{k}
=\frac{1}{2 i}\int_{-\infty }^{\infty } \log (1-\frac{1}{E}\epsilon_{ \gamma k'/2})\coth \left(\pi (k'-k) \right)dk'
+F_{k}
,
$$
where the analytic function $F_{k}$ has no branch cuts or singularities, so it must be a polynomial. In order to find this polynomial, we notice, that the first term in the right hand side does not change if we shift $k\rightarrow k+in$, while $B_{k}$ must acquire a factor of $(-1)^{n}$ and  $\log B_{k}$ must acquire extra  $\pm i\pi n$. There are only two polynomials that have this property, they are are $\pm \pi k$, so there are two solutions:
\begin{equation}\label{eq:solutions}
B_{k}=e^{\pm \pi k+\frac{1}{2i}\int_{-\infty }^{\infty } \log \left( 1-\frac{1}{E}\epsilon_{ \gamma k'/2}\right)\coth \left(\pi (k'-k) \right)dk'}  ,
\end{equation}
Correspondingly, there are two solutions for $b_{k}$
\begin{equation}\label{eq:bSolutionsPM}
b_{k}^{\pm } =B_{k-i0}=e^{\pm \pi k+\frac{1}{2i}\int_{-\infty }^{\infty } \log \left( 1-\frac{1}{E}\epsilon_{ \gamma k'/2}\right)\coth \left(\pi (k'-k+i0) \right)dk'}  ,
\end{equation}
where $k$ is now real.

Notice, that we have two solutions. One can check, that they are indeed complex conjugated.

Now we need to
check whether
 the solutions \eqref{eq:bSolutionsPM} are normalizable. From \eqref{eq:bSolutionsPM} we find
\begin{eqnarray}
&&b_{k}b_{k}^{\ast}  =e^{\pm 2\pi k}e^{\frac{1}{2i}\int_{-\infty }^{\infty } \log \left( 1-\frac{1}{E}\epsilon_{ \gamma k'/2}\right)\left[ \coth \left(\pi (k'-k+i0) \right)-\coth \left(\pi (k'-k-i0) \right)\right]dk'}\nonumber\\
&&
=e^{\pm 2\pi k}e^{-\frac{1}{2i}\int_{-\infty }^{\infty } \log \left( 1-\frac{1}{E}\epsilon_{ \gamma k'/2}\right)2i\delta (k-k')dk'}=
\cosh^{2}(\pi k)e^{-\log \left( 1-\frac{1}{E}\epsilon_{ \gamma k/2}\right)}
\nonumber
\end{eqnarray}

As the consequence,
$$
  \frac{F[b _{k }^{E }]}{N_{0}\bar{g}^{2}(1-\gamma )^{-2/\gamma }\gamma }=-\frac{E(1-E)}{\pi }\int e^{2\pi k} \epsilon _{\gamma k /2}  dk
$$
The last integral diverges. The divergence implies that the solutions with $E > \epsilon_0$ are not normalizable and are not the part of the spectrum.

\noindent {\em Case (ii). $0<E<\epsilon_{0}$.}
In this case there are two points \eqref{eq:kE} where the r.h.s of \eqref{eq:RH} changes sign.
To avoid ambiguity of $\log (-1)$,
we first rewrite \eqref{eq:RH} as
$$
B_{k+in+i0 }=\Theta (k-k_{E})\Theta (k+k_{E})\left| 1-\frac{1}{E}\epsilon_{ \gamma k/2}\right|B_{k+in-i0} ,
$$
where $\Theta $ is the
step function  ($\Theta (x<0)=-1$ and $\Theta (x>0) =1$)
 Now we take the log of both sides
$$
\log B_{k+in+i0 }-\log B_{k+in-i0}=\log \left| 1-\frac{1}{E}\epsilon_{ \gamma k/2}\right|+\log \Theta (k-k_{E})  +\log \Theta (k+k_{E})  .
$$
The $\log \Theta (k-k_{E})$ is zero for $k>k_{E}$ and either $+i\pi $ or $-i\pi $ for $k<k_{E}$. Let us introduce two  yet unknown functions $\chi_{n}$ and $\xi_{n}$, which have values $\pm 1$, depending on $n$, and  rewrite the equation above as
$$
  \log B_{k+in+i0 }-\log B_{k+in-i0}=\log \left| 1-\frac{1}{E}\epsilon_{ \gamma k/2}\right|+i\pi \chi_{n} \frac{1}{2}(1-\Theta (k-k_{E}))  +i\pi \xi_{n} \frac{1}{2}(1-\Theta (k+k_{E}))  .
$$
The Sokhotski-Plemelj theorem now gives
\begin{eqnarray}
&& \log B_{k}
=\frac{1}{2\pi i}\sum_{n=-\infty }^{\infty }\int_{-\infty }^{\infty } \frac{\log |1-\frac{1}{E}\epsilon_{ \gamma k'/2}|}{k'-k+in}dk'\nonumber\\
&&
+ \frac{1}{2} \sum_{n=-\infty }^{\infty }\chi_{n} \int_{-\infty }^{\infty } \frac{\frac{1}{2}(1-\Theta (k-k_{E}))}{k'-k+in}dk'
+ \frac{1}{2} \sum_{n=-\infty }^{\infty }\xi_{n} \int_{-\infty }^{\infty } \frac{\frac{1}{2}(1-\Theta (k+k_{E}))}{k'-k+in}dk'
+F_{k}.
\nonumber
\end{eqnarray}
This can be re-expressed as
\begin{eqnarray}
&&  \log B_{k}
=\frac{1}{2 i}\int_{-\infty }^{\infty } \log |1-\frac{1}{E}\epsilon_{ \gamma k'/2}|\coth \left(\pi (k'-k) \right)dk' \nonumber\\
&&
+ \frac{1}{2} \int_{-\infty  }^{k_{E} }\sum_{n=-\infty }^{\infty }  \frac{\chi_{n}}{k'-k+in}dk'
+ \frac{1}{2} \int_{-\infty  }^{-k_{E} }\sum_{n=-\infty }^{\infty }  \frac{\xi_{n}}{k'-k+in}dk'
+F_{k}       .     \nonumber
\label{app_C5}
\end{eqnarray}

According to  \eqref{eq:periodicity}, the difference $\log B_{k+im}-\log B_{k}$ should be equal to $i\pi m$
independent on $k$. For $B_k$ from (\ref{app_C5}) we
have
$$
 \log B_{k+im}-\log B_{k}=   \frac{1}{2} \int_{-\infty  }^{k_{E} }\sum_{n=-\infty }^{\infty }  \frac{\chi_{n+m}-\chi_{n}}{k'-k+in}dk'
+\frac{1}{2} \int_{-\infty }^{-k_{E} }\sum_{n=-\infty }^{\infty }  \frac{\xi_{n+m}-\xi_{n}}{k'-k+in}dk'
+F_{k+im}-F_{k} .
$$
In order for this expression to be independent of $k$, the functions $\chi_{n}$ and $\xi_{n}$ must be independent of $n$ and $F_{k}$ must be equal to $\pm \pi k+c$, where $c$ is an arbitrary complex constant. We then denote  $\chi_{n} = \chi $, $\xi_{n}=\xi$, where $\chi =\pm 1$,  $\xi =\pm 1$.  The signs of $\chi$ and $\xi $ can be chosen independently. After this, we obtain
\begin{eqnarray}
&& \log B_{k}
=\frac{1}{2 i}\int_{-\infty }^{\infty } \log |1-\frac{1}{E}\epsilon_{ \gamma k'/2}|\coth \left(\pi (k'-k) \right)dk' \nonumber\\
&&
+\chi   \frac{\pi }{2} \int_{-\infty  }^{k_{E} }  \coth (\pi (k'-k))dk'
+\xi \frac{\pi }{2} \int_{-\infty  }^{-k_{E} }  \coth (\pi (k'-k))dk'
\pm \pi k +c
\end{eqnarray}
Now, we can find the function $b_{k}$
\begin{eqnarray}
&&b_{k}= B_{k-i0}=\exp \Big[
\frac{1}{2 i}\int_{-\infty }^{\infty } \log |1-\frac{1}{E}\epsilon_{ \gamma k'/2}|\coth \left(\pi (k'-k+i0) \right)dk' \pm \pi k +c
\nonumber\\
&&
+\chi  \frac{\pi }{2} \int_{-\infty  }^{k_{E} }  \left[ \coth (\pi (k'-k+i0))-\coth (\pi (k'+i0))\right]dk'\nonumber\\
&&
+\xi\frac{\pi }{2} \int_{-\infty }^{-k_{E} }  \left[ \coth (\pi (k'-k+i0))-\coth (\pi k')\right]dk'
\Big],
\label{eq:b2Solution2}
\end{eqnarray}
(we redefined the constant $c$)

Now, we need to check if the solution \eqref{eq:b2Solution2} is normalizable.
For this, we
need to verify whether the integral in \eqref{eq:norm} converges at large $|k|$. This requires us to
compute the
 real part of the exponent in \eqref{eq:b2Solution2} for
 $|k|\gg k_{E}$.
We do this on term by term basis
\begin{eqnarray}
&&
\Re \left[ \frac{1}{2 i}\int_{-\infty }^{\infty } \log |1-\frac{1}{E}\epsilon_{ \gamma k'/2}|\coth \left(\pi (k'-k+i0) \right)dk' \pm \pi k +c\right]=-\frac{1}{2}\log |1-\frac{1}{E}\epsilon_{ \gamma k/2}|\pm \pi k+\Re c,
\nonumber\\
&&
\Re \left[\chi  \frac{\pi }{2} \int_{-\infty  }^{k_{E} }  \left[ \coth (\pi (k'-k+i0))-\coth (\pi (k'+i0))\right]dk'\right]
=\frac{\chi }{2}\log \left|\frac{\sinh (\pi (k'-k))}{\sinh (\pi k')} \right|_{-\infty }^{k_{E}}\nonumber\\
&&
=\frac{\chi }{2}\log \left|\frac{\sinh (\pi (k_{E}-k))}{\sinh (\pi k_{E})} \right|-\chi \frac{\pi }{2}k
\rightarrow \chi \frac{\pi}{2}\left(|k|-k \right),\qquad \mbox{for $|k|\rightarrow \infty $},
\nonumber\\
&&
\Re \left[\xi\frac{\pi }{2} \int_{-\infty }^{-k_{E} }  \left[ \coth (\pi (k'-k+i0))-\coth (\pi k')\right]dk'\right]
=\frac{\xi }{2}\log \left|\frac{\sinh (\pi (k'-k))}{\sinh (\pi k')} \right|_{-\infty }^{-k_{E}}\nonumber\\
&&
=\frac{\xi }{2}\log \left|\frac{\sinh (\pi (k_{E}+k))}{\sinh (\pi k_{E})} \right|-\xi \frac{\pi }{2}k
\rightarrow \xi \frac{\pi}{2}\left(|k|-k \right),\qquad \mbox{for $|k|\rightarrow \infty $},
\nonumber
\end{eqnarray}
Combining, we find
that the real part of the exponent in \eqref{eq:b2Solution2} is
$$
-\frac{1}{2}\log |1-\frac{1}{E}\epsilon_{ \gamma k/2}|\pm \pi k+\pi  \frac{\chi+\xi}{2}\left(|k|-k \right)
$$
We  see that if we chose $\chi =\xi=-1$, and the minus sign for the term $\pm \pi k$, we get
$$
-\frac{1}{2}\log |1-\frac{1}{E}\epsilon_{ \gamma k/2}|-\pi|k|.
$$
This
 guarantees the convergence of the integral in \eqref{eq:norm}.

With this choice of the constants, the function $b_{k}$ becomes
\begin{eqnarray}
&&b_{k}=\exp \Big[
\frac{1}{2 i}\int_{-\infty }^{\infty } \log |1-\frac{1}{E}\epsilon_{ \gamma k'/2}|\coth \left(\pi (k'-k+i0) \right)dk'
\nonumber\\
&&
-\pi k
- \frac{\pi }{2} \int_{-k_{E}  }^{k_{E} }  \left[ \coth (\pi (k'-k+i0))-\coth (\pi (k'+i0))\right]dk'\nonumber\\
&&
-\pi  \int_{-\infty }^{-k_{E} }  \left[ \coth (\pi (k'-k+i0))-\coth (\pi k')\right]dk'
\Big],
\label{eq:b2Solution}
\end{eqnarray}
where, we remind,
 $k_{E}$ is defined in \eqref{eq:kE}.

Notice the following:
 (i) The normalizability of $b_k$  is due to the $e^{-\pi |k|}$ asymptotic. This absolute value $|k|$ comes from the contribution from the branch cut of the log. This holds only
 for $E<\epsilon_{0}$; (ii)
  There exists only one normalizable solution. This means that this $b_k$
    must satisfy the  condition \eqref{eq:reality}. One can check explicitly that it is indeed so.

\subsection{Pairing vertex $\Phi (\omega_m)$}\label{sec:frequency}

We now use $b_k$ from Eq. \eqref{eq:b2Solution} to compute the function $\Phi (\omega_m )$.
We recall that $\Phi (\omega_m ) \equiv \Phi (\omega)$ is expressed via $b_k$ as
\begin{equation}\label{eq:phi0}
\Phi (\omega )=2\int_{-\infty }^{\infty }a_{\beta }\Phi_{\beta }^{\ast}(\omega )\frac{d\beta}{2\pi }=\frac{\gamma}{2\pi }\int_{-\infty }^{\infty }b_{k}\epsilon_{\gamma k/2}\Phi_{\gamma k/2}^{\ast}(\omega )dk.
\end{equation}
We show that
this expression is equivalent to
\begin{equation}\label{eq:phi}
 \Phi (\omega ) =\frac{1+|\omega |^{\gamma }}{|\omega |^{\gamma /2}}f(\log |\omega |^{\gamma })
\end{equation}
where
\beq
 f(x)=\int_{-\infty}^{\infty }b_{k}e^{-ikx}dk.
\label{eq:phi_1}
\eeq
To prove this, we employ the following trick:  use
$$
\epsilon_{\beta }\Phi_{\beta }(\omega )=\frac{2}{1-\gamma }\int_{-\infty}^{\infty }\frac{\Phi_{\beta }(\Omega )d\Omega }{|\Omega |^{1-\gamma }|\Omega -\omega |^{\gamma }}
$$
and write
$$
\Phi (\omega )=\frac{1}{\pi}\frac{\gamma}{1-\gamma }\int_{-\infty}^{\infty }\frac{d\Omega}{|\Omega |^{1-\gamma }|\Omega -\omega |^{\gamma }}\int b_{k}\Phi_{\gamma k/2}^{\ast}(\Omega )dk
$$
or
\begin{equation}
 \Phi (\omega )=\frac{1}{\pi}\frac{\gamma}{1-\gamma }\int_{-\infty}^{\infty }\frac{d\Omega}{|\Omega |^{1-\gamma/2 }|\Omega -\omega |^{\gamma }}f(\log (|\Omega |^{\gamma }),\quad  \mbox{where}\quad f(x)=\int_{-\infty}^{\infty }b_{k}e^{-ikx}dk.
\nonumber
\end{equation}

Comparing this expression with Eq. (\ref{eq:eqAlpha}), we recover
 (\ref{eq:phi}).  Notice, that this proof does not use an explicit form of the function $b_{k}$.
It does, however, explicitly uses that fact that $\Phi (\omega )$ is the solution of \eqref{eq:eq1}.\footnote{There is another way to prove \eqref{eq:phi_1} which does not use the fact that $\Phi (\omega )$ is a solution of \eqref{eq:eq1}, instead it uses the analytical properties of the function $B_{k}$.}

 The gap function $\Delta (\omega)$ is expressed as
 \begin{equation}\label{eq:phi_2}
 \Delta (\omega ) =|\omega |^{\gamma /2} f(\log |\omega |^{\gamma })
\end{equation}

\subsection{Summary: The steps to obtain the pairing vertex $\Phi (\Omega )$ and the gap function $\Delta (\omega)$  for $E < \epsilon_0$.}\label{sec:phiOmega}

The strategy is the following:
\begin{enumerate}
\item
Evaluate the integral
\begin{equation}\label{eq:I1}
I(k)=\frac{1}{2}\int_{-\infty }^{\infty } \log |1-\frac{1}{E}\epsilon_{ \gamma k'/2}|\tanh \left(\pi (k'-k) \right)dk'
\end{equation}
Notice, that it is real and antisymmetric.
\item
Construct the function
\begin{equation}\label{eq:btilda}
b_k =\frac{\sinh (\pi k_{E})e^{-iI (k)}}{\sqrt{\cosh (\pi (k-k_{E})\cosh ( \pi (k+k_{E}))}}.
\end{equation}
\item
Compute
\begin{equation}\label{eq:fTildeB}
 f(x)=\int_{-\infty }^{\infty } b_k e^{-ikx}dk.
\end{equation}
\item
The exact solutions of the linearized equations for the
pairing vertex and the gap function  are
\bea\label{eq:phiFtilda}
&& \Phi_{\text{ex}} (\omega_m)=\left(1+|\omega |^{-\gamma }\right)  f (\log |\Omega |^{\gamma }) \nonumber \\
&& \Delta_{\text{ex}} (\omega_m)= |\omega |^{\gamma/2}  f (\log |\Omega |^{\gamma })
\eea
\end{enumerate}

\section{Asymptotic expansion at small and large frequencies}
\label{app:asymptotic}

To understand the structure of  $\Delta_{ex} (\omega_m)$ it is instructive to expand it at
 small and large $\omega_m$.  One can do this in two ways: expand in the formulas for the exact solution, or perturbatively expand the gap equation (\ref{eq:lineargap_1}) order by order in $\omega_m$ or $1/\omega_m$.

 For shortness, we present here the details of the direct perturbative expansion.
 The point of departure for the expansion in $\omega_m$ is the  solution of (\ref{eq:lineargap_1})
 without the last term $1/(1 + |\bo^{'}_m|^\gamma)$:  $\Phi (\omega_m) =$ Re$ \Phi_0 (\omega_m)$, where
 \beq
 \Phi_0 (\omega_m) = \frac{C^{<}_0}{\sqrt{|\bo_m|^\gamma}} e^{i \phi + \beta_N \log{|\bo_m|^\gamma}}
 \label{dd_2}
 \eeq
  and, we  remind, $\bo_m =\omega_m/\omega_0$ and  $\beta_N$ is the solution of $\epsilon (\beta) =N$, where $\epsilon (\beta)$ is given by (\ref{su_15_2}).

 Let us expand the r.h.s. of (\ref{eq:lineargap_1}) in $\omega_m$ and express $ \Phi (\omega_m)$ as Re$[\Phi_0 (\omega_m) + \delta \Phi (\omega_m)]$.
The equation for $\delta \Phi (\omega_m)$ is
 \bea
 && \delta \Phi (\omega_m) - \frac{1-\gamma}{2N} \int d \omega'_m \frac{\delta \Phi(\omega'_m)}{|\omega'_{m}| ^{1-\gamma} |\omega_m-\omega'_{m}|^\gamma} = \nonumber \\
 &&-\frac{1-\gamma}{2N \omega^\gamma_0}  \int d \omega'_m \frac{\Phi_0 (\omega'_m)}{|\omega'_{m}| ^{1-2\gamma}} \left(\frac{1}{|\omega_m-\omega'_{m}|^\gamma} - \frac{1}{|\omega'_{m}|^\gamma}\right) +
 \frac{1-\gamma}{2N \omega^\gamma_0}  \int d \omega'_m \frac{\Phi_0 (\omega'_m)}{|\omega'_{m}| ^{1-\gamma}}
 \label{dd_5}
 \eea
   There are two types of terms in  the r.h.s of (\ref{dd_5}).  The first term is local in the sense that the integral over $\omega_m'$ is ultra-violet and infra-red convergent and is determined by $\omega'_m$ comparable to $\omega_m$. Evaluating this term, we obtain
 \beq
-\frac{1-\gamma}{2\omega^\gamma_0 N }  \int d \omega'_m \frac{\Phi_0 (\omega'_m)}{|\omega'_{m}| ^{1-2\gamma}} \left(\frac{1}{|\omega_m-\omega'_{m}|^\gamma} - \frac{1}{|\omega'_{m}|^\gamma}\right) =
- |\bo|^\gamma \Phi_0 (\omega_m) I_1
\label{dd_6}
\eeq
 where
\begin{widetext}
 \bea
&& I_1 = \frac{1-\gamma}{2N} \int_{-\infty}^\infty dy y^{(3\gamma/2-1 + i{\beta_N}\gamma)} \left(\frac{1}{|1-y|^\gamma} -\frac{1}{|y|^\gamma}\right)
=
\frac{1-\gamma}{2N}
\frac{\Gamma(3\gamma/2 +i{\beta_N}\gamma) \Gamma(-\gamma/2 -i{\beta_N}\gamma)}{\Gamma(\gamma)}
+
\nonumber \\
 &&
\Gamma(1-\gamma)\frac{1-\gamma}{2N}  \left(\frac{\Gamma(3\gamma/2 +i{\beta_N}\gamma)}{\Gamma(1+\gamma/2 +i{\beta_N}\gamma)} + \frac{\Gamma(-\gamma/2 -i{\beta_N}\gamma)}{\Gamma(1-3\gamma/2 -i{\beta_N}\gamma)}\right)
\label{dd_7}
\eea
\end{widetext}
and the integration variable $y = \omega'_m/\omega_m$.

  \subsection{Local corrections.}

  Let us  momentarily keep only the local term in the r.h.s. of (\ref{dd_5}).  We label the corresponding portion of  $\delta \Phi (\omega_m)$ as $\delta \Phi_L (\omega_m)$.
    The functional form of $\delta \Phi_L (\omega_m)$ is determined by (\ref{dd_6}):
    $\delta \Phi_L (\omega_m) = |\bo|^\gamma \Phi_0 (\omega_m) C_1$. Substituting into (\ref{dd_5}), we find the relation between $C_1$ and $I_1$:
 \beq
 C^{<}_1 = C^{<}_0 \left(\frac{I_1}{I_1-1}\right)
 \label{dd_8}
 \eeq
 This analysis can be straightforwardly extended to higher orders in the expansion in $|\bo|^\gamma$. At each other we extract infra-red convergent contribution in the source term by subtracting from $1/|\omega'_m -\omega_m|^\gamma$  a proper number of terms ( $1/|\omega'_m|^\gamma$,
 $\omega^2_m/|\omega'_m|^{\gamma+2}$, etc) to make the remaining integral over $\omega'_m$ ultra-violet  convergent, and solve for $\delta \Phi_L (\omega_m)$ induced by such source terms.
  The result is
 \beq
 \delta \Phi_L (z) = \Phi_0 (z)~ \sum_{n=1}^\infty C^{<}_n ~z^{n}
 \label{dd_9}
 \eeq
  where, we remind, $z= |\bo_m|^\gamma$, and
  \beq
  C^{<}_n = C^{<}_0~ \left[I_n  \displaystyle\prod_{m=1}^{n} \frac{1}{I_m-1}\right]
  \label{dd_10_a}
  \eeq
  where
 \begin{widetext}
 \bea
&&I_m = \frac{1-\gamma}{2N} Q(m, \gamma, \beta_N),  \nonumber \\
&&  Q(m, \gamma, \beta_N)  =  \frac{\Gamma((m+1/2)\gamma +i{\beta_N} \gamma) \Gamma((1/2-m)\gamma -i{\beta_N} \gamma)}{\Gamma(\gamma)} +  \nonumber \\
&&\Gamma(1-\gamma) \left(\frac{\Gamma((m+1/2)\gamma +i{\beta_N} \gamma)}{\Gamma(1-(1/2-m)\gamma +i{\beta_N} \gamma)} + \frac{\Gamma((1/2-m)\gamma -i{\beta_N} \gamma)}{\Gamma(1-(m+1/2)\gamma -i{\beta_N} \gamma)}\right)
\label{dd_7aa}
\eea
\end{widetext}
  Note than $C^{<}_n$ are complex numbers.
  Combining $\Delta \Phi_L$ and $\Phi_0$,  we obtain the
   total local contribution to the pairing vertex $\Phi_L =$ Re $[\Phi_0+ \delta \Phi_L]$  as
   \beq
     \Phi_L (z) = {\text Re} \left[e^{i (\phi + {\beta_N} \log{z})} \sum_{n=0}^\infty C^{<}_n z^{n-1/2} \right]
   \label{dd_11}
   \eeq
  Expressing (\ref{dd_11}) as the equation for the local contribution to the gap function $\Delta_L (z) = z \Phi_L (z)/(1+z)$ , we obtain the first term in (\ref{eq:x<}).

  At small $\gamma$ and $N =O(1)$,   $\beta_{N} \approx  \frac{1}{\sqrt{N \gamma}} (1  -N \gamma/8) \gg 1$.
  Evaluating  $I_n$  in two orders in $1/\beta_N$, we find a simple expression
\beq
I_n = 1 + 2 i \frac{n}{\beta_N} -3 \left(\frac{n}{\beta_N}\right)^2
\label{dd_19}
\eeq
Substituting this $I_n$ into (\ref{dd_10_a}) and then into  (\ref{dd_9}), using the relation between $\Phi$ and $\Delta$, and  exponentiating the series to  order $z^3$, we obtain, up to corrections of order $z/\beta^2_N$,
 \beq
\Delta_L (z)  =  C^{<}_0 ~ \frac{z^{1/2}}{(1+ z)^{3/4}} \cos{\left(\phi + \beta_{N} Q_1 (z)\right)}
    \label{dd_20}
 \eeq
where  $Q_1(z) = \log{z} - z/2 +3 z^2/16 + ...$.
 Eq. (\ref{dd_20}) and the expansion of $Q_1 (z)$ coincide with those for $\Delta_{\text{diff}} (z)$, Eqs. (\ref{dd_1}) and (\ref{dd_1a}).   This explicitly confirms that for $\gamma \ll 1$, $\Delta (\omega_m) \approx \Delta_L (\omega_m) \approx
\Delta_{\text{diff}} (\omega_m)$.

 \subsection{Non-local corrections.}

 We now look at the other terms, which we had to add and subtract at each order to make local contributions ultra-violet convergent.
 At the leading order in the expansion the additional term in the r.h.s. of Eq. (\ref{dd_5}) is
 \beq
 \frac{1-\gamma}{2N \omega^\gamma_0}  \int d \omega'_m \frac{\Phi_0 (\omega'_m)}{|\omega'_{m}| ^{1-\gamma}} =  \frac{1-\gamma}{N\gamma} C^{<}_0 e^{i\phi} \int_0^\infty dx x^{i{\beta_N}-1/2}
\label{dd_12}
\eeq
   This integral is ultra-violet divergent, but the divergence is fictitious because the actual
    $\Phi (\omega_m)$ must drop as $1/|\omega_m|^\gamma$ at large
  $|\omega_m|$.  Accordingly, we cut ultra-violet divergencies at some  $|\omega_m| = \omega_{max}$, which, we argue below, is roughly where $x'$ changes sign.  The separation into local and non-local terms obviously holds only for $|\omega_m| < \omega_{max}$.  The non-local term (\ref{dd_12}) induces another set of corrections to the bare $\Phi (\omega_m) =$ Re$\Phi_0 (\omega_m)$. We label the corresponding non-local part of the full $\Phi (\omega_m)$ as
  $\Phi_{nL} (\omega_m)$. By construction, $\Phi_{nL} (\omega_m)$ is real.
  Substituting  (\ref{dd_12}) into (\ref{dd_5}), adding the complex conjugated term, and adding and subtracting $1/|\omega'_m|^\gamma$ to/from $1/|\omega_m - \omega'_m|^\gamma$,
    we  obtain the equation for $\Phi_{nL} (\omega_m)$ in the form
  \beq
 \Phi_{nL} (\omega_m) - \frac{1-\gamma}{2N} \int d \omega'_m \frac{ \Phi_{nL} (\omega'_m)}{|\omega'_{m}| ^{1-\gamma}} \left(\frac{1}{|\omega_m-\omega'_{m}|^\gamma} -\frac{1}{|\omega'_m|^\gamma}\right) = K
 \label{dd_12_1}
 \eeq
 where
 \beq
 K = \frac{1-\gamma}{N}  \displaystyle\int_0^{\omega_{max}} d\omega_m   \left( \frac{\Phi_{nL} (\omega_m)}{|\omega_m|^\gamma} -  \frac{{\text Re} \Phi_0 (\omega_m)}{\omega^\gamma_0} \right)
 \label{dd_14}
 \eeq
  We first notice that Eq. (\ref{dd_12_1}) without the $K$ term does have the solution $\Phi_{nL}  (\omega_m) = E^{<}_{0,0} |\omega_m|^{(b_0-\gamma/2)}$, where
 $b_0$ is the solution  of $\epsilon_{b_0/\gamma} = N$  in the interval $\left[\gamma/2, \gamma/2  + 2\right]$.  For $b_0$ within this range, the integral in the l.h.s. of (\ref{dd_12_1}) does not diverge in the  infra-red and ultra-violet limits.
  To satisfy (\ref{dd_12_1}) one needs to satisfy also the condition  $K=0$. Substituting our trial solution for
   $\Phi_{nL} (\omega_m)$  and Re$\Phi_0 (\omega_m)$ into (\ref{dd_14}), we find that the condition $K=0$ is satisfied for some particular ratio $E^{<}_{0,0}/C^{<}_0$.

 This analysis can be extended to higher orders in the perturbation theory. The non-local terms in the r.h.s. of the equation for the pairing vertex appear in the form $A_{10} + A_{11} |\bo_m|^\gamma  + A_{12} \bo_m^{2\gamma} + ... +
A_{20} \bo_m^{2} + A_{21} |\bo_m|^{2 + \gamma}  +  A_{22} |\bo_m|^{2 + 2\gamma} +... +  A_{30} \bo_m^{4} + ..$, where all $A_{ij}$ contain $C^{<}_0$ as the overall factor.
 To cancel all these non-local terms,  we choose $\Phi_{nL} (\omega_m) $ in the form
\beq
\Phi_{nL} (\omega_m) = \frac{1}{|\bo|^{\gamma/2}} \sum_{n,m=0}^\infty  D_{n,m}~  |\bo|^{b_m +\gamma n}
\label{dd_16}
\eeq
where $\gamma/2 + 2m < b_m < \gamma/2 + 2(m+1)$.  The terms with $m >0$ are solutions of the same equation (\ref{dd_14}), but
 with the proper number of terms subtracted from $1/|\omega_m - \omega'_m|^\gamma$ and added to the $K$ term in the r.h.s, to make the integral over $\omega'_m$ in the l.h.s. ultra-violet convergent.
   We see from Fig. \ref{fig:epsilon}
   that there is one $b_n$ in each interval that satisfies (\ref{dd_14}).  Substituting this form into the equation for the pairing vertex and collecting non-local terms, we find a set of coupled algebraic equations  for $D_{n,m}$ with $A_{ij}$ playing the role of source terms.  The precise forms of these equations cannot be obtained within perturbative expansion in $\omega_m$, and we just have to assume that the solution  $D_{n,m}$ exists.
  Reexpressing (\ref{dd_16})  as the equation for  $\Delta_{nL}$ instead of $\Phi_{nL}$, we reproduce $D^{<}_{n,m}$ series in (\ref{eq:x<}).

  Combining (\ref{dd_11}) and (\ref{dd_16}) we see that the r.h.s. of  Eq. (\ref{eq:x<}) is the sum of local and non-local contributions
\beq
\Delta (z) = \Delta_L (z) + \Delta_{nL} (z).
\label{y_1}
\eeq

  At the largest $\omega_m$  one can expand in $1/z$ and use $\Delta_0 (z) = C^{>}_{0}/z$ as the zero-order solution.  The expansion is straightforward and yields Eq. (\ref{eq:x>}).

\bibliography{zero_T_gamma_less_1_final}

\end{document}